\documentclass[aps,twocolumn,prx,superscriptaddress,letterpaper]{revtex4-1}
\usepackage{amssymb}
\usepackage{mathtools}
\usepackage[pdftex]{graphicx}
\usepackage[usenames, dvipsnames, svgnames, table]{xcolor}
\usepackage[colorlinks=true, citecolor=Blue,
            linkcolor=BrickRed, urlcolor=ForestGreen]{hyperref}
\usepackage{txfonts}
\usepackage{mathrsfs}
\usepackage{bm}
\usepackage{multirow}

\newcommand{\sens}{strain$/\sqrt{\rm Hz}$ }

\begin{document}  

\title{Exploring the sensitivity of gravitational wave detectors to neutron star physics} 

\author{Denis Martynov}
\affiliation{School of Physics and Astronomy, and Institute of Gravitational Wave Astronomy, 
University of Birmingham, Edgbaston, Birmingham B15 2TT, United Kingdom}

\author{Haixing Miao}
\affiliation{School of Physics and Astronomy, and Institute of Gravitational Wave Astronomy, 
University of Birmingham, Edgbaston, Birmingham B15 2TT, United Kingdom}

\author{Huan Yang} 
\affiliation{Perimeter Institute for Theoretical Physics, Waterloo, ON N2L2Y5, Canada}
\affiliation{University of Guelph, Guelph, ON N2L3G1, Canada}

\author{Francisco Hernandez Vivanco}
\affiliation{School of Physics and Astronomy, Monash University, Vic 3800, Australia}
\affiliation{OzGrav: The ARC Centre of Excellence for Gravitational Wave Discovery, Clayton VIC 3800, Australia}

\author{Eric Thrane}
\affiliation{School of Physics and Astronomy, Monash University, Vic 3800, Australia}
\affiliation{OzGrav: The ARC Centre of Excellence for Gravitational Wave Discovery, Clayton VIC 3800, Australia}

\author{Rory Smith}
\affiliation{School of Physics and Astronomy, Monash University, Vic 3800, Australia}
\affiliation{OzGrav: The ARC Centre of Excellence for Gravitational Wave Discovery, Clayton VIC 3800, Australia}

\author{Paul Lasky}
\affiliation{School of Physics and Astronomy, Monash University, Vic 3800, Australia}
\affiliation{OzGrav: The ARC Centre of Excellence for Gravitational Wave Discovery, Clayton VIC 3800, Australia}

\author{William E.\ East}
\affiliation{Perimeter Institute for Theoretical Physics, Waterloo, ON N2L2Y5, Canada}

\author{Rana Adhikari}
\affiliation{LIGO, California Institute of Technology, Pasadena, CA 91125, USA}

\author{Andreas Bauswein}
\affiliation{GSI Helmholtzzentrum für Schwerionenforschung, Planckstraße 1, 64291 Darmstadt, Germany}
\affiliation{Heidelberg Institute for Theoretical Studies, Schloss-Wolfsbrunnenweg 35, 69118 Heidelberg, Germany}

\author{Aidan Brooks}
\affiliation{LIGO, California Institute of Technology, Pasadena, CA 91125, USA}

\author{Yanbei Chen}
\affiliation{LIGO, California Institute of Technology, Pasadena, CA 91125, USA}

\author{Thomas Corbitt}
\affiliation{Department of Physics \& Astronomy, Louisiana State University, Baton Rouge, USA}

\author{Andreas Freise}
\affiliation{School of Physics and Astronomy, and Institute of Gravitational Wave Astronomy, 
University of Birmingham, Edgbaston, Birmingham B15 2TT, United Kingdom}

\author{Hartmut Grote}
\affiliation{School of Physics and Astronomy, Cardiff University, Cardiff, United Kingdom}

\author{Yuri Levin}
\affiliation{Physics Department and Columbia Astrophysics Laboratory, Columbia University, New York, NY 10027}
\affiliation{Center for Computational Astrophysics, Flatiron Institute, New York, NY 10010}

\author{Chunnong Zhao}
\affiliation{ARC Centre of Excellence for Gravitational Wave Discovery, The University of Western Australia, Perth, Australia}
\affiliation{OzGrav: The ARC Centre of Excellence for Gravitational Wave Discovery, Clayton VIC 3800, Australia}

\author{Alberto Vecchio}
\affiliation{School of Physics and Astronomy, and Institute for Gravitational Wave Astronomy, 
University of Birmingham, Edgbaston, Birmingham B15 2TT, United Kingdom
}
\begin{abstract}

The physics of neutron stars can be studied with gravitational waves emitted from coalescing binary systems. Tidal effects become significant during the last few orbits and can be visible in the gravitational-wave spectrum above 500\,Hz. After the merger, the neutron star remnant oscillates at frequencies above 1\,kHz and can collapse into a black hole. Gravitational-wave detectors with a sensitivity of $\simeq 10^{-24}$\,\sens at 2$-$4\,kHz can observe these oscillations from a source which is $\sim 100$\,Mpc away. The current observatories, such as LIGO and Virgo, are limited by shot noise at high frequencies and have a sensitivity of $\geq 2 \times 10^{-23}$\,\sens at 3\,kHz. In this paper, we propose an optical configuration of gravitational-wave detectors which can be set up in present facilities using the current interferometer topology. This scheme has a potential to reach $7 \times 10^{-25}$\,\sens at 2.5\,kHz without compromising the detector sensitivity to black hole binaries. We argue that the proposed instruments have a potential to detect similar amount of post-merger neutron star oscillations as the next generation detectors, such as Cosmic Explorer and Einstein Telescope. We also optimise the arm length of the future detectors for neutron star physics and find that the optimal arm length is $\approx 20$\,km. These instruments have the potential to observe neutron star post-merger oscillations at a rate of $\sim 30$ events  per year with a signal-to-noise ratio of 5 or more.

\end{abstract}

\maketitle 


\section{Introduction}
On August 17, 2017, the LIGO and Virgo gravitational-wave (GW) detectors observed the coalescence of a binary neutron star system \cite{LSC_GW170817}. This event triggered a remarkable follow-up observation of the post-merger electromagnetic radiation across the full spectrum~\cite{LSC_MULTI_MESSENGER_2017}. This first GW multi-messenger observation provided insight into astrophysics, dense matter, gravitation, and cosmology. In particular, combining this event with priors from the upper limit from the previous LIGO observing runs and radio pulsar surveys  
sets the astrophysical rate of binary neutron star mergers equal to $1540^{+3200}_{-1220}$\,Gpc$^{-3}$\,yr$^{-1}$. The observed event has also shown that binary neutron star mergers may be the progenitors of at least some short-hard gamma-ray bursts~\cite{LSC_Gamma_ray_2017} and an important site for rapid neutron-capture nucleosynthesis in the Universe (see, for example,~\cite{Tanvir_2017}). In addition, the detected signal set constraints on the tidal deformability of neutron stars and provided an independent measurement~\cite{LSC_GW170817, LSC_EoS:2018} of the Hubble constant with a precision of 10\,\%~\cite{LSC_Hubble_2017}.

Binary neutron star mergers may result in a promptly forming black hole or a short- or long-lived neutron star remnant which emits GW above 1\,kHz after the merger.
The remnant of the detected binary neutron star event is unknown due to the diminished response of the LIGO and Virgo instruments at high frequencies~\cite{LSC_aLIGO_2015, Martynov_Noise_2016}. These detectors are optimised to increase the observatory reach and are only sensitive to the post-merger oscillations of neutron stars which are closer than $\simeq 10$ Mpc. These oscillations contain crucial information about the neutron star equation of state, and the structure of the post-merger remnant~\cite{Bauswein2012, Bauswein2012a, Hotokezaka2013a, Takami2014, Bauswein2015}. A measured GW waveform of the binary neutron star coalescence also allows an independent determination of the Hubble constant without an electromagnetic counterpart~\cite{Messenger_2012}. More broadly, high sensitivity above the kHz band may also improve tests of General Relativity at shorter length scales by observing the ringdown phase of solar-mass binary black hole mergers~\cite{Berti:2018vdi}.

The sensitivity of the current GW detectors above 1\,kHz is determined by the sum of the quantum shot noise ($2 \times 10^{-23}$\,strain$/\sqrt{\rm Hz}$)~\cite{Martynov_Noise_2016}, and classical noises ($5 \times 10^{-25}$\,\sens)~\cite{LSC_aLIGO_2015}, such as coating thermal noise~\cite{Harry_2007} and gas phase noise~\cite{Zucker_GAS_1996}. The gap between the quantum and classical noises can be reduced by (i) optimising the configuration of the interferometer to high frequencies, (ii) increasing the input power, (iii) injecting squeezed states of light~\cite{LSC_SQUEEZING_2013} and (iv) increasing the gain-bandwidth product of the interferometer~\cite{Miao_kHz_2018}. In this paper, we focus on options (i)-(iii) since technology for (iv) is not developed yet. We propose to close the quantum-classical gap by tuning the coupled cavity resonance between the arm and the signal recycling cavities~\cite{Thuring_SRC_2007, Thuering_SRC_2009, Graf_SRC_2013, Miao_SRC_2014} to 2.5\,kHz, increasing the arm powers, and using squeezed state of light. In Sec.~\ref{4km} and Appendix~\ref{high_power}, we identify technical difficulties associated with this scheme and outline the technology which should be developed to detect neutron star post-merger oscillations. We also construct a cost function to optimise detectors for neutron star physics.

In Sec.~\ref{arm_length}, we discuss three high-frequency detector designs in the order of their complexity. The first detector (LIGO-HF) can be built in the current LIGO facilities and keeps the current LIGO test masses. The second detector (12\,km-HF) adds folding to the arm cavities and requires larger test masses. These detectors can also be implemented in the current LIGO facilities. The third detector is 20\,km long and requires a new facility. This length is a result of the optimisation of the cost function from Sec.~\ref{4km}. We find that the sensitivity of detectors improves at high frequencies for arm lengths below $\simeq 18$\,km. Above this length, the free spectral range of the interferometers significantly reduces the response of the observatories to the sources at particular sky locations~\cite{Essick_2017}, and therefore the overall sensitivity at high frequencies.

In Sec~\ref{science_case} and Sec~\ref{pre_merger}, we discuss post- and pre-merger neutron star physics. We find that the current 4\,km facilities have a potential to detect neutron star post-merger oscillations with similar signal-to-noise ratio as proposed third-generation detectors, such as Einstein Telescope~\cite{Punturo_ET_2010} and Cosmic Explorer~\cite{LSC_FUTURE_2017}. Therefore, these facilities will have a long-term impact as neutron star observatories in the era of the next generation detectors.
We also consider distinguishability of different equations of state, measurement of neutron star tidal deformability and Hubble constant, and tidal disruption in black hole-neutron star binary systems. Additionally, trying to extend the scope of possible science cases,  in Appendices\ref{stochastic_background}, ~\ref{superradiant}, and~\ref{bbh_spectroscopy}, 
we also consider the sensitivity of the proposed detectors to the stochastic
background, superradiant instability of ultralight bosons, and mode spectroscopy for binary black holes. 


\section{General issues}
\label{4km}

\begin{figure*}[tb]
\includegraphics[width=0.85\columnwidth]{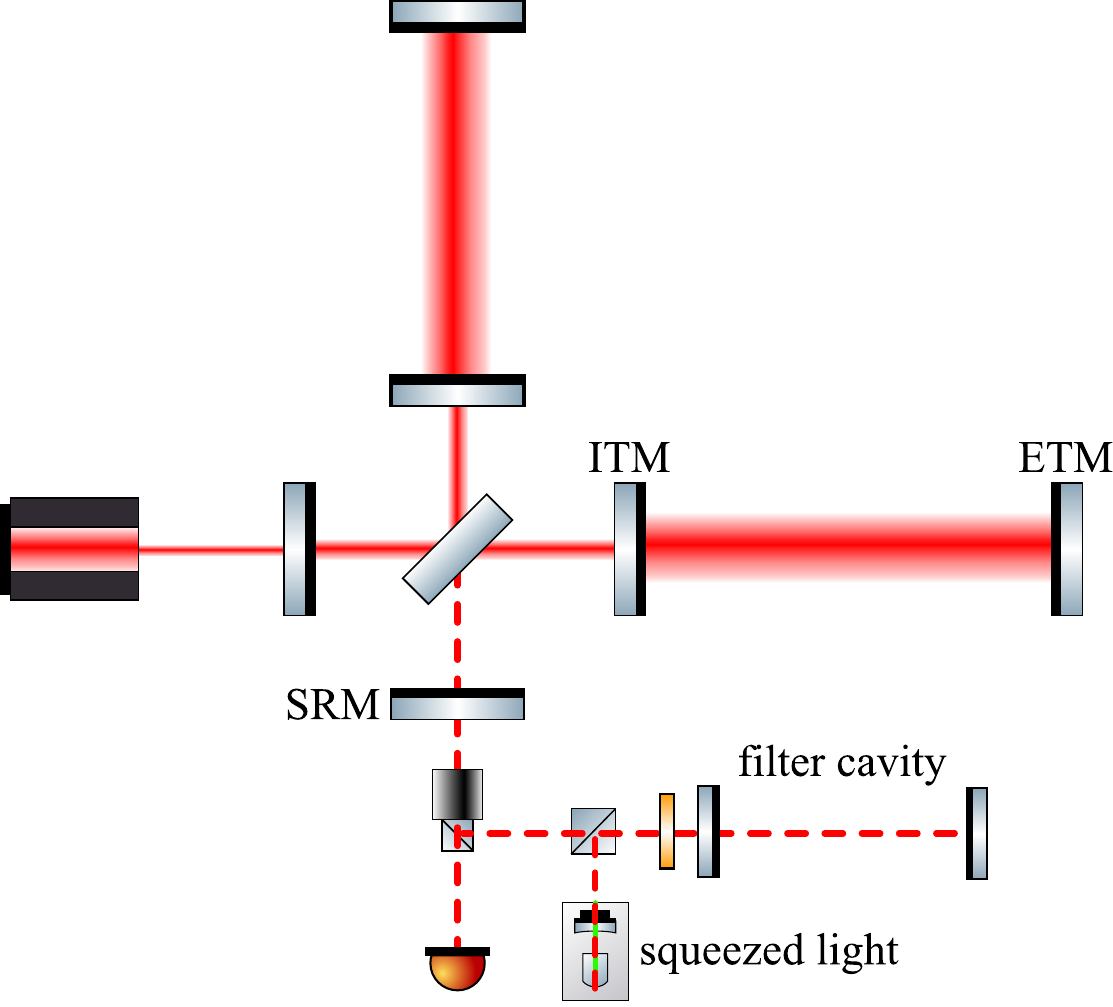}
\quad\quad
\includegraphics[width=\columnwidth]{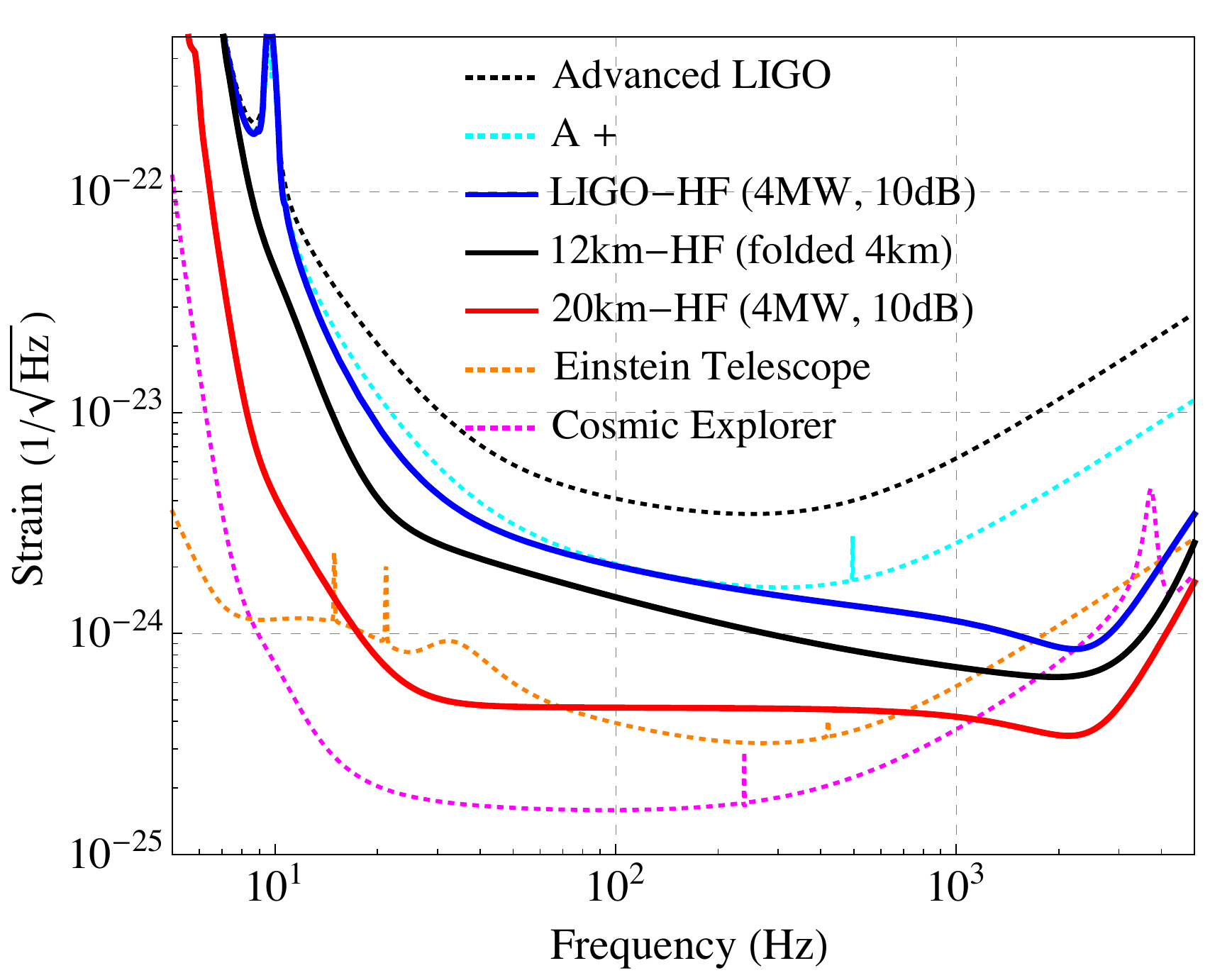}
\caption{Schematics showing the detector design  (left) and the resulting 
sensitivity curves (right). Sensitivities of Advanced LIGO design, Advanced LIGO plus (A+)\,\cite{LSC_White_2014}, 
 Einstein Telescope\,\cite{Punturo_ET_2010}  
and Cosmic Explorer\,\cite{LSC_FUTURE_2017} are also shown as references. 
The LIGO Voyager\,\cite{Adhikari_VOYAGER_2014} sensitivity is is not shown here since this upgrade focuses on the improvement around 100\,Hz.}
\label{fig:config}
\end{figure*} 

\begin{table*}
\caption{Summary of the key parameters\label{tab:App_sum}}
\label{table_parameters}
\begin{ruledtabular}
\begin{tabular}{lcccc}
Parameter & LIGO & LIGO-HF & 12\,km-HF & 20\,km-HF \\
\hline
Mirror mass & 40\,kg & 40\,kg & 87\,kg & 100\,kg\\
Arm gain & 270 & 270 & 100& 100 \\ 
Power recycling gain & 50 &  60 & 160 & 160 \\
Signal recycling mirror transmissivity & 0.32 & 0.030 & 0.04 & 0.016 \\
Signal recycling length & 56\,m & 356 \,m & 200\,m &100\,m\\
Coupled cavity resonance, $\omega_s/2\pi$ & 6.1\,kHz &  2.5\,kHz & 3.1\,kHz & 3.4\,kHz \\
Coupled resonance bandwidth, $\gamma_s/2\pi$ & 68\,kHz & 1.0\,kHz & 2.4\,kHz & 1.9\,kHz \\
Arm cavity bandwidth & 45\,Hz & 45\,Hz & 40\,Hz & 24\,Hz\\
Input power &125\,W &  500\,W  & 1500\,W & 670\,W\\
Power on beam splitter & 6.2\,kW & 30\,kW & 80\,kW & 80\,kW \\
Arm power & 0.8\,MW & 4.0\,MW & 4\,MW & 4.0\,MW \\
Squeezing level (observed) & --- & 10\,dB & 10\,dB & 10\,dB \\
Filter cavity (bandwidth=detuning) & --- &  28\,Hz & 13.5\,Hz & 7.9\,Hz\\
Static loss in the signal recycling cavity, $\epsilon_{\rm st}$ & 500\,ppm & 200\,ppm & 100\,ppm & 100\,ppm\\
Suppression of ITM distortion, $\kappa_{\rm itm}$ & 30 & 70 & 70 & 70 \\
Suppression of BS distortion, $\kappa_{\rm bs}$ &  1&  3 & 10 & 10\\
Heating loss on the input test masses, $\epsilon_{\rm itm}$ & 1000\,ppm & 735\,ppm & 735\,ppm & 735\,ppm \\
Coating absorption, $\alpha_{x,y}$ & 0.5\,ppm & 0.25\,ppm & 0.25\,ppm &0.25\,ppm \\
Beam splitter absorption, $\alpha_{\rm bs}$ & 1\,ppm & 1\,ppm & 1\,ppm & 1\,ppm \\
Heating loss on the beam splitter, $\epsilon_{\rm bs}$ & 225\,ppm & 694\,ppm & 444\,ppm & 444\,ppm
\end{tabular}
\end{ruledtabular}
\end{table*}

In this section, we discuss the topology of the proposed high-frequency detectors, thermal lenses in the mirrors and quantum shot noise. We also define the figure of merit for the sensitivity optimisation, of which the result is described in Sec.~\ref{arm_length}. Throughout this paper, we apply the current room temperature LIGO technology for the proposed high-frequency detectors: 1064\,nm lasers and fused silica mirrors with Ta$_2$O$_5$/SiO$_2$ optical coatings. The techniques for improving 
high-frequency sensitivity described here can be equally applied to the proposed 
cryogenic upgrade of LIGO---the LIGO Voyager\,\cite{Adhikari_VOYAGER_2014}, which assumes different laser wavelength, test mass and coating material. 

\subsection{Optical layout}
The layout of the proposed detector, shown in Fig.~\ref{fig:config}, is similar to the one of the Advanced LIGO detectors. It consists of two 4\,km-long perpendicular arm cavities, a Michelson interferometer, and power and signal recycling cavities. The GW signal is extracted from the difference between the two arms. The common arm length is a frequency reference for the laser. The Michelson interferometer splits and recombines the laser beam from the two arms. The power recycling cavity passively filters laser noise and the signal recycling cavity shapes the response of the detector. The parameters of the current and proposed detectors are summarised in Table~\ref{table_parameters}.

The key feature of the proposed layout is an optical resonance arising from the coupling between the signal recycling and arm cavities. We tune parameters of the detector to achieve the frequency and bandwidth of the resonance equal to $\approx 2.5$\,kHz and $\approx 1.5$\,kHz. This resonance enhances the response of the interferometer to GW from neutron star oscillations. The frequency $\omega_s$ and bandwidth $\gamma$ of the resonance are given by the equations
\begin{equation}\label{omegas_gamma}
\omega_s = \frac{c \sqrt{T_{\rm ITM}}}{2\sqrt{L_{\rm arm} L_{\rm src}}}\hspace{1cm}
\gamma = \frac{c\, T_{\rm SRM}}{4 L_{\rm src}}\,,  
\end{equation}
where $T_{\rm ITM}$ and $T_{\rm SRM}$ are the power transmissivity of input test mass (ITM) and signal recycling mirror (SRM), 
$L_{\rm arm}$ is the arm cavity length, and $L_{\rm src}$ is the length of signal recycling cavity.

Previous studies proposed to run the interferometer with the detuned signal recycling~\cite{Hild_SRC_2007}. Such an interferometer has an improved high-frequency sensitivity, but is difficult to control in practice~\cite{Ward_THESIS_2010}. The detuned design also requires an additional filter cavity~\cite{Kimble_FILCAV_2001} for squeezed states of light since upper and lower signal sidebands acquire different phases in the interferometer.
In the detector proposed in this paper (LIGO-HF), the interferometer control is similar to the current Advanced LIGO scheme~\cite{Martynov_THESIS_2015, Staley_LOCK_2014} and no filter cavities are required in addition to the one for reducing the low-frequency radiation pressure noise. Eq.~\eqref{omegas_gamma} sets two constraints on the three  parameters: $L_{\rm src}$, $T_{\rm SRM}$ and $T_{\rm ITM}$. One more constraint is set by the quantum noise caused by the optical loss in the signal recycling cavity. This loss arises from thermal lenses in the input test masses and is discussed in Sec.~\ref{thermal_heating} and Sec.~\ref{quantum_noise}.

\subsection{High power effects}
\label{thermal_heating}
By design, the Advanced LIGO detectors resonate $\simeq 0.8$\,MW of optical power in the arm cavities. We propose to further increase this number to improve the response of the detector at high frequencies. However, high power operation triggers a number of technical challenges such as (i) thermal lenses in the mirror substrates, (ii) angular instabilities in the arm cavities, and (iii) parametric instabilities in the arm cavities. We discuss (ii) and (iii) in the Appendix~\ref{high_power} since these problems complicate the interferometer control but do not influence the detector sensitivity at high frequencies. Item (i) is a more significant challenge since vacuum fluctuations couple to the instrument through optical losses in the mirror substrates as discussed in Sec.~\ref{quantum_noise}.

Thermal lenses arise from temperature gradients inside the input test masses. These gradients are created by the absorption of a small fraction ($<1$\,ppm) of the resonating laser power in the coatings and substrates of the mirrors. 
In Advanced LIGO, thermal lenses are suppressed by a factor of $\kappa_{\rm itm}=30$ by heating the mirrors near their edges using ring heaters and CO$_2$ lasers with a special beam profile~\cite{Zhao_TCS_2006, Brooks_TCS_2016}.
Unsuppressed thermal gradients lead to the wavefront distortion and effective scattering of the fundamental mode into higher-order optical modes.
This process introduces an additional optical loss in the signal recycling cavity (SRC). The total loss $\epsilon_{\rm src}$ is given by
\begin{equation}
\epsilon_{\rm src} = \epsilon_{\rm static} + \frac{\epsilon_{\rm itmx} + \epsilon_{\rm itmy} + \epsilon_{\rm bs}}{2}\,,
\end{equation}
where $\epsilon_{\rm static}$ is a power-independent loss in the cavity due to scattering from the coatings, $\epsilon_{\rm itmx,y}$ and $\epsilon_{\rm bs}$ are losses due to the wavefront distortion in the two input test masses (ITM) and the beam splitter (BS). These terms can be approximated by (cf. Ref.\,\cite{Brooks_TCS_2016})
\begin{equation}
\label{eq:thermal_loss}
\begin{split}
	\epsilon_{\rm itmx,y} &\approx 1000\,{\rm ppm} \times \left(\frac{P_{\rm arm}}{1\,{\rm MW}}
	\frac{\alpha_{x,y}}{0.5\,{\rm ppm}} \frac{30}{\kappa_{\rm itm}} \right)^2\,, \\
	\epsilon_{\rm bs} &\approx 250\,{\rm ppm} \times \left(\frac{P_{\rm bs}}{6\,{\rm kW}}
	\frac{\alpha_{\rm bs}}{1\,{\rm ppm}}\frac{1}{\kappa_{\rm bs}}\right)^2\,,
\end{split}
\end{equation}
where $\alpha_{x,y}$ is an absorption coefficient of the coating of the x- and y- test masses, $\alpha_{\rm bs}$ is total absorption coefficient of the beam splitter and $\kappa_{\rm bs}$ is a suppression factor of the beam splitter wave-front distortion. In these equations we neglect absorption in the substrates of the input test masses since this power is significantly smaller than $\alpha_{x, y}P_{\rm arm}$~\cite{LSC_aLIGO_2015}.

In this paper, we consider only the uniform absorption of the laser beam by the optical coatings. Point defects on the mirror surfaces can lead to the non-uniform absorption pattern of the laser light and significantly increase optical losses in the power and signal recycling cavities. Compensating for the non-uniform absorption will require an advanced design of the thermal compensation system that allows correction of the higher order spatial distortions.

\subsection{Quantum noise and optical loss}
\label{quantum_noise}

Quantum shot noise is caused by the vacuum fields which enter the interferometer through
the antisymmetric port and optical losses in the interferometer. Phase quadrature fluctuations of the former vacuum field can be suppressed by using non-classical squeezed state of light~\cite{LSC_SQUEEZING_2013, LSC_SQUEEZING_2011, Grote_SQUEEZING_2013}.
The spectral density of the shot noise from the squeezed vacuum field at frequencies smaller than the free spectral range of the arm cavity 
can be approximated as 
\begin{equation}
\label{eq:shot}
S_{hh}^{\rm AS}(\Omega)= \frac{ \hbar\,c  
[\gamma^2 \Omega^2+(\Omega^2-\omega_s^2)^2]}
{4 L_{\rm arm}  \omega_0 P_{\rm arm} \gamma \, \omega_s^2 } e^{-2r_{\rm sqz}},
\end{equation}
where $\hbar$ is the reduced Planck constant, $r_{\rm sqz}$ is the squeezing factor of the injected squeezed state of light~\cite{Caves_1981}. Similar to the A+ upgrade~\cite{LSC_FUTURE_2017}, the proposed detector will use squeezed state of light with one 300\,m-long filter cavity for creating the frequency-dependent 
squeezing to reduce the low-frequency radiation pressure noise.
At high frequencies, we assume that phase fluctuations of the vacuum field are suppressed by 10\,dB ($e^{-2r_{\rm sqz}}=0.1$). At $\Omega/2\pi=\omega_s/2\pi = 2.5$\,kHz we have
\begin{equation}
\label{eq_as_shot_value}
\sqrt{S_{hh}^{\rm AS}(\omega_s)} \approx 6.1 \times 10^{-25} \sqrt{ \frac{4\, {\rm km}}{L_{\rm arm}} \frac{4\, {\rm MW}}{P_{\rm arm}} \frac{e^{-2r_{\rm sqz}}}{0.1}}\,\frac{\rm strain}{\sqrt{\rm Hz}},
\end{equation}
where we choose the bandwidth of the couple cavity resonance equal to 1.6\,kHz.

Vacuum fields also enter the interferometer through optical losses in the mirrors, in particular in the arm cavities and in the signal recycling cavity~\cite{Miao_LOSS_2018}. The former noise is not significant in our design, 
and the spectral density of the shot noise caused by the latter is
\begin{equation}
\label{eq_src_loss_shot}
S_{hh}^{\rm src}(\Omega)=\frac{\hbar
(\gamma_{\rm arm}^2+\Omega ^2)\epsilon_{\rm src}}{\omega_0 P_{\rm arm} T_{\rm ITM}}\,,
\end{equation}
where $\gamma_{\rm arm}=c\,T_{\rm ITM}/(4 L_{\rm arm})$ is the arm cavity bandwidth. Eq.~\eqref{eq_src_loss_shot} shows that shot noise from the loss in the signal recycling cavity is independent from the arm length $L_{\rm arm}$ at high frequencies ($\Omega \gg \gamma_{\rm arm}$) and grows with $\Omega$. At $\Omega/2\pi=\omega_s/2\pi = 2.5$\,kHz we have
\begin{equation}
\label{eq_src_shot_value}
\sqrt{S_{hh}^{\rm src}(\omega_s)} \approx 4.4 \times 10^{-25} \sqrt{ \frac{0.0148}{T_{\rm ITM}} \frac{4\, {\rm MW}}{P_{\rm arm}} \frac{\epsilon_{\rm src}}{10^{-3}}}\,\frac{\rm strain}{\sqrt{\rm Hz}}.
\end{equation}
Eqs.~\eqref{eq_as_shot_value} and \eqref{eq_src_shot_value} imply that quantum shot noise caused by the squeezed vacuum field from the antisymmetric port and vacuum field that couples through loss in the signal recycling cavity are comparable in the proposed detector.

\subsection{Sensitivity optimisation}
\label{cost_function}

We optimise the detector parameters by converting the sensitivity curve into a single figure of merit $X$ and by maximising it. Our approach is to (i) incorporate the antenna response $R$ to an astrophysical source located in a particular point on the sky with angular coordinates $(\theta, \phi)$~\cite{Essick_2017}, (ii) divide the interferometer noise spectrum $N(\Omega)$ by the antenna response function and get a sensitivity curve of the detector $\sqrt{S_{hh}(\Omega, \theta, \phi)} = N(\Omega) / R(\Omega, \theta, \phi)$ in units of strain/$\sqrt{\rm Hz}$, (iii) average $1 / S_{hh}$ over sky locations $(\theta, \phi)$, and (iv) average the resulting frequency dependent curve from 2\,kHz up to 4\,kHz. The resulting figure of merit $X$ is
\begin{equation}\label{eq:cost_length}
	X = \frac{1}{2\pi}\int_0^\pi d\theta \sin \theta \int_{0}^{2\pi} d\phi \sqrt{\int^{4\,{\rm kHz}}_{2\,{\rm kHz}}df \frac{h_0^2}{S_{hh}(f, \theta, \phi)}}\,,
\end{equation}
where $S_{hh}$ is the total noise in the GW channel, $h_0 = 10^{-25}$\,strain is a normalisation factor, $f = \Omega/2\pi$ is the frequency. The cost function does not include inclination angles of the sources since they change only the normalisation factor $h_0$. We average $1/S_{hh}$ since this quantity is proportional to the SNR of the detected signal. The chosen frequency band follows from the uncertainty in the peak frequency $f_{\rm peak}$ of the post-merger oscillations of neutron stars. According to the recent constraints~\cite{LSC_GW170817}, we expect $f_{\rm peak} = 2.5-3.5$\,kHz for $1.35-1.35\,M_\odot$ binaries~\cite{Bauswein_NS_2018}. Since the quality factor of the dominant mode is predicted to be in the range from 10 to 30~\cite{Yang:2017xlf}, we increase the frequency band by 0.5\,kHz from each side. Broadening the frequency band is also motivated by the mass distribution of neutron stars~\cite{Kiziltan_NS_2013}.

\section{Different designs and optimisations} 
\label{arm_length}

In this section, we discuss the details of the three detector designs: LIGO-HF, 12km-HF, and 20km-HF, which 
are results of optimising the figure of merit $X$ mentioned earlier. The first two designs can also be applied to the 3-km facilities, such as Virgo~\cite{Acernese_aVIRGO_2015} and KAGRA~\cite{Somiya_Kagra_2012}.  

\subsection{LIGO-HF detector}

We start with a conservative upgrade of the LIGO detectors. This implies that we optimise parameters of the detector under constraints of the current infrastructure and optical configuration. In particular, we keep the arm build-up in the LIGO-HF detector the same as in the current LIGO detectors to recycle the test masses for the proposed upgrade. We also have a discrete set of choices of the signal recycling cavity length: 56\,m, 356\,m and 656\,m to consider the possibility of sharing the vacuum envelop of the 300-m filter cavity for A+. Eq.~\eqref{eq_src_shot_value} implies that a long signal recycling cavity improves the sensitivity of the detector. However, from Eq.~\eqref{omegas_gamma} we get $\omega_s/2\pi = 1.8$\,kHz for $L_{\rm src} = 656$\,m, and the dip in the sensitivity curve due to the coupled cavity resonance shifts outside of the optimisation window. Therefore, we choose $L_{\rm src} = 356$\,m and get $\omega_s/2\pi = 2.5$\,kHz.

\begin{figure}[!b]
\includegraphics[width=\columnwidth]{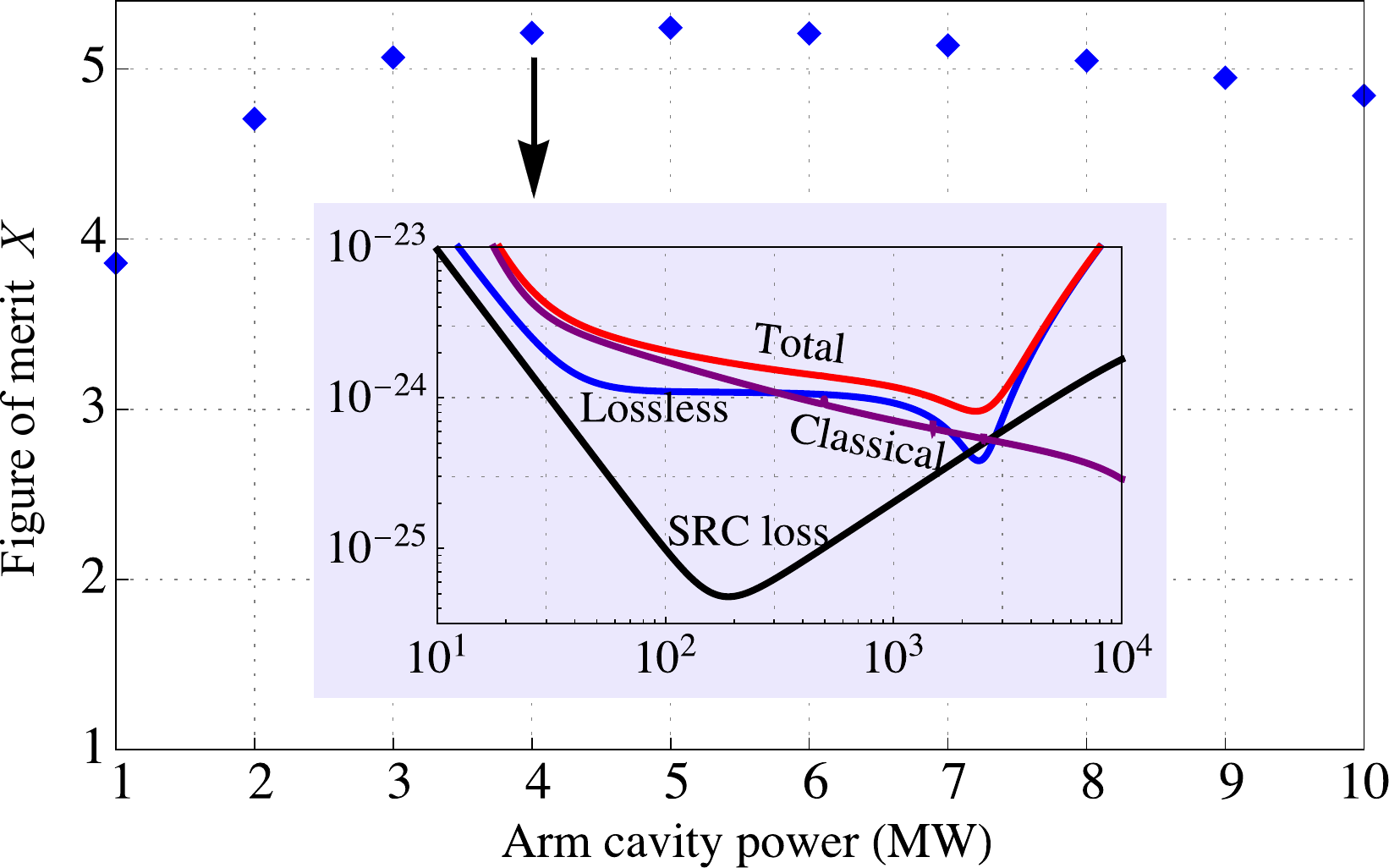}
\caption{The power dependence of the figure of merit $X$ defined in Eq.\,\eqref{eq:cost_length}. 
The inset shows the sensitivity curve for the arm cavity power being 4.0\,MW.}
\label{fig:power_scan}
\end{figure}

Next, we optimise the power resonating in the arms and finesse of the signal recycling cavity. Shot noise from the squeezed vacuum field given by Eq.~\eqref{eq:shot} is inversely proportional to $P_{\rm arm}$. However, shot noise given by Eq.~\eqref{eq_src_loss_shot} grows with $P_{\rm arm}$ since $\epsilon_{\rm src} \sim P_{\rm arm}^2$ according to Eq.~\eqref{eq:thermal_loss}. The optimal value depends on the cancellation factors of thermal gradients in the input test masses and the beam splitter, $\kappa_{\rm itmx,y}$ and $\kappa_{\rm bs}$. For the parameters shown in Table~\ref{table_parameters}, the optimal arm power is around $4-5$\,MW. Fig.~\ref{fig:power_scan} shows the value of the figure of merit for different arm powers.

The proposed detectors require more powerful lasers compared to the Advanced LIGO detectors. For a critically coupled interferometer the input power is given by the equation $P_{\rm in} = 2 Y P_{\rm arm} / (1-\eta)$, where $Y = 50$\,ppm is a round trip loss in the arms, and $\eta \approx 0.2$ is the power loss between the laser and interferometer input. For $P_{\rm arm} = 4$\,MW, we require an input interferometer power of 500\,W. Such power can be achieved by combining laser beams from four 150\,W lasers using, for example, Mach-Zehnder interferometers.

\begin{figure*}[tb]
\includegraphics[width=0.92\columnwidth]{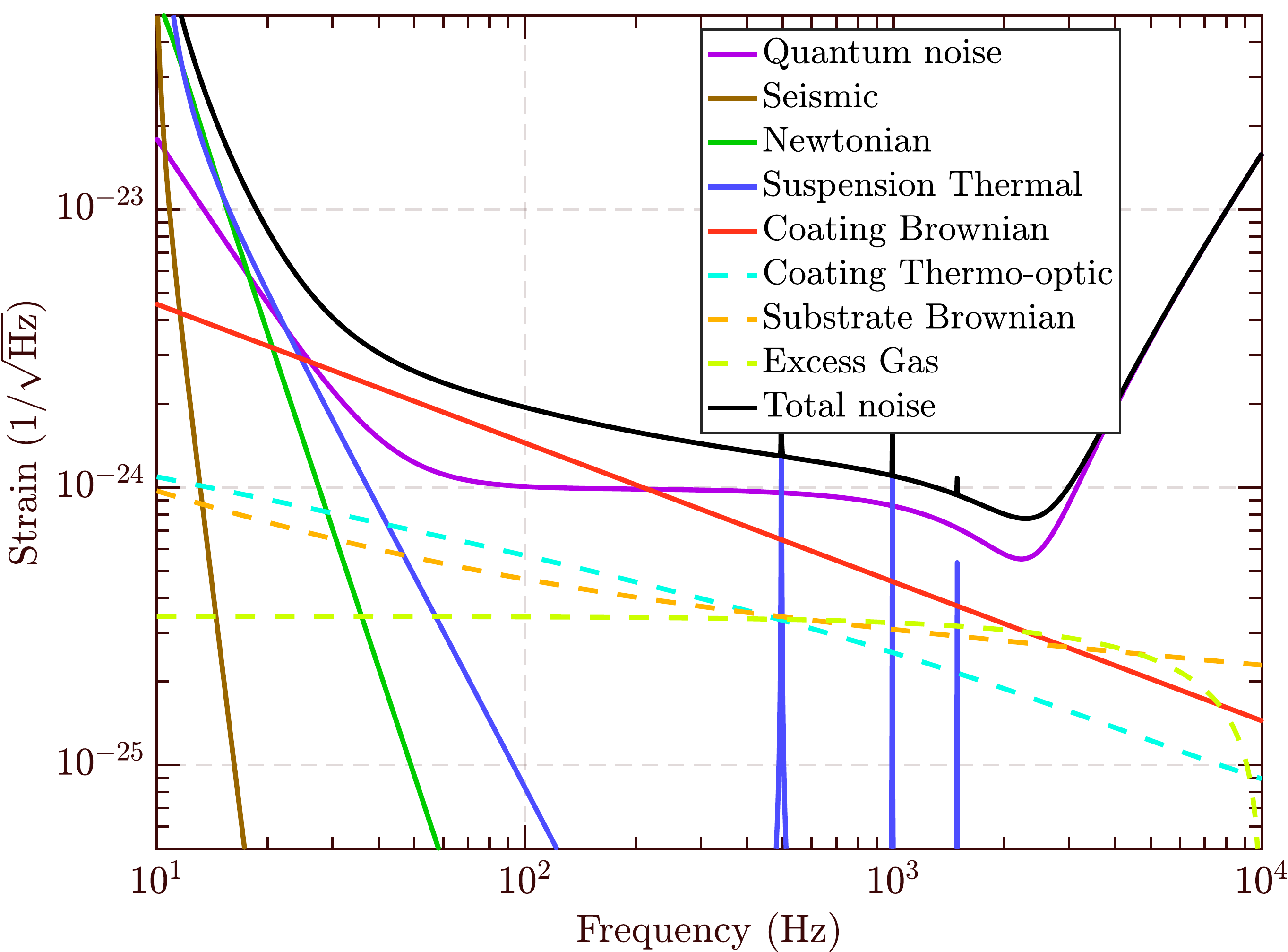}
\quad\quad
\includegraphics[width=0.92\columnwidth]{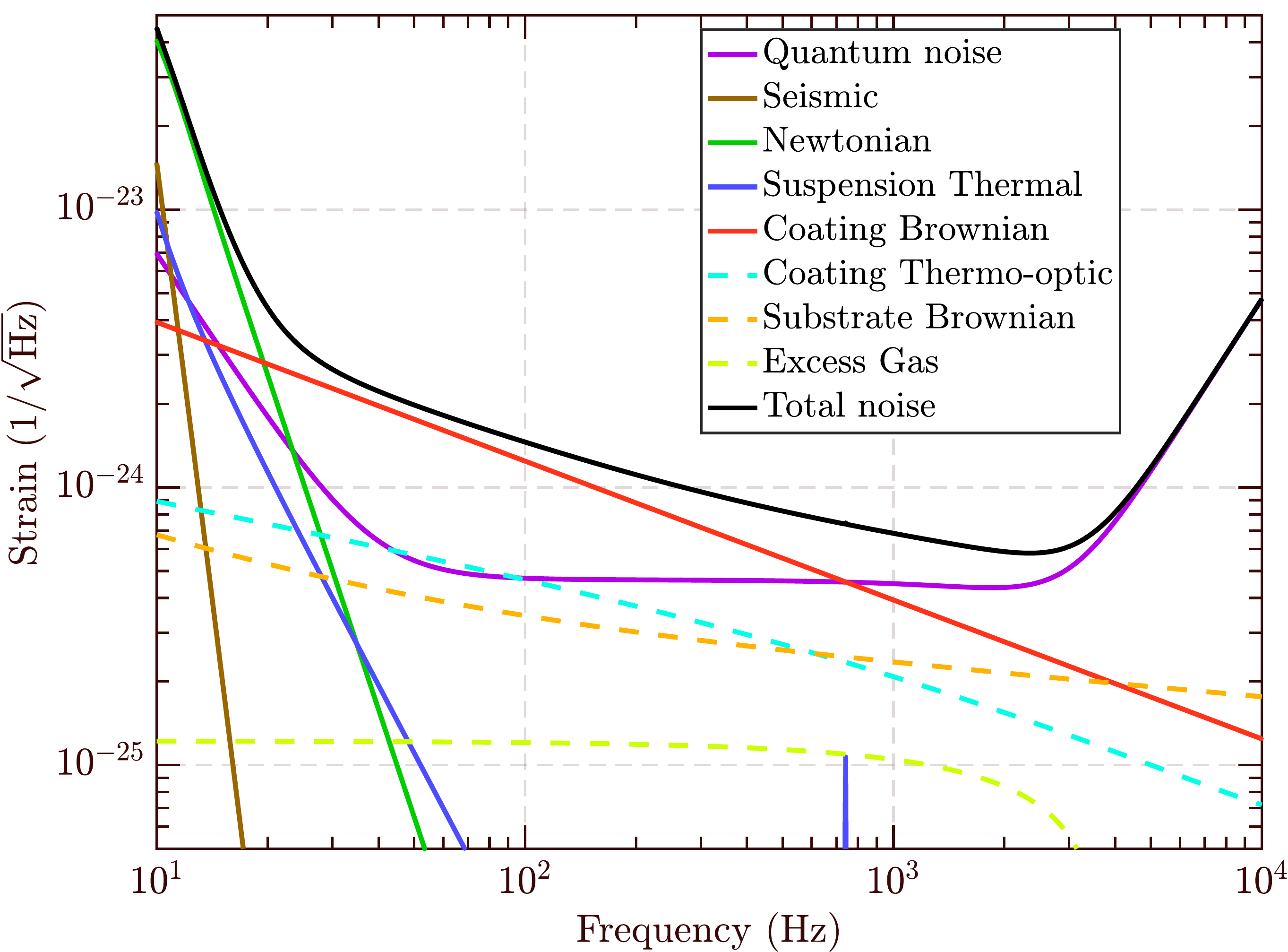}
\caption{Noise budget of the proposed upgrade to the LIGO detectors---LIGO-HF (left). The gap between quantum and classical noise is significantly reduced above 1\,kHz. In our design, sum of classical noises equals to the quantum noise in the frequency range from 1\,kHz up to 4\,kHz. Noise budget of the folded 4\,km detector--12km-HF (right). Folding the arm cavities can improve classical noises and reduce quantum shot noise. The resonance dip at high frequency is however
 smeared out due to the optical loss in the signal recycling cavity. }
\label{fig:4km_NB}
\end{figure*}

Our design focuses on improving the quantum noise of the detector. Classical noises in the LIGO-HF detector are similar to the ones in the A+ proposal~\cite{LSC_FUTURE_2017}. At high frequencies, the sensitivity is limited by shot noise, discussed in Sec~\ref{quantum_noise}, gas phase noise and thermal noises. Gas phase noise is induced by the stochastic transit of molecules through the laser beam in the arm cavities~\cite{Zucker_GAS_1996, Martynov_QUCORR_2017}. We calculate this noise assuming the residual pressure in the arms equal to 3\,nTorr and is dominated by hydrogen. Thermal noises come from the thermal heat flows and Brownian motion of atoms in the substrate~\cite{Harry_Thermal_2012} and coating of the mirrors~\cite{Harry_2007, Gras_CTN_2017}. The level of classical noises at 3\,kHz is $\approx 5 \times 10^{-25}$\,strain/$\sqrt{\rm Hz}$.
At low frequencies,  the sensitivity is limited by seismic noise, gravity-gradient (Newtonian) noise~\cite{Driggers_NN_2012} and suspension thermal noise~\cite{Gonzalez_SUS_2000}. Fig.~\ref{fig:4km_NB} shows the noise budget of the proposed LIGO-HF detector. Technical noises, such as actuator noise, scattered light, controls of auxiliary degrees of freedom, laser frequency, amplitude and pointing noises, also couple to the GW readout but are suppressed below the fundamental noises~\cite{Martynov_THESIS_2015, Martynov_Noise_2016}.

\subsection{12\,km-HF folded detector}

We can further increase the high-frequency sensitivity of the proposed detectors by introducing two folding mirrors in each arm cavity. The total arm lengths and optical loss in these cavities increases by a factor of 3. Therefore, we need to triple the input power to sustain the same power in the arms as in the case of linear cavities. However, thermal lenses and other high power effects are similar to the current LIGO design. Therefore, using folding we can achieve a higher sensitivity to neutron star post-merger oscillations while keeping the similar requirements to the thermal compensation system for the linear cavities. We optimise parameters of the folded detector to minimise shot noise at high frequencies according to the Eq.~\eqref{eq:cost_length}. In contrast to the previous section, we treat the arm gain as a free parameter. The only constraint we impose to the folded design is $\epsilon_{\rm bs} \leq G_{\rm arm} \times 5$\,ppm. It is required to minimise the arm imbalance and reduce the coupling of technical noises, such as laser frequency and intensity noises, to the GW channel. Fig.~\ref{fig:4km_NB} (right) shows that the resulting shot noise at 2.5\,kHz is around $5.0 \times 10^{-25}$\,\sens. This also implies that with folding we can achieve the sensitivity of $10^{-24}$\,\sens at kHz even with less than 1\,MW of arm power. 

The folded design also improves classical noises which limit the detector sensitivity at high frequencies, such as gas phase noise and thermal noises~\cite{Ballmer_FOLD_2013}. Gas phase noise improves by a factor of 3 due to folding since the beam size and effective arm lengths are increased. Thermal noises improve due to a longer arm length and larger beam sizes on the test masses. We assume g-factor of the arm cavities equal to 0.11 to simplify interferometer angular control. In this case, the amplitude spectral density of the coating thermal noise is given by
\begin{equation}
	\sqrt{S_{\rm ctn}} = 2.27 \times 10^{-25}\,\sqrt{\frac{3\,{\rm kHz}}{f}}\,\frac{\rm strain}{\sqrt{\rm Hz}}.
\end{equation}
This noise spectrum is a factor of 1.16 smaller compared to the coating thermal noise in the proposed A+ upgrade due to the longer arm length and larger beam sizes. This factor takes intro account amplification of the coating thermal noise on the folding mirrors~\cite{Heinert_FOLD_2014}. The factor of 1.16 can be further improved if we increase g-factor of the arm cavities up to 0.5. The noise budget of the folded design is shown in Fig~\ref{fig:4km_NB}. The peak sensitivity of this detector which can be implemented in the current facilities is $6.0 \times 10^{-25}$\,\sens. We note that the proposed folded detector requires test masses with radii of curvature equal to 30\,km. Similar to the Cosmic Explorer design~\cite{LSC_FUTURE_2017}, we propose to use flat mirrors and tune radii of curvature by a thermal compensation system~\cite{Brooks_TCS_2016}. The beam size is 6.4\,cm on the folded mirrors and 7.1\,cm on the input and end mirrors and the mirror radius is 22\,cm. This radius increases the mirror mass from the current value of 40\,kg up to 87\,kg.

\subsection{20\,km-HF detector}
\begin{figure*}[tb]
\includegraphics[width=0.9\columnwidth]{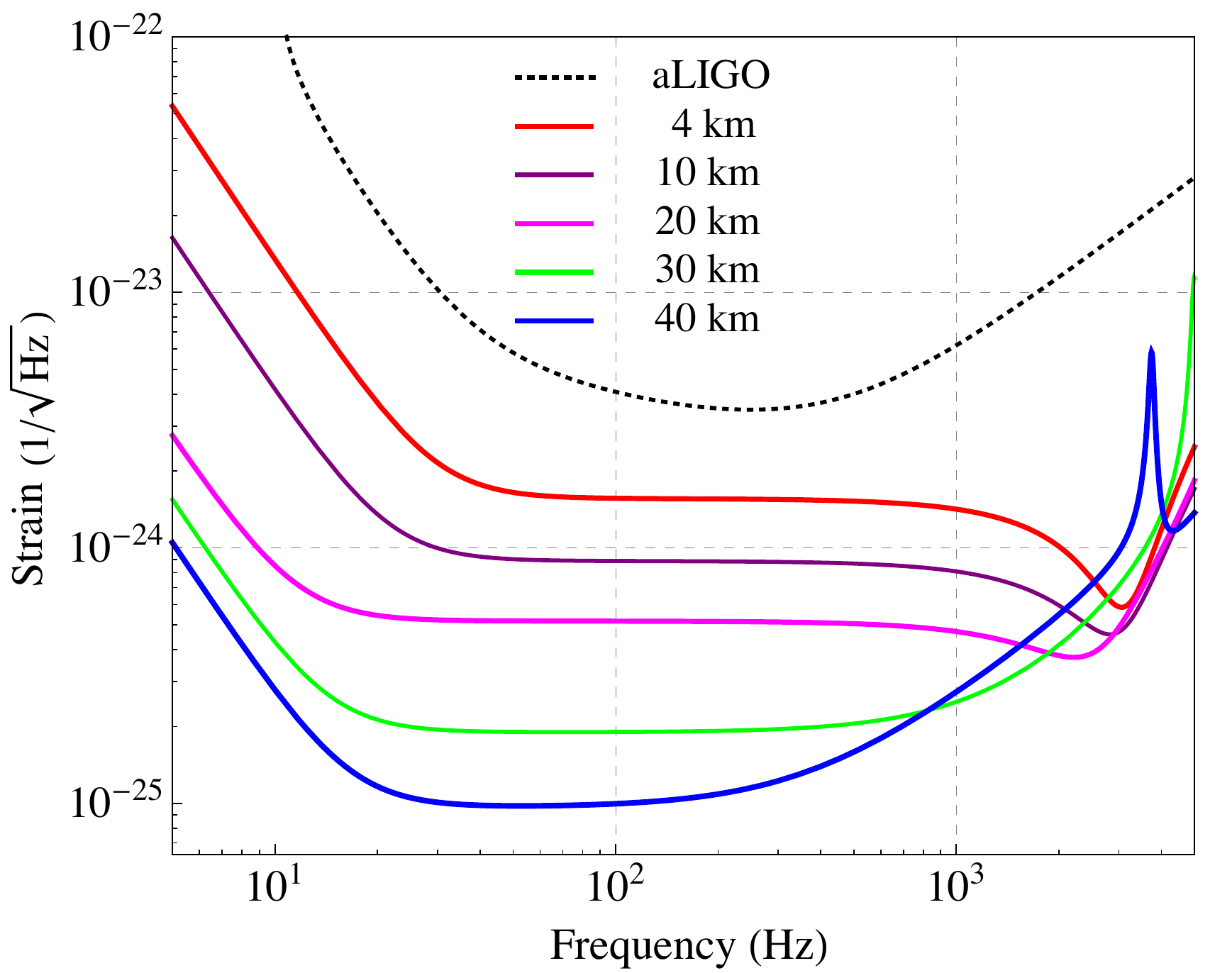}
\quad\quad
\includegraphics[width=0.88\columnwidth]{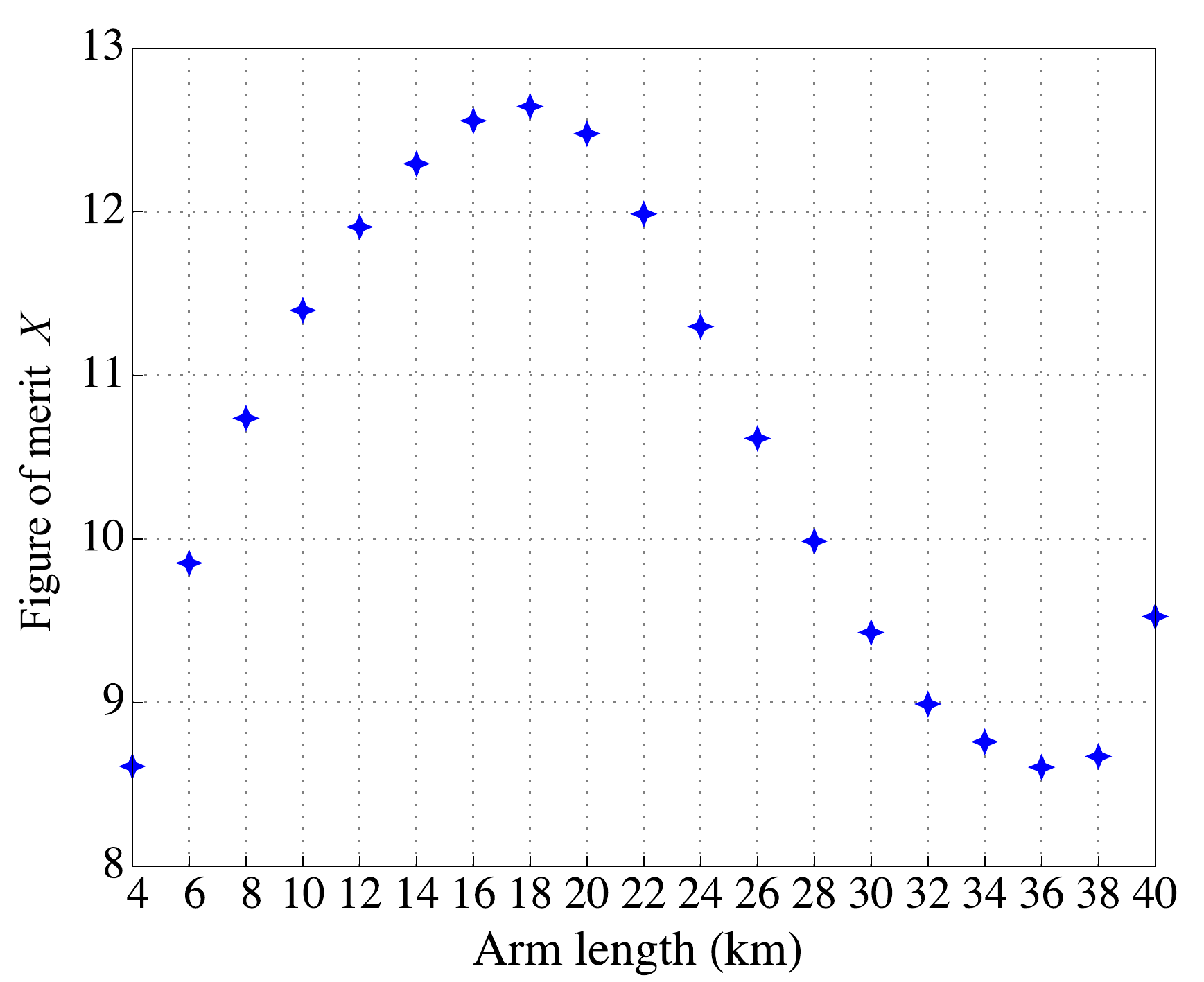}
\caption{Quantum noise curves for the optimised configurations with different 
arm lengths. We have included the antenna response of the normal incidence (left).  Value of the figure of merit for different arm lengths (right).}
\label{fig:arm_opt}
\end{figure*} 

\label{sens_length}
At low frequencies (below 100\,Hz), the sensitivity of the Advanced LIGO detectors is limited by two types of noise. The first type is related to the motion of the mirror centre of mass and includes ground vibrations, suspension thermal noises, gravity-gradient noises, scattered light noise, and actuation electronics. The contribution of these noises to the detector sensitivity scales as $1/L_{\rm arm}$. The second type of noise comes from the relative motion between the mirror centre of mass and the surface of the mirror probed by the beam. Noise of this type, such as coating and substrate thermal noise, scale in inverse proportion to the beam size, which increases with arm length as $1/\sqrt{L_{\rm arm}}$. Therefore, the contribution of the second type of noise to the GW channel in units of strain scales as $1/L_{\rm arm}^{3/2}$. Scaling factors of both types of noises imply higher sensitivity at low frequencies with longer arm lengths~\cite{LSC_FUTURE_2017}. However, this conclusion does not hold for the high-frequency sensitivity for the following reasons.

Firstly, a squeezed vacuum field from the antisymmetric port of the interferometer leads to the shot noise which is inversely proportional to $\sqrt{L_{\rm arm}}$ rather than $L_{\rm arm}$ in units of \sens cf. Eq.~\eqref{eq:shot}.
This result is a direct consequence of a tradeoff between the peak sensitivity and the
bandwidth of a shot-noise-limited interferometer. Secondly, losses in the signal recycling cavity are responsible for the shot noise which is independent of $L_{\rm arm}$, cf. Eq.~\eqref{eq_src_loss_shot}. Thirdly, the antenna response is suppressed above 1\,kHz for the detectors with arm lengths comparable to the GW wavelength at high frequencies~\cite{Essick:2017}. Finally, the resonant frequency of the coupled signal recycling and arm cavities is inversely proportional to the arm length, cf. Eq.~\eqref{omegas_gamma}. Since the shot noise increases as $\Omega^2$ at frequencies $\Omega \gg \omega_s$, it is important to keep $\omega_s/2\pi$ above $2$\,kHz by reducing the signal recycling cavity length compared with the LIGO-HF design.
However, since an optical telescope inside the signal recycling cavity~\cite{Arain_Recycling_2008}  reduces the beam size from $\simeq 10$\,cm down to $\simeq 1$\,mm, decreasing the signal recycling cavity length leads to optical losses. In particular, 
small defects in the curvature of the telescope mirrors create mode mismatch between the coupled signal recycling and arm cavities. Furthermore,
the beam sees different horizontal and vertical curvatures of the spherical mirrors in the folded recycling cavities. These two losses due to the curvature mismatch $\epsilon_{\rm curv}$ and folding in the recycling cavities $\epsilon_{\rm fold}$ can be approximated by 
\begin{equation}\label{eq:loss_length}
\begin{split}
	&\epsilon_{\rm curv} = 120\, \left( \frac{56\,{\rm m}}{L_{\rm src}} \right) ^{2.6}  \left( \frac{L_{\rm arm}}{4\,{\rm km}} \right)^{1.8} \left( \frac{\zeta}
	{10^{-4}} \right)^{2} \,{\rm ppm}\,, \\
	&\epsilon_{\rm fold} = 11\, \left( \frac{56\,{\rm m}}{L_{\rm src}} \right) ^7  \left( \frac{L_{\rm arm}}{4\,{\rm km}} \right)^{7/2} \,{\rm ppm}\,,
\end{split}
\end{equation}
where $\zeta$ is a relative defect in the radius of curvature in one of the telescope mirrors in the signal recycling cavity. 

We maximise the figure of merit $X$ by optimising the finesse of the arm cavity and the parameters of the signal recycling cavity, such as its finesse and length. Table~\ref{tab:opt_parameters} summarises the optimal parameters for different arm lengths. We calculate losses in the signal recycling cavity according to Eq.~\eqref{eq:thermal_loss}, and assume cancellation of the wave-front distortion by a factor of $\kappa_{\rm itmx,y}=70$ for the input test masses and $\kappa_{\rm bs}=10$ for the beam splitter. We minimise the coupling of the laser noises to the GW channel by setting a constraint on the imbalance between the two arms caused by thermal lensing $\epsilon_{\rm MI} <  G_{\rm arm}\times 5\,{\rm ppm}$. During the optimisation process, we keep the arm power equal to 4\,MW and the observed level of squeezing to 10\,dB. We do not include the classical 
noises to provide an upper bound to the figure of merit. 

\begin{table}
\begin{ruledtabular}
\begin{tabular}{lccccc}
   Parameters              & 4\,km & 10\,km  & 20\,km & 30\,km & 40\,km \\
 \hline
Mirror mass  & 40\,kg & 40\,kg  & 111\,kg  & 205\,kg & 316\,kg \\[0.1cm]
ITM transmission   & 0.04 & 0.04  & 0.04 & 0.04 & 0.04  \\[0.1cm]
SRM transmission & 0.045 & 0.022  & 0.016 & 0.054 & 0.11  \\[0.1cm] 
SRC length & 563\,m & 215\,m  & 94\,m & 105\,m & 137\,m  \\
 \end{tabular}
 \end{ruledtabular}
  \caption{Values of the optimisation parameters for different arm lengths. Parameters for the 4\,km detector are different from the ones used in Table~\ref{table_parameters} because we do not impose any constraints related to the existing facilities in the current optimisation process.}
  \label{tab:opt_parameters}
 \end{table}

The optimisation result is shown in Fig.\,\ref{fig:arm_opt}. The detector sensitivity to high-frequency GW improves approximately as $L_{\rm arm}^{1/4}$ below 18\,km according to the discussion in Sec.~\ref{sens_length}. For longer facilities, diminished antenna response from the free spectral range limits the sensitivity at high frequencies. The free spectral range equals 4\,kHz for $L_{\rm arm} = 37.5$\,km. Above this length, the sensitivity improves again since the shot noise decreases and the damage from the free spectral range does not get worse; the sensitivity is actually improved for some sky locations. We find that 18\,km-long instruments maximise the figure of merit $X$. Such detectors can be implemented using linear arm cavities or shorter folded cavities. In the next section, we study astrophysical reach of the proposed detectors and confirm that the optimal arm length to study neutron star physics is $\simeq 20$\,km.
\section{Post-merger neutron star physics}
\label{science_case}

In this and the next section, we discuss the science case for the proposed detectors and compare their
performance with existing and proposed future instruments. This section 
focuses on the neutron star post-merger emission, which contains information 
about the dense nuclear matter in its ``hot" state. 
A multi-messenger detection of the post-merger GW and electromagnetic emission 
will help settle many important questions related to the engine of 
short gamma-ray bursts, the properties of kilonova ejecta,
the structure of neutron stars, and the nuclear equation of state. 
Specifically, we discuss the resulting signal-to-noise ratio (SNR) for detecting 
the post-merger emission in Sec.\,\ref{subsec:SNR_postmerger} for different detectors,
and distinguishability of different equations of state in
Sec.\,\ref{subsec:EOS_dist}.

\subsection{SNR for detecting post-merger emission}
\label{subsec:SNR_postmerger}
To show the capability of detecting the post-merger GW emissions for different detectors, 
we use the Monte Carlo simulation to sample $1.35M_\odot-1.35M_\odot$ binary neutron star events with random sky positions, inclinations, polarization angles, and distances using a merger rate of $1540^{+3200}_{-1220}$ $\textnormal{Gpc}^{-3}\textnormal{yr}^{-1}$~\cite{LSC_GW170817}. We compute the 50th percentile (median) for both the number of detections (events with $\rm SNR>5$~\cite{Bose:2017jvk,Yang:2017xlf,Katerina_2017}) and the SNR of the loudest event for 
one-year observation time. The SNR is defined as 
\begin{equation} \label{eq:SNR}
\textnormal{SNR} = 2\sqrt{\int^{\infty}_{f_{\rm contact}}df \frac{|\tilde{h}(f)|^2}{S_{hh}(f)}}\,.
\end{equation}
Here $\tilde{h}(f)$ is the GW waveform in the frequency domain and $f_{\rm contact}$ is the contact frequency which depends on the equation of state, which is obtained using the fitting formula derived in Refs.\,\cite{Damour:2012yf,Takami_Merger_2015}. We choose three representative equations of state: SFHo~\cite{Steiner2013, Hempel2012}, Sly~\cite{Douchin2001}, and APR4~\cite{Akmal1998} to cover a wide range of stiffness of the nuclear matter. These equations of state satisfy the most recent tidal-Love number constraint from GW170817, and result in maximum neutron star mass above $2 M_\odot$. The corresponding numerical post-merger waveforms and star compactness for these equations of state are presented in Refs.~\cite{Takami_Merger_2015, Palenzuela_EOS_2015}.

\begin{figure}[!t]
  \includegraphics[width=\columnwidth]{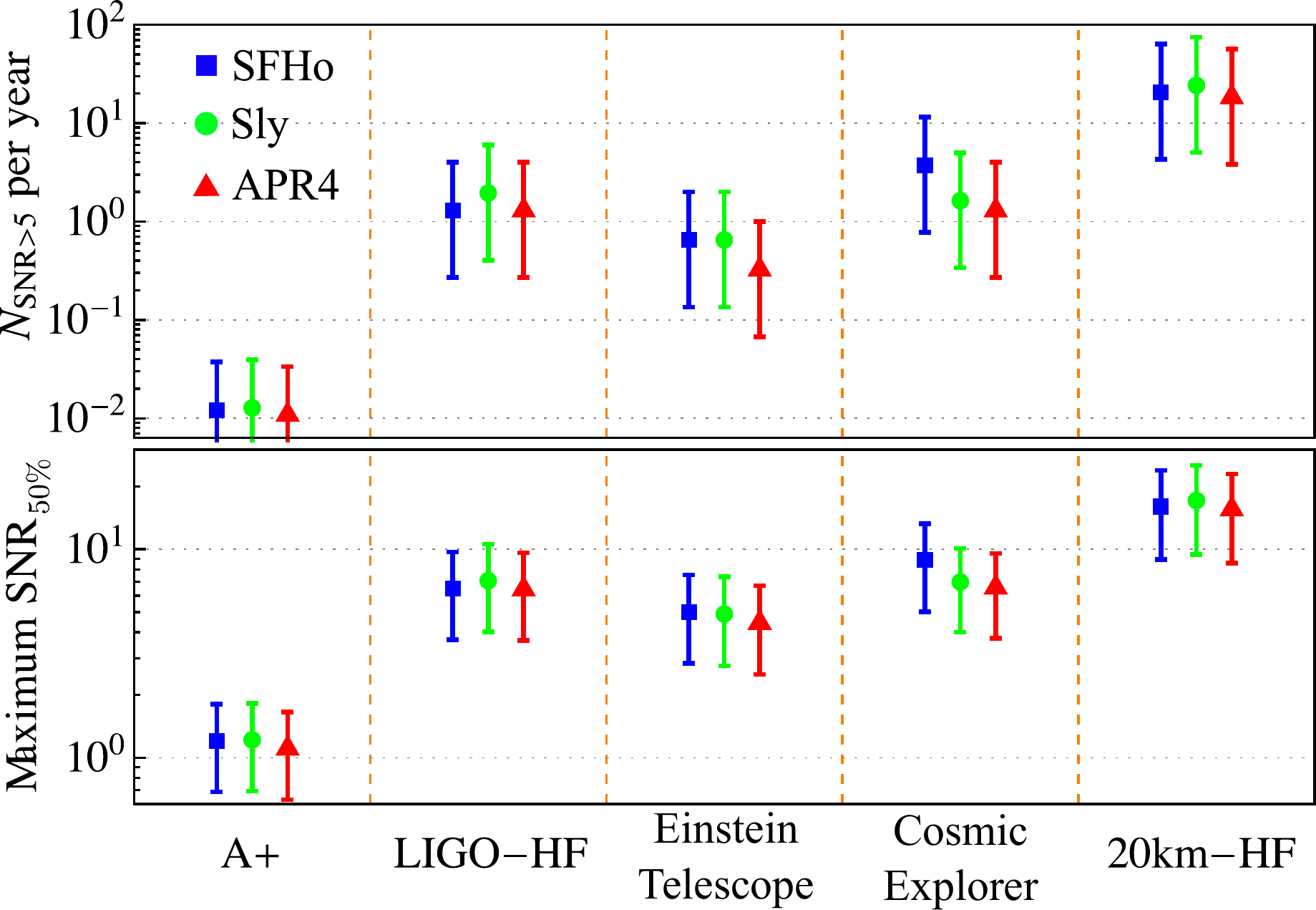}
    \caption{Number of detected post-merger oscillations per year with SNR more than 5 (top) and maximum SNR (loudest events) for the post-merger GW emissions assuming a one-year observation (bottom). These correspond to the 50th percentile (median) of the distribution obtained using the Monte Carlo simulation. The detector sensitivities are shown in Fig.~\ref{fig:config}. The range of merger rates are assumed to be within $320 -4540\, {\rm Gpc^{-3} yr^{-1}}$, with the filled symbols associated with the most probable merger rate $1540\,{\rm Gpc^{-3} yr^{-1}}$. The equations of state are chosen to cover a range of the stiffness of nuclear matter and to obtain a maximum neutron star mass above $2 M_\odot$.}
\label{fig:snr}
\end{figure}

\begin{figure}
\centering
   \includegraphics[width=0.85\columnwidth]{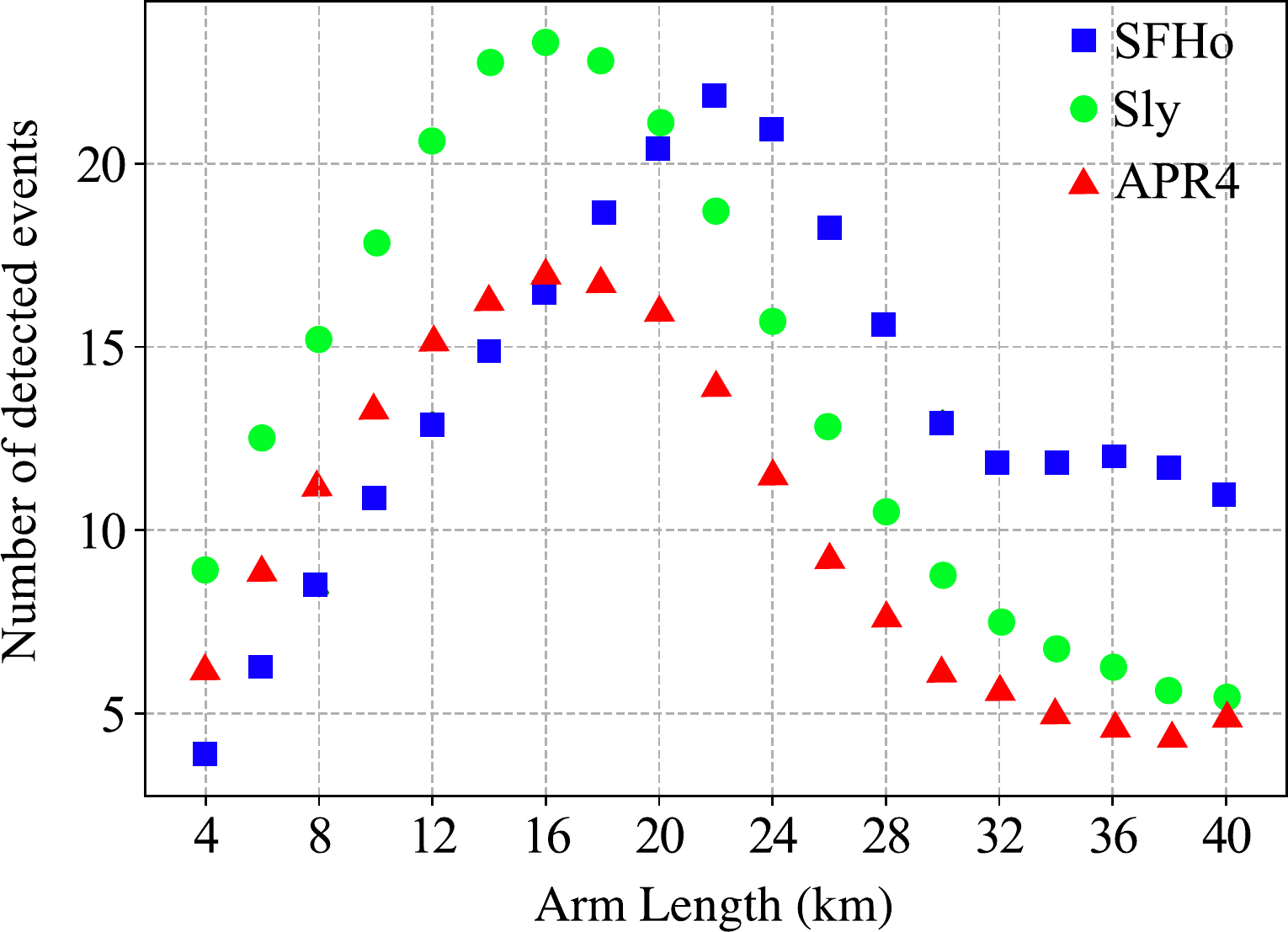}
   \caption{Number of detected post-merger oscillations as a function of arm length after one year of observation assuming a merger rate of 1540\,Gpc$^{-3}$yr$^{-1}$ .}
   \label{fig:opt_minmax} 
\end{figure}

The simulation result is shown in Fig.~\ref{fig:snr}. The upper (and lower) ends of the error bars are associated with the upper (and lower) limit of the merger rate. We find that LIGO-HF has a performance comparable to Einstein Telescope, and Cosmic Explorer in measuring post-merger waveforms, with the SNR of the loudest event $\ge 3$ even for the lower limit of the merger rate. The expected number of detections  with LIGO-HF, Einstein Telescope, Cosmic Explorer, and 20\,km-HF is roughly between 1 and 100. A+ will detect a loud event in a few years. 
We apply a similar procedure to calculate the number of detected sources for different arm lengths ranging from 4 km to 40\,km. The results are shown in Fig.~\ref{fig:opt_minmax}. An optimal arm length to detect post-merger remnants is around 16 km to 22 km, depending on the equation of state. 

In addition to the above mentioned quadrupole deformation which dominates the early post-merger
GW emission, there is also the possibility of a longer-lived GW signal such as
from the post-merger neutron star undergoing the one-armed spiral, or $m=1$,
instability.  The one-armed spiral instability was originally found in studies
of isolated differentially rotating
stars~\cite{Centrella2001,Saijo2003,Watts2005,Ou2006,Corvino2010}, and later in
hydrodynamic simulations of neutron star cores arising from the
core collapse~\cite{Ott2005,Ott2007PhRvL,Kuroda2014}.  It has also been found
to arise in post-merger neutron stars for some binary
parameters~\cite{East:2015vix,East:2016zvv,Radice:2016gym,Lehner:2016wjg},
sourcing GW radiation that is narrowly peaked in the range $\sim1$--$2$ kHz (at
a lower frequency than those coming from $m=2$ azimuthal density oscillations),
and that can be long-lived, potentially lasting 100s of milliseconds. Hence,
this could provide not only another probe of the neutron star equation of
state, but also information about the longer-term survival and dynamics of the
post-merger remnant. Though the GW signal is weaker than the initial
post-merger component, the increased sensitivity of the instrument proposed
here in the 1 to 2 kHz range would significantly increase the detectability.
This instability has only been studied for a limited number of parameters, and
the possible impact of magnetic fields, and other microphysical effects on the
long-term evolution is not known. For such
a signal at 1 kHz, lasting $100$ ms, the horizon for an ${\rm SNR}=5$
threshold detection by LIGO-HF is around 20 to 90 Mpc for nearly equal mass
binaries~\cite{Radice:2016gym,East:2016zvv}, or possibly even 100s of Mpc for
binaries with disparate mass ratio~\cite{Lehner:2016wjg}.

\subsection{Distinguishing different equations of state}\label{subsec:EOS_dist}
Complementary to the number of detected events, we present a study of how well different detector designs can distinguish between different equations of state using 
the post-merger signal. For this purpose, we perform a model selection analysis using Bayes factors. Given some data $\vec{d}$ and post-merger waveform parameters $\vec{\theta}$, the Bayes factor is calculated as
$B = Z_1/Z_2$,
where $Z_1$ and $Z_2$ are Bayesian evidences defined as
\begin{equation} \label{eq:comp_eos}
Z \equiv \int d\vec{\theta} \, L(\vec{d}|\vec{\theta},\mathcal{H}_{s}) \, \pi(\vec{\theta}),
\end{equation}
where $L(\vec{d}|\vec{\theta},\mathcal{H}_{s})$ is the likelihood probability function under the hypothesis of having a signal and $\pi(\vec{\theta})$ is the prior probability function~\cite{Jeffreys:61} defined by the equation of state. For post-merger waveforms, we assume a Dirac delta function prior: $\pi(\vec{\theta})=\delta(\vec{\theta}-\vec{\theta}_{0})$ over all parameters $\vec{\theta}$ at the waveform true values $\vec{\theta}_0$. This assumption comes from the lack of post-merger waveform approximates to marginalise over all waveform parameters efficiently. Therefore, our results should be treated as upper limits. Similar to the analysis in the previous section, we perform Monte-Carlo simulations (assuming random sky location, polarisation angles, and distances) and calculate the Bayes factor for each simulation. Only events with $\log B \geq 8$ are considered as distinguishable. The results are shown in Table \ref{tab:compare_eos}.

The 20\,km-HF detector has more distinguishable events than Cosmic Explorer and Einstein Telescope for all considered equations of state. In addition, the detectors dedicated for high frequencies and Cosmic Explorer can distinguish between APR4, SLY and SFHo, because these three equations of state have similar spectral features---the frequency of the dominant mode is $\simeq 3.2$ kHz in this parameter space. The 20\,km-HF detector is more sensitive than Einstein Telescope around the mode frequency. 

 
 \begin{table}
 \begin{ruledtabular}
 \begin{tabular}{lccc}
                      & SLY / APR4 & SLY / SFHo  & APR4 / SFHo \\
 \hline
LIGO-HF   & $0.53^{+1.4}_{-0.41}$    & $2.22^{+4.22}_{-1.82}$   & $1.21^{+3.5}_{-1}$ \\
Einstein Telescope  & $0.15^{+0.13}_{-0.12}$ &$0.42^{+0.8}_{-0.37}$  &$0.27^{+0.65}_{-0.2}$ \\
Cosmic Explorer & $1.44^{+3}_{-1.18}$ & $4.84^{+10.01}_{-3.88}$ & $3.94^{+8.17}_{-3.15}$ \\
20\,km-HF &$9.18^{+18.22}_{-7.24}$ &$31.37^{+65.25}_{-24.39}$ & $22.27^{+45.13}_{-17.71}$ \\
 \end{tabular}
 \end{ruledtabular}
  \caption{Number of distinguishable post-merger detections between different equations of state after one-year observation. The results shown are obtained from Monte Carlo simulations by only considering events with $\log B \geq 8$ as distinguishable. We assume the perfect knowledge of a $1.35M_\odot - 1.35 M_\odot$ binary neutron star post-merger waveform to provide the upper limit.}
  \label{tab:compare_eos}
 \end{table}

\section{Pre-merger neutron star physics}
\label{pre_merger}
This section focuses on the late-inspiral part of binary neutron star and black hole-neutron star mergers
when matter effects start to take place through the tidal interactions between 
the compact objects. This allows us to study the physics of nuclear matter
in its ``cold" state. Specifically, we 
discuss the tidal deformability in Sec.\,\ref{subsec:tidal}, binary neutron star as standard sirens for 
cosmology in Sec.\,\ref{subsec:cosm}, and black hole-neutron star binaries in Sec.\,\ref{subsec:BHNS}. 

\subsection{Measurement of the tidal deformability}
\label{subsec:tidal}

During the late binary neutron star inspiral, the information about the equation of state can be quantified by the tidal deformability parameter $\lambda$. To the leading order, the quadrupole moment tensor $Q_{ij}$ is related to the tidal field tensor $\mathcal{E}_{ij}$ by $Q_{ij} = \lambda\,\mathcal{E}_{ij}$, where $\lambda=(2/3)k_2R^5/G$ \cite{Flanagan_2008}. The value of $\lambda$ depends on the second Love number $k_2$ and the radius $R$ of the neutron star, where both of these quantities depend on the equation of state. For this reason, if the parameter $\lambda$ is constrained, the equation of state can be constrained as well. For the purpose of our analysis, we focus our attention on the dimensionless tidal deformability parameter $\Lambda = G\lambda[c^2/(Gm)]^5$. Moreover, since the tidal deformability for each neutron star $\Lambda_1$ and $\Lambda_2$ are highly correlated, it is convenient to parametrise the tidal deformability in terms of the weighted average parameters $\tilde{\Lambda}$ and $\delta \tilde{\Lambda}$ defined in equations (5) and (6) of Ref.~\cite{Wade_2014}.

We  estimate the expected error in the measurement $\tilde{\Lambda}$ using the Fisher matrix analysis~\cite{Hinderer_2010}. The relevant parameters are
$\vec{\theta} = (\mathcal{M},\eta,\tilde{\Lambda},t_c,\phi_c,\mathcal{A})$ where $\mathcal{M}$ is the chirp mass, $t_c$ is the time of coalescence, $\phi_c$ is the phase of coalescence, and $\mathcal{A}$ is the waveform amplitude which depends on sky position, inclination, GW polarization angle and distance. We assume a $1.35M_\odot - 1.35 M_\odot$ binary neutron star located at a distance of 100 Mpc and study three different equation of states: APR4 ($\tilde{\Lambda}=321.7$), SLY ($\tilde{\Lambda}=390.2$), and SFHo ($\tilde{\Lambda}=387$) \cite{Takami_Merger_2015} averaging over sky position, inclination and GW polarization angle. The waveform IMRPhenomD\_NRTidal~\cite{Husa:2016,Khan:2016,Dietrich:2017} is used in all calculations  starting at 20 Hz.  The errors are shown in Table \ref{tab:lambda_error0}, where we find that for the considered equations of state, LIGO-HF can constrain $\tilde{\Lambda}$ approximately two times better than A+. 
Interestingly, 20\,km-HF will perform similar to Cosmic Explorer since tidal effects become significant around 1\,kHz, where the sensitivity of
these two detectors are comparable. 

\begin{table}[ht!]
 \begin{ruledtabular}
 \begin{tabular}{lccc}
        		   & SLY  & APR4  & SFHo      \\
 \hline
 A+            &0.147 &0.173  &0.148 \\
  LIGO-HF      &0.055 &0.064  &0.056\\
  Einstein Telescope           &0.038 & 0.043 &0.038\\
Cosmic Explorer            &0.027 &0.032  &0.03 \\
 20\,km-HF       &0.017 &0.020  &0.017 \\
 \end{tabular}
 \end{ruledtabular}
 \caption{Fractional errors in the tidal deformability $\Delta \tilde{\Lambda} / \tilde{\Lambda}$ for a $1.35M_\odot - 1.35 M_\odot$ binary neutron star located at a distance of 100 Mpc averaged over sky position, inclination and GW polarization angle.}
  \label{tab:lambda_error0}
 \end{table}

\subsection{Cosmology}\label{subsec:cosm}
Another interesting aspect of binary neutron star is their role as standard sirens for cosmology. Messenger and Read showed that if the equation of state is known, it is possible to measure the redshift $z$ even without an electromagnetic counterpart \cite{Messenger_2012}. The main idea behind this work is that if tidal effects are included during the inspiral phase, the rest frame mass $\mathcal{M}_r$ can be measured independently from the red-shifted mass $\mathcal{M}_z$, hence the degeneracy between these two quantities related by $\mathcal{M}_z = \mathcal{M}_r (1+z)$ is broken~\cite{Messenger_2012}. With future detectors, the equation of state will be well constrained, and we can obtain an uncertainty in $z$. 
Similar to the previous section, the waveform used in our calculation is IMRPhenomD\_NRTidal and we assume a $1.35M_\odot - 1.35 M_\odot$ binary with the same list of equations of state. To calculate the error in redshift $z$, we again use the Fisher Matrix analysis, the same as the previous section. The parameter space is $\vec{\theta} = (\mathcal{M},\eta,\mathcal{A},z,t_c,\phi_c)$. The spin is assumed to be zero for both neutron stars and the errors are calculated at redshift $z=0.01$ starting at a frequency of 20 Hz, where we have assumed the standard cosmological model~\cite{Planck:2015}. The results are shown in Table \ref{tab:zeta}, where one can approximate the error in redshift to the error in the Hubble constant if the luminosity distance is well constrained. 

In contrast, if a binary neutron star detection is accompanied by an electromagnetic counterpart (without knowing the equation of state), the main error in the Hubble constant will be dominated by the error in the luminosity distance. For GW170817, the 1$\sigma$ error in the luminosity distance is  $\simeq 11 \% $ implying a 1$\sigma$ error in the Hubble constant of $\simeq 14 \%$~\cite{Vitale:2018,LSC_GW170817}. Although the error for the case with an electro-magnetic counterpart are smaller compared to the one without, it is expected that in the future most of binary neutron star detections will not have electro-magnetic counterparts~\cite{Abbott:2017xzu}.
The error of estimating $z$ for the Einstein Telescope ranges from 25\% to 31\%, while for Cosmic Explorer and 20\,km-HF it ranges from around 11\% to 15\% for a sky-averaged binary neutron star merger located at $z=0.01$.  

\begin{table}[h]
 \begin{ruledtabular}
 \begin{tabular}{lccc}
         & SLY  & APR4  & SFHo   \\
 \hline
 A+      &1.009 &1.206 &1.008 \\
 LIGO-HF &0.384 &0.448 &0.383 \\
 Einstein Telescope      &0.259 &0.310 &0.259 \\
Cosmic Explorer      &0.187 &0.195 &0.186 \\
 20\,km-HF &0.119 &0.139 &0.119 \\
 \end{tabular}
 \end{ruledtabular}
  \caption{Errors in the redshift $\Delta z / z$ for a $1.35M_\odot-1.35M_\odot$ binary neutron star located at redshift $z=0.01$ averaged over sky position, inclination and GW polarization angle. We assume that the equation of state is known without an electromagnetic counterpart.}
   \label{tab:zeta}
 \end{table}
 
 \subsection{Tidal disruption in black hole-neutron star binaries}\label{subsec:BHNS}
 So far we have discussed the tidal deformability in binary neutron star mergers. A coalescence of a low mass black hole and a neutron star can also provide an interesting case for the high-frequency detectors. If the ratio between the mass of a black hole $M_{\rm BH}$ and the mass of the neutron star $M_{\rm NS}$ is small enough, the neutron star will be tidally disrupted before the merger. Along with the GW radiation, this process can produce copious electromagnetic emission and offer a new opportunity to probe nuclear physics with multi-messenger signals. 
 
Using numerical relativity simulations, Shibata \textit{et al.} \cite{Shibata_BHNS_2009} showed that black hole-neutron star merger can be categorised into three different kinds of waveforms.
 For the first type (type I), the tidal disruption occurs during the inspiral, outside the Inner-Most-Stable-Circular orbit. The GW amplitude decreases rapidly after the disruption, which makes the post-merger GW signal difficult to detect if the detector sensitivity is not high enough. For type II, the tidal disruption occurs during the plunging phase, and the ringdown of the system is still significant. Because the ringdown is significantly affected by the disrupted matter, the merger-ringdown 
signal is different from those of binary black holes. 
Finally, if the mass ratio between the black hole and neutron star is beyond $\sim 5$, the black hole swallows the neutron star during the merger without tidal disruption. The GW from the merger and ringdown phase is similar to the one from a binary black hole system.

Considering these three types of events, we calculate the corresponding SNR given a fixed neutron star mass $M_{\rm NS} = 1.35 M_\odot$ and different black hole masses at a fixed source  distance of 100 Mpc. The post-merger waveform for computing the SNR starts from the tidal-disruption frequency, which  depends on the mass ratio and the equation of state; for type III binaries, such a frequency is approximately equal to the quasinormal mode frequency of the final black hole. We parametrise the waveforms and fit them to the numerical ones presented in Ref.~\cite{Shibata_BHNS_2009}. The resulting SNRs are shown in Table~\ref{tab:tidal_def}, where we only show detectors that achieve SNR $\geq$ 1.

 
 \begin{table}[h]
 \begin{ruledtabular}
 \begin{tabular}{lccc}
  & Type I & Type II & Type III \\
   \hline
LIGO-HF & 1.59 &3.65 &4.05 \\
Einstein Telescope & 1.37 &2.44 &2.98 \\
Cosmic Explorer & 2.00 &3.27 &4.18 \\
20\,km-HF & 4.86  & 10.98 & 12.61 \\
 \end{tabular}
 \end{ruledtabular}
\caption{SNRs for detecting different types of black hole-neutron star post-merger signals. The source is assumed to be at 100 Mpc and having a polytropic equation of state $\Gamma =2$ (GAM2). We consider mass ratios of $M_{\rm BH} / M_{\rm NS}=1.5$ for type I, $M_{\rm BH} / M_{\rm NS}=3$ for type II and $M_{\rm BH} / M_{\rm NS}=5$ for type III waveforms.}
  \label{tab:tidal_def}
 \end{table}



\section{Conclusions}

Reducing the quantum shot noise at high frequencies is essential to the detection of GWs from neutron star post-merger oscillations. Precise post-merger GW observations will lead answers to important questions regarding the structure of neutron stars, the equation of state of nuclear matter, and the role of the remnants as the central engine for energetic electromagnetic emissions. This calls for an improved optical configuration of the GW detectors operating at high optical power. In particular, the proposed upgrade of the 4\,km facilities requires (i) development of 500\,W lasers, (ii) increasing the signal recycling cavity length and finesse, and (iii) cancellation of thermal wavefront distortion in the input test masses and the beam splitter. The LIGO-HF has a similar or better sensitivity for probing neutron star physics compared with other proposed future detectors, such as Cosmic Explorer and Einstein Telescope. Therefore, the existing 3\,km and 4\,km facilities could still have a long-term impact as neutron star observatories in the era of the next generation detectors. 

Our study also shows that the optimal arm length for observing neutron star post-merger oscillations 
is $\simeq 20$\,km, and 20km-HF allows us to detect $\simeq 30$ events per year with maximum SNR of $\simeq 20$. The detector performance improves approximately as $L_{\rm arm}^{1/4}$ below 18\,km. For longer facilities, diminished antenna response from the free spectral range limits the sensitivity at high frequencies. Since the gap between the shot noise and classical noise increases proportionally to the arm length, which is faster than $L_{\rm arm}^{1/4}$, we may take advantage of this fact by using folded arms to reduce the construction costs without compromising the high-frequency sensitivity. 


\section{Acknowledgements}
We would like to thank members of the LSC AIC, MQM, and QN groups for fruitful discussions and Valery Frolov for reviewing the paper during the internal LIGO review. We acknowledge the support of the National Science Foundation and the LIGO Laboratory. 
LIGO was constructed by the California Institute
 of Technology and Massachusetts Institute of Technology with funding
 from the National Science Foundation and operates under cooperative
 agreement PHY-0757058.
D.M. and H.M. acknowledge the support of the Institute for Gravitational Wave Astronomy at University of Birmingham. 
H.M. is supported by UK STFC Ernest Rutherford Fellowship (Grant No. ST/M005844/11). 
H.Y. is supported by the Natural Sciences and Engineering Research Council of Canada. 
This research was supported in part by Perimeter Institute for Theoretical Physics.
Research at Perimeter Institute is supported by the Government of Canada through Industry Canada and by the Province of Ontario through the Ministry of Research and Innovation. 
AB acknowledges support by the European Research Council (ERC) under the European Union’s Horizon 2020 research and innovation programme under grant agreement No. 759253 and the Klaus-Tschira Foundation.
This work is supported through Australian Research Council (ARC) Centre of Excellence CE170100004. ET is supported through ARC Future Fellowship FT150100281.
PDL is supported through ARC Future Fellowship FT160100112 and ARC Discovery Project DP180103155.

\section{Appendix}

\subsection{High-power effects in the interferometer}
\label{high_power}
Apart from the thermal lenses considered in Sec.~\ref{thermal_heating} of the main text, optical power in the arm cavities also creates angular and parametric instabilities~\cite{Dooley_ASC_2013, Braginsky_PI_2002}. In this subsection we consider how these instabilities influence the detector performance at high frequencies.
Small beam off-centering results in the radiation pressure torque given by the equation~\cite{Hirose_ASC_2010}
\begin{equation}
	T = \frac{2 P_{\rm arm}}{c} \xi \theta,
\end{equation}
where $\xi$ is the beam motion on the cavity mirrors due to the tilt $\theta$ of these mirrors. Arm cavities have two angular modes, known as hard ($\xi = -2.2 \times 10^4$\,m/rad) and soft ($\xi = 10^3$\,m/rad) which correspond to the tilt and shift of the optical axis~\cite{Barsotti_ASC_2010}. Since the radiation torque is proportional to the tilt angle of the mirrors, it modifies their dynamics. If $\xi < 0$ then the mechanical resonance of suspended mirrors shifts up, and if $\xi > 0$ then the resonance shifts down and the mode can become unstable. The instability occurs when $T_{\rm soft} = I \Omega_{\rm ang}^2$, where $I$ is moment of inertia and $\Omega_{\rm ang} \approx 2 \pi \times 0.5$ rad/s is the mechanical angular resonance frequency of the test masses.

\begin{figure}[h]
\includegraphics[width=\linewidth]{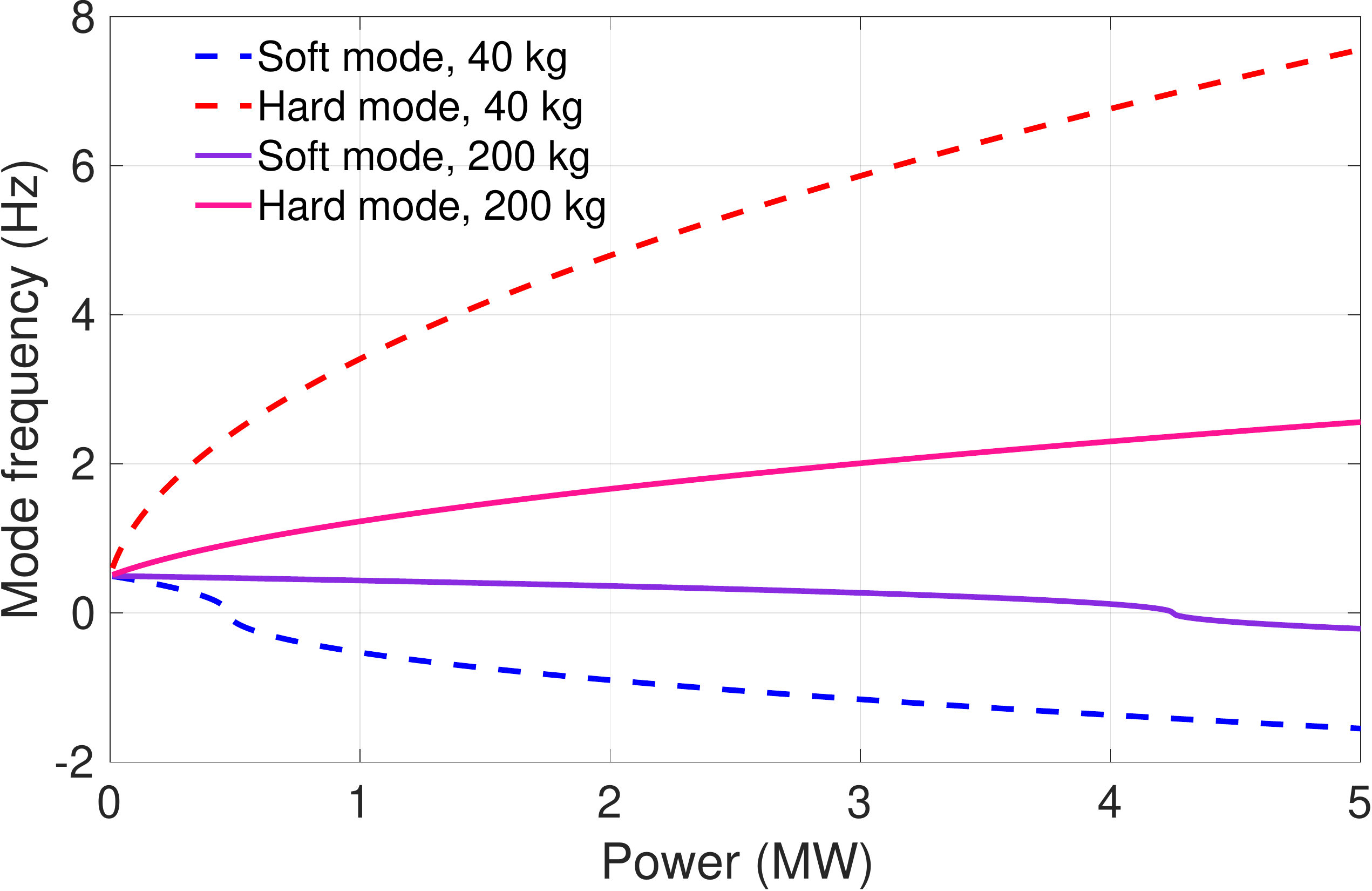}
\caption{Shift of the angular mode frequency due to the radiation torque. For the current mirror mass of 40\,kg the soft mode becomes unstable at the arm power of 0.5\,MW. This power can be increased up to 4.2\,MW by using 200\,kg test masses.}
\label{fig:asc_modes}
\end{figure} 

In Advanced LIGO, soft modes of the test masses become unstable at the threshold power of $P_{\rm arm} = 0.5$\,MW, as shown in Fig.~\ref{fig:asc_modes}. However, the unstable mode frequency does not exceed 2\,Hz for arm powers less than 5\,MW. The tilt angle $\theta$ of the test masses is measured using wave-front sensors and the instability can be suppressed in a feedback loop with bandwidth of 5\,Hz. Therefore, angular instabilities do not influence the detector sensitivity at high frequencies and inject angular control noise below 20\,Hz. We can suppress this noise by increasing the mirror mass up to 200\,kg~\cite{Yu_5Hz_2018}. In this case, angular modes are stable up to the arm power of 4.2\,MW as shown in Fig.~\ref{fig:asc_modes}.


 \begin{figure*}[tb]
\includegraphics[width=0.95\columnwidth]{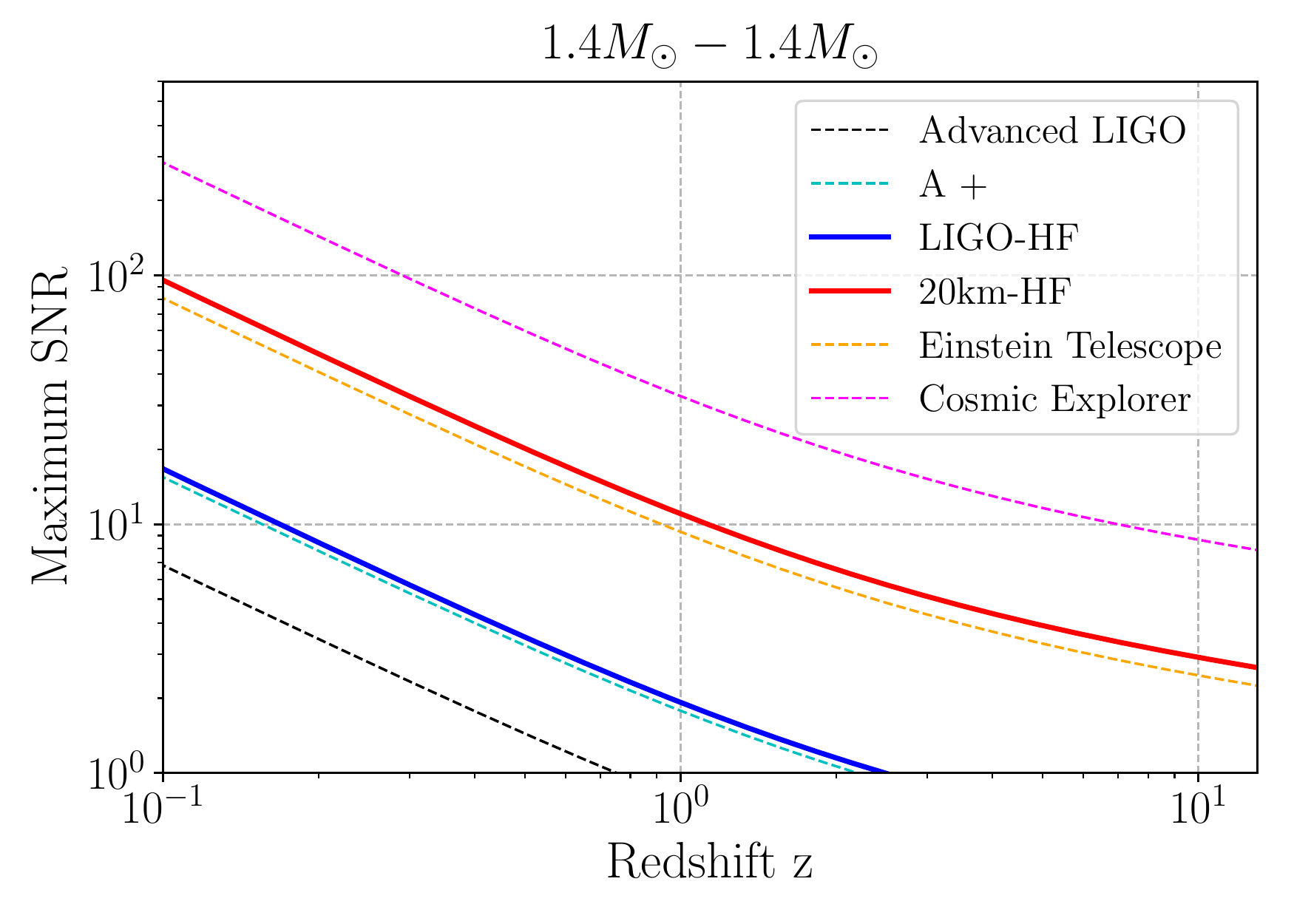}
\quad\quad
\includegraphics[width=0.95\columnwidth]{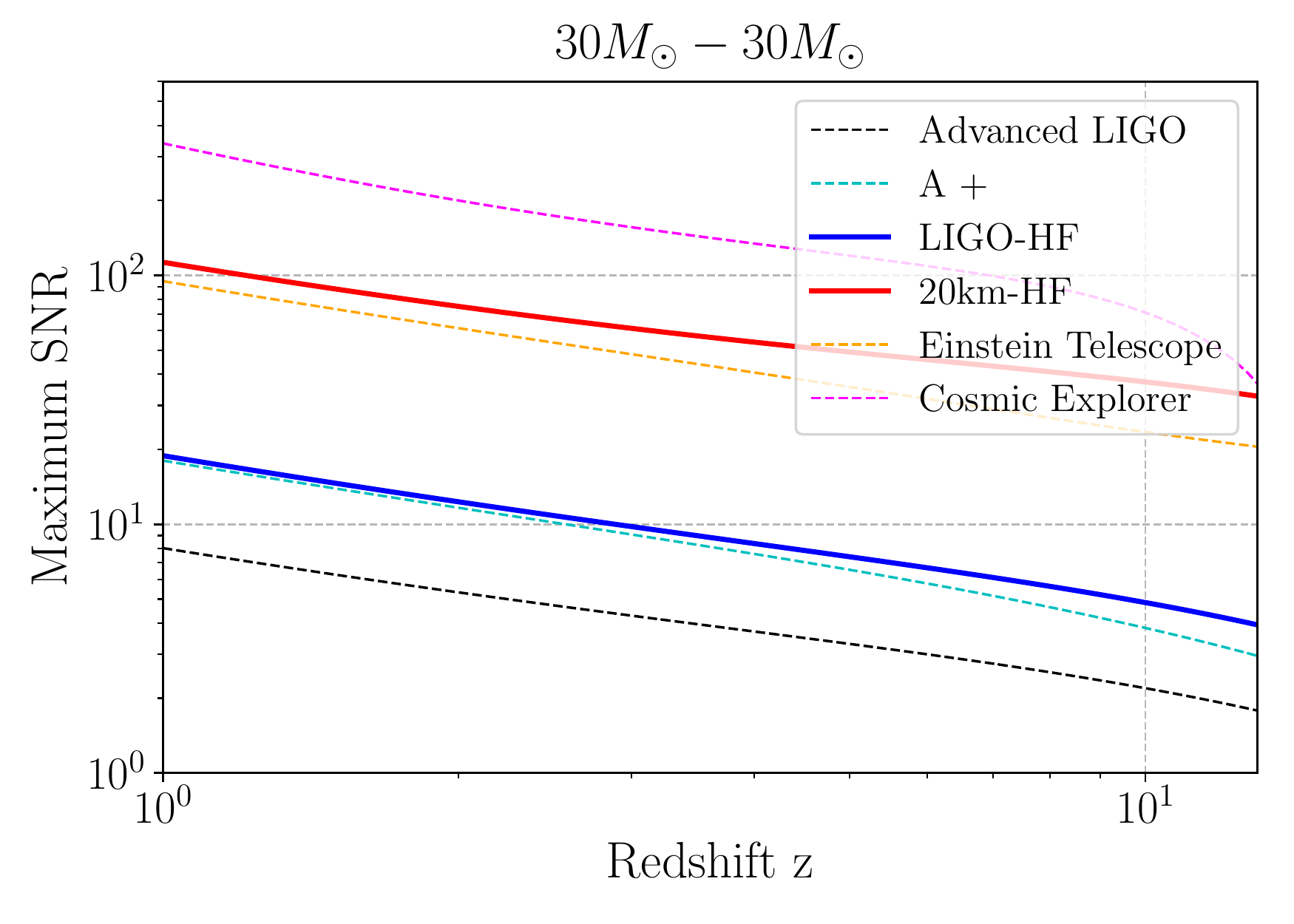}
\caption{Horizon distance as a function of redshift for a $1.4M_\odot-1.4M_\odot$ binary neutron system (left). Horizon distance as a function of redshift for a $30M_\odot-30M_\odot$ (right). For both systems the spin is equal to zero. The proposed LIGO-HF detector has similar low frequency sensitivity to the A+ upgrade.}
\label{fig:maximum_snr}
\end{figure*} 

Parametric instabilities are excitations of the mirror body modes in an unstable feedback loop due ti radiation pressure force from higher order optical modes~\cite{Braginsky_PI_2002}. Quantitatively, the mode growth rate is characterised by the parametric gain which is given by the equation~\cite{Evans_PI_15}
\begin{equation}
R_m = \frac{8 \pi Q_m P_{\rm arm}}{M \Omega_m^2 c \lambda} \sum_{n=0}^\infty \Re[G_n] B_{m,n}^2,
\end{equation}
where $\Omega_m$ and $Q_m$ are the frequency and quality factor of the mechanical mode, $c$ is the speed of light, $\lambda = 1064$\,nm is the laser wavelength, $\Re[G_n]$ is the real part of the optical gain, and $B_{m,n}$ is the spacial overlap between the mechanical mode m and optical mode n. If the parametric gain $R_m > 1$, the mode can become unstable and grow exponentially. For Advanced LIGO operating at full power, the largest expected parametric gain is $R_m \simeq 10$ and the number of unstable modes is $\approx 40$ in the frequency range 10$-$50\,kHz. In the proposed detector, we plan to increase $P_{\rm arm}$ up to 4\,MW and have maximum parametric gain of $R_m \simeq 50$ and see unstable modes up to 80\,kHz. These modes can be passively and actively suppressed by reducing their Q-factors~\cite{Blair_PI_2017}. Passive dampers reduce the quality factor of the modes by an order of magnitude~\cite{Gras_PI_2015} and are installed perpendicular to the beam direction. Therefore, thermal noise of the test masses is not compromised and does not reduce sensitivity of the detector at high frequencies.

 \subsection{Black hole and neutron star science}
 \label{detector_ranges}
 
Another important comparison that can be studied is how well the proposed detector will perform at detecting binary black holes and binary neutron stars during the inspiral, merger and ringdown. In this section we determine the range distance $R$  following the similar procedure defined by Chen, \textit{et al.}~\cite{Chen_2017}. The range distance $R$ is calculated as the radius of the redshifted detectable volume defined by the equation
\begin{equation} \label{eq:antenna}
 V_z = \frac{\int_{D_c<d_h} \frac{D_c^2}{1+z(D_c)} dD_cd\Omega \sin\iota d\iota d\psi}{\int \sin\iota d\iota d\psi},
\end{equation}
where $D_c$ is the comoving distance, $d_h$ is the horizon distance, $\Omega$ is the solid angle, $\iota$ is the inclination of the binary system and $\psi$ is the polarisation angle. Equation \ref{eq:antenna} takes into account the interferometer's antenna response and the dependence of the merger rate reduction as a function of redshift. The results are shown in Tab. \ref{tab:ns_bh}. In addition, plots of the maximum SNR as a function of redshift are shown in figure \ref{fig:maximum_snr}.

\begin{table}[t]
 \begin{ruledtabular}
 \begin{tabular}{lcc}
 & $1.4 M_{\odot}-1.4 M_{\odot}$ [Gpc] & $30 M_{\odot}-30 M_{\odot}$ [Gpc]\\
 \hline
 Advanced LIGO      &0.16 &1.5	\\
 A+                 &0.35 &2.6	\\
 LIGO-HF            &0.38 &2.8	\\
 Einstein Telescope &2.10 &5.81\\
 Cosmic Explorer    &4.23 &6.1	\\
 20km-HF            &1.91 &5.6\\
 \end{tabular}
 \end{ruledtabular}
  \caption{Values for the range distance of a $1.4 M_{\odot}-1.4 M_{\odot}$ binary neutron star and $30 M_{\odot}-30 M_{\odot}$ binary black hole.}
 \label{tab:ns_bh}
 \end{table}

 \subsection{Stochastic background}
\label{stochastic_background}

\begin{figure}[ht!]
\includegraphics[width=\columnwidth]{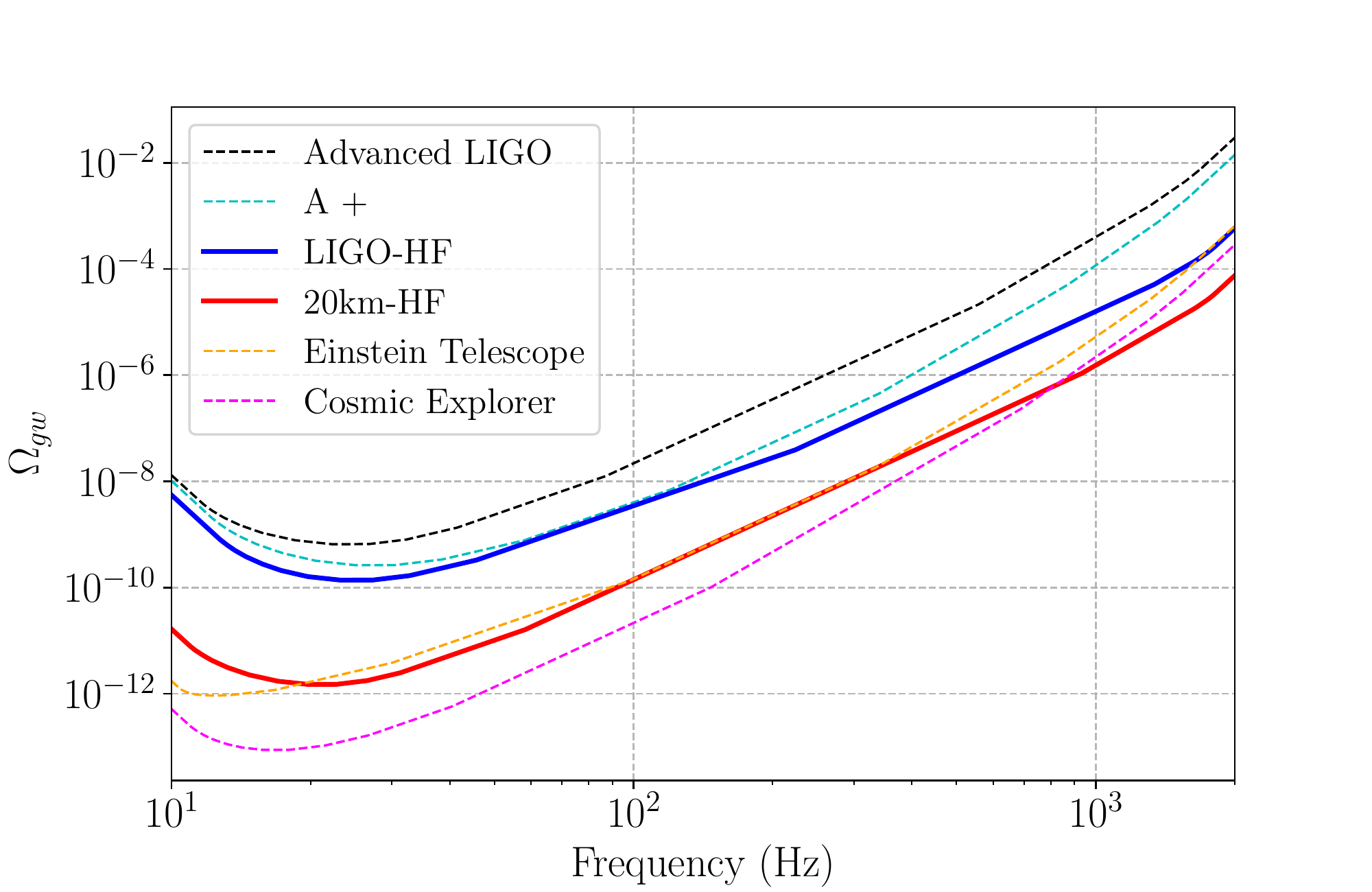}
\caption{Sensitivity curves for the gravitational wave background $\Omega_{gw}$. We have assumed that two interferometers with the same characteristics are built in the current LIGO Hanford and Livingston facilities and an observation period of 1 year.}
\label{fig:background}
\end{figure} 

 \begin{figure*}[t!]
\includegraphics[width=0.95\columnwidth]{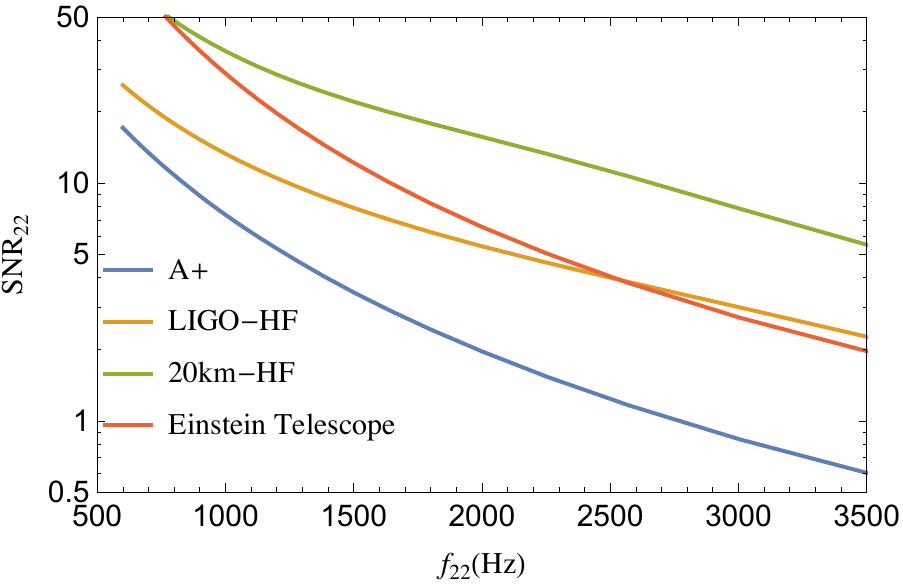}
\quad\quad
\includegraphics[width=0.95\columnwidth]{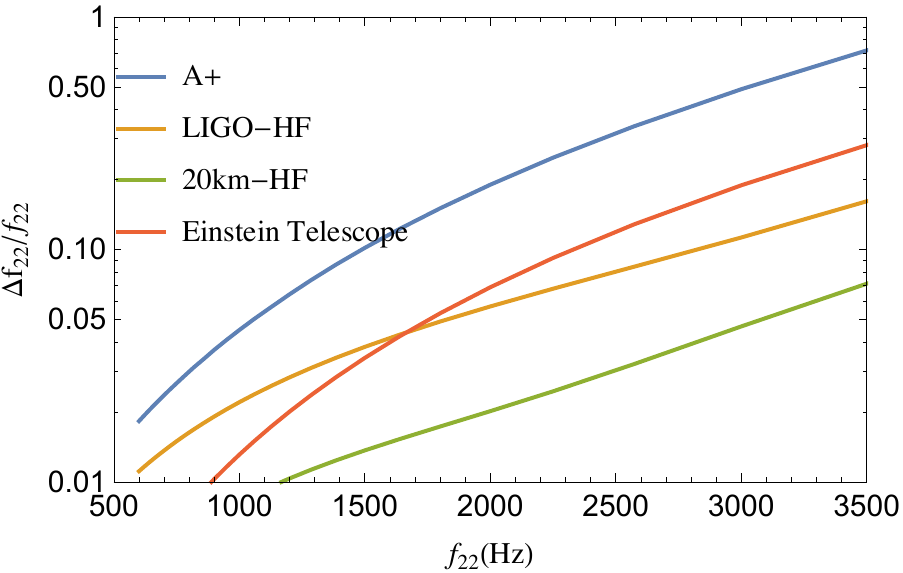}
\caption{Black hole spectroscopy at high frequencies. The top panel present the SNR for the $22$ mode, during an equal-mass binary black hole coalescence at $200$ Mpc. The bottom panel shows the measurement accuracy of the $22$ mode frequency given such events.}
\label{fig:bhsnr}
\end{figure*} 

The stochastic gravitational wave background is usually searched by cross-correlating data from two different interferometers. To show the sensitivity to the gravitational wave background, the fractional energy density of gravitational waves $\Omega_{gw}(f)$ is often plotted instead of the usual power spectral density $S_n(f)$ shown in figure \ref{fig:config}. We limit our study to gravitational wave backgrounds that follow a power law distribution given by $\Omega_{gw}(f)= (f/f_{ref})^\beta$, where $\beta$ is a spectral index and equals to 2/3 for binary coalescences  and $f_{ref}$ is a reference frequency set to 100 Hz for ground-based detectors. For a detailed analysis of the method used in our study refer to Thrane and Romano 2014 \cite{Thrane_Stochastic_2013}, where a method to the increase sensitivity by integrating gravitational backgrounds in frequency and time is used. The results are shown in figure \ref{fig:background}, noting that the sensitivity curves we show are analogous to the sensitivity curves shown in Fig.~\ref{fig:config}, but for $\Omega_{gw}(f)$. In addition, we have assumed that two interferometers with the same characteristics are built in the current LIGO Hanford and Livingston facilities and that the results shown in figure \ref{fig:background} assume an observation period of one year.
 
 \subsection{Superradiant instability of ultralight bosons}
 \label{superradiant}

Another application of GW observations to fundamental physics of  
recent interest is the possibility of searching for ultralight
bosons with Compton wavelength comparable to the radius of stellar mass black
holes
~\cite{Arvanitaki:2014wva,Arvanitaki:2016qwi,Baryakhtar:2017ngi,Brito:2017zvb,Brito:2017wnc,East:2017mrj,East:2018glu}
. If such ultralight bosons exist, spinning black holes are unstable to
superradiance, and develop massive boson clouds---up to $~\sim10\%$ of the
black hole's mass---that grow at the expense of the black hole's rotational
energy~\cite{Arvanitaki:2010sy,Dolan:2012yt,2015CQGra..32m4001B,East:2017ovw}.
These oscillating bosons clouds would produce nearly monochromatic GW signals,
allowing GW detectors to look for
axions~\cite{Weinberg:1977ma,Arvanitaki:2009fg,Arvanitaki:2010sy}, dark
photons~\cite{Holdom:1985ag,Cicoli:2011yh}, or other types of dark bosonic
matter that is weakly coupled to the Standard Model. 
This could include searches that target newly
formed black holes (e.g., arising from mergers)~\cite{Isi:2018pzk},
as well as all-sky continuous wave searches~\cite{Goncharov:2018ufi,Pierce:2018xmy}, 
or searches for a possible stochastic background
due to superradiance~\cite{Brito:2017zvb,Brito:2017wnc}. 

The frequency of the cloud oscillations, and hence GW signals, is roughly
proportional to the boson mass, and there is theoretical interest in probing a
wide range of possible boson masses.  A GW detector with improved sensitivity
in the frequency range of $300$ Hz to $5$ kHz would probe boson masses of
$6\times10^{-13}$ to $10^{-11}$ eV, which could efficiently grow through
superradiance around black holes with masses in the $\sim 1$--$100$ $M_{\odot}$
range (with lower black hole masses corresponding to higher boson masses 
and GW frequencies).  It should be noted, however, that this boson mass range is already
disfavoured by X-ray measurements of black hole
spins~\cite{Arvanitaki:2014wva,Baryakhtar:2017ngi}, so such GW observations
would provide an independent check of these spin measurement models.

The above refers to the so-called ``annihilation" GW signals from an
oscillating boson cloud that is dominated by single unstable mode. In some
special cases, multiple modes may be populated, leading to a GW signal due to
the beating of the different frequency modes (referred to as the ``transition"
GW signal)~\cite{Arvanitaki:2014wva}, which can be used to probe higher boson
masses.  For example, a $2\times10^{-11}$ eV axion (at the upper edge of the
range probed by X-ray observations) around a $2.7$ solar mass black hole would
produce a $f\sim 300$ Hz GW signal from the most favorable transition (the
$6g\rightarrow 5g$; see~\cite{Arvanitaki:2014wva,Yoshino:2015nsa}).  In this
scenario, improved sensitivity at 300 Hz and above could enhance GW probes of
superradiance occurring around light black holes---as might arise from
binary neutron star mergers---not probed by X-ray spin measurements.

\subsection{Black hole spectroscopy}
\label{bbh_spectroscopy}

Black hole ringdown encodes critical information of the black hole spacetime and its progenitors (i.e., merging binary black holes) that lead to the ringing black hole.
As modified gravity theories generically predict a different ringing black hole spectrum from the Kerr  spectrum at certain length scale, measuring quasinormal modes of black holes provides an important opportunity to constrain modified gravity theories and search for new fundamental physics  \cite{Berti:2018vdi}. Many modified gravity theories that contain high-order temporal and spatial derivatives in the action (such as the dynamical Chern-Simon theory \cite{Jackiw:2003pm} and the Einstein-Dilaton-Gauss-Bonet theory \cite{Moura:2006pz}) preferably deviate from General Relativity at short length scales and high frequencies.
Because of the superior sensitivity of the high frequency detector beyond $500$ Hz, it is reasonable to apply it for the black hole spectroscopy measurement. 

In Fig.~\ref{fig:bhsnr} we present the SNR of the $22$ mode of a ringing black hole, originated from an equal-mass binary black hole merger located at $200$ Mpc. The black holes within the binary are assumed to be not spinning, and the detection is assumed to be made in the maximally emitting direction. As expected, LIGO-HF outperforms Advanced LIGO plus with higher 22 mode SNR. In addition, as shown in the top panel of Fig.~\ref{fig:bhsnr}, Einstein Telescope always outperforms LIGO-HF below $\sim 3500$ Hz. However, a detailed analysis reveals that most of the mode SNR for Einstein Telescope comes from the tail of quasinormal mode spectrum below $600$ Hz. As the tail part is insensitive of the quasinormal mode frequency, it has less impact on the parameter estimation. For example, we have performed a Fisher analysis for the ringdown 22 mode, with the mode amplitude $A$, frequency $f_{22}$, damping rate $\gamma_{22}$ and phase $\phi$ being the unknown parameters. The corresponding measurement uncertainty associated with each detector is shown in the bottom panel of Fig.~\ref{fig:bhsnr}. We find that LIGO-HF outperforms Einstein Telescope in such parameter estimation task beyond $\sim 1500$ Hz. Precision measurement of mode frequencies from the ringdown waveform alone is important, as they may be compared with the General Relativity implication using the inspiral parameters \cite{Ghosh:2016qgn}.

\bibliographystyle{apsrev4-1}
\bibliography{Bibliography}

\begin{thebibliography}{117}%
\makeatletter
\providecommand \@ifxundefined [1]{%
 \@ifx{#1\undefined}
}%
\providecommand \@ifnum [1]{%
 \ifnum #1\expandafter \@firstoftwo
 \else \expandafter \@secondoftwo
 \fi
}%
\providecommand \@ifx [1]{%
 \ifx #1\expandafter \@firstoftwo
 \else \expandafter \@secondoftwo
 \fi
}%
\providecommand \natexlab [1]{#1}%
\providecommand \enquote  [1]{``#1''}%
\providecommand \bibnamefont  [1]{#1}%
\providecommand \bibfnamefont [1]{#1}%
\providecommand \citenamefont [1]{#1}%
\providecommand \href@noop [0]{\@secondoftwo}%
\providecommand \href [0]{\begingroup \@sanitize@url \@href}%
\providecommand \@href[1]{\@@startlink{#1}\@@href}%
\providecommand \@@href[1]{\endgroup#1\@@endlink}%
\providecommand \@sanitize@url [0]{\catcode `\\12\catcode `\$12\catcode
  `\&12\catcode `\#12\catcode `\^12\catcode `\_12\catcode `\%12\relax}%
\providecommand \@@startlink[1]{}%
\providecommand \@@endlink[0]{}%
\providecommand \url  [0]{\begingroup\@sanitize@url \@url }%
\providecommand \@url [1]{\endgroup\@href {#1}{\urlprefix }}%
\providecommand \urlprefix  [0]{URL }%
\providecommand \Eprint [0]{\href }%
\providecommand \doibase [0]{http://dx.doi.org/}%
\providecommand \selectlanguage [0]{\@gobble}%
\providecommand \bibinfo  [0]{\@secondoftwo}%
\providecommand \bibfield  [0]{\@secondoftwo}%
\providecommand \translation [1]{[#1]}%
\providecommand \BibitemOpen [0]{}%
\providecommand \bibitemStop [0]{}%
\providecommand \bibitemNoStop [0]{.\EOS\space}%
\providecommand \EOS [0]{\spacefactor3000\relax}%
\providecommand \BibitemShut  [1]{\csname bibitem#1\endcsname}%
\let\auto@bib@innerbib\@empty
\bibitem [{\citenamefont {{The {LIGO} Scientific Collaboration}}\ and\
  \citenamefont {{The Virgo
  Collaboration}}(2017{\natexlab{a}})}]{LSC_GW170817}%
  \BibitemOpen
  \bibfield  {author} {\bibinfo {author} {\bibnamefont {{The {LIGO} Scientific
  Collaboration}}}\ and\ \bibinfo {author} {\bibnamefont {{The Virgo
  Collaboration}}} (\bibinfo {collaboration} {LIGO Scientific Collaboration and
  Virgo Collaboration}),\ }\href {\doibase 10.1103/PhysRevLett.119.161101}
  {\bibfield  {journal} {\bibinfo  {journal} {Phys. Rev. Lett.}\ }\textbf
  {\bibinfo {volume} {119}},\ \bibinfo {pages} {161101} (\bibinfo {year}
  {2017}{\natexlab{a}})}\BibitemShut {NoStop}%
\bibitem [{\citenamefont {{The {LIGO} Scientific Collaboration}}\ and\
  \citenamefont {{The Virgo
  Collaboration}}(2017{\natexlab{b}})}]{LSC_MULTI_MESSENGER_2017}%
  \BibitemOpen
  \bibfield  {author} {\bibinfo {author} {\bibnamefont {{The {LIGO} Scientific
  Collaboration}}}\ and\ \bibinfo {author} {\bibnamefont {{The Virgo
  Collaboration}}} (\bibinfo {collaboration} {LIGO-Virgo Scientific
  Collaboration}),\ }\href {http://stacks.iop.org/2041-8205/848/i=2/a=L12}
  {\bibfield  {journal} {\bibinfo  {journal} {The Astrophysical Journal
  Letters}\ }\textbf {\bibinfo {volume} {848}},\ \bibinfo {pages} {L12}
  (\bibinfo {year} {2017}{\natexlab{b}})}\BibitemShut {NoStop}%
\bibitem [{\citenamefont {{The {LIGO} Scientific Collaboration}}\ and\
  \citenamefont {{The Virgo
  Collaboration}}(2017{\natexlab{c}})}]{LSC_Gamma_ray_2017}%
  \BibitemOpen
  \bibfield  {author} {\bibinfo {author} {\bibnamefont {{The {LIGO} Scientific
  Collaboration}}}\ and\ \bibinfo {author} {\bibnamefont {{The Virgo
  Collaboration}}} (\bibinfo {collaboration} {LIGO-Virgo Scientific
  Collaboration}),\ }\href {http://stacks.iop.org/2041-8205/848/i=2/a=L13}
  {\bibfield  {journal} {\bibinfo  {journal} {The Astrophysical Journal
  Letters}\ }\textbf {\bibinfo {volume} {848}},\ \bibinfo {pages} {L13}
  (\bibinfo {year} {2017}{\natexlab{c}})}\BibitemShut {NoStop}%
\bibitem [{\citenamefont {Tanvir}\ \emph {et~al.}(2017)\citenamefont {Tanvir},
  \citenamefont {Levan}, \citenamefont {GonzÃ¡lez-FernÃ¡ndez},
  \citenamefont {Korobkin}, \citenamefont {Mandel}, \citenamefont {Rosswog},
  \citenamefont {Hjorth}, \citenamefont {D?Avanzo}, \citenamefont {Fruchter},
  \citenamefont {Fryer}, \citenamefont {Kangas}, \citenamefont
  {Milvang-Jensen}, \citenamefont {Rosetti}, \citenamefont {Steeghs},
  \citenamefont {Wollaeger}, \citenamefont {Cano}, \citenamefont {Copperwheat},
  \citenamefont {Covino}, \citenamefont {D?Elia}, \citenamefont
  {de~Ugarte~Postigo}, \citenamefont {Evans}, \citenamefont {Even},
  \citenamefont {Fairhurst}, \citenamefont {Jaimes}, \citenamefont {Fontes},
  \citenamefont {Fujii}, \citenamefont {Fynbo}, \citenamefont {Gompertz},
  \citenamefont {Greiner}, \citenamefont {Hodosan}, \citenamefont {Irwin},
  \citenamefont {Jakobsson}, \citenamefont {JÃ¸rgensen}, \citenamefont
  {Kann}, \citenamefont {Lyman}, \citenamefont {Malesani}, \citenamefont
  {McMahon}, \citenamefont {Melandri}, \citenamefont {O?Brien}, \citenamefont
  {Osborne}, \citenamefont {Palazzi}, \citenamefont {Perley}, \citenamefont
  {Pian}, \citenamefont {Piranomonte}, \citenamefont {Rabus}, \citenamefont
  {Rol}, \citenamefont {Rowlinson}, \citenamefont {Schulze}, \citenamefont
  {Sutton}, \citenamefont {ThÃ¶ne}, \citenamefont {Ulaczyk}, \citenamefont
  {Watson}, \citenamefont {Wiersema},\ and\ \citenamefont
  {Wijers}}]{Tanvir_2017}%
  \BibitemOpen
  \bibfield  {author} {\bibinfo {author} {\bibfnamefont {N.~R.}\ \bibnamefont
  {Tanvir}}, \bibinfo {author} {\bibfnamefont {A.~J.}\ \bibnamefont {Levan}},
  \bibinfo {author} {\bibfnamefont {C.}~\bibnamefont
  {GonzÃ¡lez-FernÃ¡ndez}}, \bibinfo {author} {\bibfnamefont
  {O.}~\bibnamefont {Korobkin}}, \bibinfo {author} {\bibfnamefont
  {I.}~\bibnamefont {Mandel}}, \bibinfo {author} {\bibfnamefont
  {S.}~\bibnamefont {Rosswog}}, \bibinfo {author} {\bibfnamefont
  {J.}~\bibnamefont {Hjorth}}, \bibinfo {author} {\bibfnamefont
  {P.}~\bibnamefont {D?Avanzo}}, \bibinfo {author} {\bibfnamefont {A.~S.}\
  \bibnamefont {Fruchter}}, \bibinfo {author} {\bibfnamefont {C.~L.}\
  \bibnamefont {Fryer}}, \bibinfo {author} {\bibfnamefont {T.}~\bibnamefont
  {Kangas}}, \bibinfo {author} {\bibfnamefont {B.}~\bibnamefont
  {Milvang-Jensen}}, \bibinfo {author} {\bibfnamefont {S.}~\bibnamefont
  {Rosetti}}, \bibinfo {author} {\bibfnamefont {D.}~\bibnamefont {Steeghs}},
  \bibinfo {author} {\bibfnamefont {R.~T.}\ \bibnamefont {Wollaeger}}, \bibinfo
  {author} {\bibfnamefont {Z.}~\bibnamefont {Cano}}, \bibinfo {author}
  {\bibfnamefont {C.~M.}\ \bibnamefont {Copperwheat}}, \bibinfo {author}
  {\bibfnamefont {S.}~\bibnamefont {Covino}}, \bibinfo {author} {\bibfnamefont
  {V.}~\bibnamefont {D?Elia}}, \bibinfo {author} {\bibfnamefont
  {A.}~\bibnamefont {de~Ugarte~Postigo}}, \bibinfo {author} {\bibfnamefont
  {P.~A.}\ \bibnamefont {Evans}}, \bibinfo {author} {\bibfnamefont {W.~P.}\
  \bibnamefont {Even}}, \bibinfo {author} {\bibfnamefont {S.}~\bibnamefont
  {Fairhurst}}, \bibinfo {author} {\bibfnamefont {R.~F.}\ \bibnamefont
  {Jaimes}}, \bibinfo {author} {\bibfnamefont {C.~J.}\ \bibnamefont {Fontes}},
  \bibinfo {author} {\bibfnamefont {Y.~I.}\ \bibnamefont {Fujii}}, \bibinfo
  {author} {\bibfnamefont {J.~P.~U.}\ \bibnamefont {Fynbo}}, \bibinfo {author}
  {\bibfnamefont {B.~P.}\ \bibnamefont {Gompertz}}, \bibinfo {author}
  {\bibfnamefont {J.}~\bibnamefont {Greiner}}, \bibinfo {author} {\bibfnamefont
  {G.}~\bibnamefont {Hodosan}}, \bibinfo {author} {\bibfnamefont {M.~J.}\
  \bibnamefont {Irwin}}, \bibinfo {author} {\bibfnamefont {P.}~\bibnamefont
  {Jakobsson}}, \bibinfo {author} {\bibfnamefont {U.~G.}\ \bibnamefont
  {JÃ¸rgensen}}, \bibinfo {author} {\bibfnamefont {D.~A.}\ \bibnamefont
  {Kann}}, \bibinfo {author} {\bibfnamefont {J.~D.}\ \bibnamefont {Lyman}},
  \bibinfo {author} {\bibfnamefont {D.}~\bibnamefont {Malesani}}, \bibinfo
  {author} {\bibfnamefont {R.~G.}\ \bibnamefont {McMahon}}, \bibinfo {author}
  {\bibfnamefont {A.}~\bibnamefont {Melandri}}, \bibinfo {author}
  {\bibfnamefont {P.~T.}\ \bibnamefont {O?Brien}}, \bibinfo {author}
  {\bibfnamefont {J.~P.}\ \bibnamefont {Osborne}}, \bibinfo {author}
  {\bibfnamefont {E.}~\bibnamefont {Palazzi}}, \bibinfo {author} {\bibfnamefont
  {D.~A.}\ \bibnamefont {Perley}}, \bibinfo {author} {\bibfnamefont
  {E.}~\bibnamefont {Pian}}, \bibinfo {author} {\bibfnamefont {S.}~\bibnamefont
  {Piranomonte}}, \bibinfo {author} {\bibfnamefont {M.}~\bibnamefont {Rabus}},
  \bibinfo {author} {\bibfnamefont {E.}~\bibnamefont {Rol}}, \bibinfo {author}
  {\bibfnamefont {A.}~\bibnamefont {Rowlinson}}, \bibinfo {author}
  {\bibfnamefont {S.}~\bibnamefont {Schulze}}, \bibinfo {author} {\bibfnamefont
  {P.}~\bibnamefont {Sutton}}, \bibinfo {author} {\bibfnamefont {C.~C.}\
  \bibnamefont {ThÃ¶ne}}, \bibinfo {author} {\bibfnamefont {K.}~\bibnamefont
  {Ulaczyk}}, \bibinfo {author} {\bibfnamefont {D.}~\bibnamefont {Watson}},
  \bibinfo {author} {\bibfnamefont {K.}~\bibnamefont {Wiersema}}, \ and\
  \bibinfo {author} {\bibfnamefont {R.~A. M.~J.}\ \bibnamefont {Wijers}},\
  }\href@noop {} {\bibfield  {journal} {\bibinfo  {journal} {The Astrophysical
  Journal Letters}\ }\textbf {\bibinfo {volume} {848}},\ \bibinfo {pages} {L27}
  (\bibinfo {year} {2017})}\BibitemShut {NoStop}%
\bibitem [{\citenamefont {Abbott}\ \emph {et~al.}(2018)\citenamefont {Abbott}
  \emph {et~al.}}]{LSC_EoS:2018}%
  \BibitemOpen
  \bibfield  {author} {\bibinfo {author} {\bibfnamefont {B.~P.}\ \bibnamefont
  {Abbott}} \emph {et~al.} (\bibinfo {collaboration} {The LIGO Scientific
  Collaboration and the Virgo Collaboration}),\ }\href {\doibase
  10.1103/PhysRevLett.121.161101} {\bibfield  {journal} {\bibinfo  {journal}
  {Phys. Rev. Lett.}\ }\textbf {\bibinfo {volume} {121}},\ \bibinfo {pages}
  {161101} (\bibinfo {year} {2018})}\BibitemShut {NoStop}%
\bibitem [{\citenamefont {{The {LIGO} Scientific Collaboration}}\ \emph
  {et~al.}(2017)\citenamefont {{The {LIGO} Scientific Collaboration}},
  \citenamefont {{The Virgo Collaboration}}, \citenamefont {{The 1M2H
  Collaboration}}, \citenamefont {{The Dark Energy Camera GW-EM Collaboration
  and the DES Collaboration}}, \citenamefont {{The DLT40 Collaboration}},
  \citenamefont {{The Las Cumbres Observatory Collaboration}}, \citenamefont
  {{The VINROUGE Collaboration}},\ and\ \citenamefont {{The MASTER
  Collaboration}}}]{LSC_Hubble_2017}%
  \BibitemOpen
  \bibfield  {author} {\bibinfo {author} {\bibnamefont {{The {LIGO} Scientific
  Collaboration}}}, \bibinfo {author} {\bibnamefont {{The Virgo
  Collaboration}}}, \bibinfo {author} {\bibnamefont {{The 1M2H
  Collaboration}}}, \bibinfo {author} {\bibnamefont {{The Dark Energy Camera
  GW-EM Collaboration and the DES Collaboration}}}, \bibinfo {author}
  {\bibnamefont {{The DLT40 Collaboration}}}, \bibinfo {author} {\bibnamefont
  {{The Las Cumbres Observatory Collaboration}}}, \bibinfo {author}
  {\bibnamefont {{The VINROUGE Collaboration}}}, \ and\ \bibinfo {author}
  {\bibnamefont {{The MASTER Collaboration}}},\ }\href {\doibase
  10.1038/nature24471} {\bibfield  {journal} {\bibinfo  {journal} {Nature}\
  }\textbf {\bibinfo {volume} {551}},\ \bibinfo {pages} {85?88} (\bibinfo
  {year} {2017})}\BibitemShut {NoStop}%
\bibitem [{\citenamefont {{The LIGO Scientific
  Collaboration}}(2015)}]{LSC_aLIGO_2015}%
  \BibitemOpen
  \bibfield  {author} {\bibinfo {author} {\bibnamefont {{The LIGO Scientific
  Collaboration}}},\ }\href {http://stacks.iop.org/0264-9381/32/i=7/a=074001}
  {\bibfield  {journal} {\bibinfo  {journal} {Classical and Quantum Gravity}\
  }\textbf {\bibinfo {volume} {32}},\ \bibinfo {pages} {074001} (\bibinfo
  {year} {2015})}\BibitemShut {NoStop}%
\bibitem [{\citenamefont {Martynov}\ \emph {et~al.}(2016)\citenamefont
  {Martynov}, \citenamefont {Hall}, \citenamefont {Abbott} \emph
  {et~al.}}]{Martynov_Noise_2016}%
  \BibitemOpen
  \bibfield  {author} {\bibinfo {author} {\bibfnamefont {D.~V.}\ \bibnamefont
  {Martynov}}, \bibinfo {author} {\bibfnamefont {E.~D.}\ \bibnamefont {Hall}},
  \bibinfo {author} {\bibfnamefont {B.~P.}\ \bibnamefont {Abbott}},  \emph
  {et~al.},\ }\href {\doibase 10.1103/PhysRevD.93.112004} {\bibfield  {journal}
  {\bibinfo  {journal} {Phys. Rev. D}\ }\textbf {\bibinfo {volume} {93}},\
  \bibinfo {pages} {112004} (\bibinfo {year} {2016})}\BibitemShut {NoStop}%
\bibitem [{\citenamefont {{Bauswein}}\ and\ \citenamefont
  {{Janka}}(2012)}]{Bauswein2012}%
  \BibitemOpen
  \bibfield  {author} {\bibinfo {author} {\bibfnamefont {A.}~\bibnamefont
  {{Bauswein}}}\ and\ \bibinfo {author} {\bibfnamefont {H.-T.}\ \bibnamefont
  {{Janka}}},\ }\href {\doibase 10.1103/PhysRevLett.108.011101} {\bibfield
  {journal} {\bibinfo  {journal} {\prl}\ }\textbf {\bibinfo {volume} {108}},\
  \bibinfo {eid} {011101} (\bibinfo {year} {2012})},\ \Eprint
  {http://arxiv.org/abs/1106.1616} {arXiv:1106.1616 [astro-ph.SR]} \BibitemShut
  {NoStop}%
\bibitem [{\citenamefont {{Bauswein}}\ \emph {et~al.}(2012)\citenamefont
  {{Bauswein}}, \citenamefont {{Janka}}, \citenamefont {{Hebeler}},\ and\
  \citenamefont {{Schwenk}}}]{Bauswein2012a}%
  \BibitemOpen
  \bibfield  {author} {\bibinfo {author} {\bibfnamefont {A.}~\bibnamefont
  {{Bauswein}}}, \bibinfo {author} {\bibfnamefont {H.-T.}\ \bibnamefont
  {{Janka}}}, \bibinfo {author} {\bibfnamefont {K.}~\bibnamefont {{Hebeler}}},
  \ and\ \bibinfo {author} {\bibfnamefont {A.}~\bibnamefont {{Schwenk}}},\
  }\href {\doibase 10.1103/PhysRevD.86.063001} {\bibfield  {journal} {\bibinfo
  {journal} {\prd}\ }\textbf {\bibinfo {volume} {86}},\ \bibinfo {eid} {063001}
  (\bibinfo {year} {2012})},\ \Eprint {http://arxiv.org/abs/1204.1888}
  {arXiv:1204.1888 [astro-ph.SR]} \BibitemShut {NoStop}%
\bibitem [{\citenamefont {{Hotokezaka}}\ \emph {et~al.}(2013)\citenamefont
  {{Hotokezaka}}, \citenamefont {{Kiuchi}}, \citenamefont {{Kyutoku}},
  \citenamefont {{Muranushi}}, \citenamefont {{Sekiguchi}}, \citenamefont
  {{Shibata}},\ and\ \citenamefont {{Taniguchi}}}]{Hotokezaka2013a}%
  \BibitemOpen
  \bibfield  {author} {\bibinfo {author} {\bibfnamefont {K.}~\bibnamefont
  {{Hotokezaka}}}, \bibinfo {author} {\bibfnamefont {K.}~\bibnamefont
  {{Kiuchi}}}, \bibinfo {author} {\bibfnamefont {K.}~\bibnamefont {{Kyutoku}}},
  \bibinfo {author} {\bibfnamefont {T.}~\bibnamefont {{Muranushi}}}, \bibinfo
  {author} {\bibfnamefont {Y.}~\bibnamefont {{Sekiguchi}}}, \bibinfo {author}
  {\bibfnamefont {M.}~\bibnamefont {{Shibata}}}, \ and\ \bibinfo {author}
  {\bibfnamefont {K.}~\bibnamefont {{Taniguchi}}},\ }\href@noop {} {\bibfield
  {journal} {\bibinfo  {journal} {\prd}\ }\textbf {\bibinfo {volume} {88}},\
  \bibinfo {eid} {044026} (\bibinfo {year} {2013})}\BibitemShut {NoStop}%
\bibitem [{\citenamefont {Takami}\ \emph {et~al.}(2014)\citenamefont {Takami},
  \citenamefont {Rezzolla},\ and\ \citenamefont {Baiotti}}]{Takami2014}%
  \BibitemOpen
  \bibfield  {author} {\bibinfo {author} {\bibfnamefont {K.}~\bibnamefont
  {Takami}}, \bibinfo {author} {\bibfnamefont {L.}~\bibnamefont {Rezzolla}}, \
  and\ \bibinfo {author} {\bibfnamefont {L.}~\bibnamefont {Baiotti}},\ }\href
  {\doibase 10.1103/PhysRevLett.113.091104} {\bibfield  {journal} {\bibinfo
  {journal} {\prl}\ }\textbf {\bibinfo {volume} {113}},\ \bibinfo {eid}
  {091104} (\bibinfo {year} {2014})},\ \Eprint {http://arxiv.org/abs/1403.5672}
  {arXiv:1403.5672 [gr-qc]} \BibitemShut {NoStop}%
\bibitem [{\citenamefont {{Bauswein}}\ and\ \citenamefont
  {{Stergioulas}}(2015)}]{Bauswein2015}%
  \BibitemOpen
  \bibfield  {author} {\bibinfo {author} {\bibfnamefont {A.}~\bibnamefont
  {{Bauswein}}}\ and\ \bibinfo {author} {\bibfnamefont {N.}~\bibnamefont
  {{Stergioulas}}},\ }\href {\doibase 10.1103/PhysRevD.91.124056} {\bibfield
  {journal} {\bibinfo  {journal} {\prd}\ }\textbf {\bibinfo {volume} {91}},\
  \bibinfo {eid} {124056} (\bibinfo {year} {2015})},\ \Eprint
  {http://arxiv.org/abs/1502.03176} {arXiv:1502.03176 [astro-ph.SR]}
  \BibitemShut {NoStop}%
\bibitem [{\citenamefont {Messenger}\ and\ \citenamefont
  {Read}(2012)}]{Messenger_2012}%
  \BibitemOpen
  \bibfield  {author} {\bibinfo {author} {\bibfnamefont {C.}~\bibnamefont
  {Messenger}}\ and\ \bibinfo {author} {\bibfnamefont {J.}~\bibnamefont
  {Read}},\ }\href {\doibase 10.1103/PhysRevLett.108.091101} {\bibfield
  {journal} {\bibinfo  {journal} {Phys. Rev. Lett.}\ }\textbf {\bibinfo
  {volume} {108}},\ \bibinfo {pages} {091101} (\bibinfo {year}
  {2012})}\BibitemShut {NoStop}%
\bibitem [{\citenamefont {Berti}\ \emph {et~al.}(2018)\citenamefont {Berti},
  \citenamefont {Yagi}, \citenamefont {Yang},\ and\ \citenamefont
  {Yunes}}]{Berti:2018vdi}%
  \BibitemOpen
  \bibfield  {author} {\bibinfo {author} {\bibfnamefont {E.}~\bibnamefont
  {Berti}}, \bibinfo {author} {\bibfnamefont {K.}~\bibnamefont {Yagi}},
  \bibinfo {author} {\bibfnamefont {H.}~\bibnamefont {Yang}}, \ and\ \bibinfo
  {author} {\bibfnamefont {N.}~\bibnamefont {Yunes}},\ }\href {\doibase
  10.1007/s10714-018-2372-6} {\bibfield  {journal} {\bibinfo  {journal} {Gen.
  Rel. Grav.}\ }\textbf {\bibinfo {volume} {50}},\ \bibinfo {pages} {49}
  (\bibinfo {year} {2018})},\ \Eprint {http://arxiv.org/abs/1801.03587}
  {arXiv:1801.03587 [gr-qc]} \BibitemShut {NoStop}%
\bibitem [{\citenamefont {Harry}\ \emph {et~al.}(2007)\citenamefont {Harry},
  \citenamefont {Abernathy}, \citenamefont {Becerra-Toledo}, \citenamefont
  {Armandula}, \citenamefont {Black}, \citenamefont {Dooley}, \citenamefont
  {Eichenfield}, \citenamefont {Nwabugwu}, \citenamefont {Villar},
  \citenamefont {Crooks}, \citenamefont {Cagnoli}, \citenamefont {Hough},
  \citenamefont {How}, \citenamefont {MacLaren}, \citenamefont {Murray},
  \citenamefont {Reid}, \citenamefont {Rowan}, \citenamefont {Sneddon},
  \citenamefont {Fejer}, \citenamefont {Route}, \citenamefont {Penn},
  \citenamefont {Ganau}, \citenamefont {Mackowski}, \citenamefont {Michel},
  \citenamefont {Pinard},\ and\ \citenamefont {Remillieux}}]{Harry_2007}%
  \BibitemOpen
  \bibfield  {author} {\bibinfo {author} {\bibfnamefont {G.~M.}\ \bibnamefont
  {Harry}}, \bibinfo {author} {\bibfnamefont {M.~R.}\ \bibnamefont
  {Abernathy}}, \bibinfo {author} {\bibfnamefont {A.~E.}\ \bibnamefont
  {Becerra-Toledo}}, \bibinfo {author} {\bibfnamefont {H.}~\bibnamefont
  {Armandula}}, \bibinfo {author} {\bibfnamefont {E.}~\bibnamefont {Black}},
  \bibinfo {author} {\bibfnamefont {K.}~\bibnamefont {Dooley}}, \bibinfo
  {author} {\bibfnamefont {M.}~\bibnamefont {Eichenfield}}, \bibinfo {author}
  {\bibfnamefont {C.}~\bibnamefont {Nwabugwu}}, \bibinfo {author}
  {\bibfnamefont {A.}~\bibnamefont {Villar}}, \bibinfo {author} {\bibfnamefont
  {D.~R.~M.}\ \bibnamefont {Crooks}}, \bibinfo {author} {\bibfnamefont
  {G.}~\bibnamefont {Cagnoli}}, \bibinfo {author} {\bibfnamefont
  {J.}~\bibnamefont {Hough}}, \bibinfo {author} {\bibfnamefont {C.~R.}\
  \bibnamefont {How}}, \bibinfo {author} {\bibfnamefont {I.}~\bibnamefont
  {MacLaren}}, \bibinfo {author} {\bibfnamefont {P.}~\bibnamefont {Murray}},
  \bibinfo {author} {\bibfnamefont {S.}~\bibnamefont {Reid}}, \bibinfo {author}
  {\bibfnamefont {S.}~\bibnamefont {Rowan}}, \bibinfo {author} {\bibfnamefont
  {P.~H.}\ \bibnamefont {Sneddon}}, \bibinfo {author} {\bibfnamefont {M.~M.}\
  \bibnamefont {Fejer}}, \bibinfo {author} {\bibfnamefont {R.}~\bibnamefont
  {Route}}, \bibinfo {author} {\bibfnamefont {S.~D.}\ \bibnamefont {Penn}},
  \bibinfo {author} {\bibfnamefont {P.}~\bibnamefont {Ganau}}, \bibinfo
  {author} {\bibfnamefont {J.-M.}\ \bibnamefont {Mackowski}}, \bibinfo {author}
  {\bibfnamefont {C.}~\bibnamefont {Michel}}, \bibinfo {author} {\bibfnamefont
  {L.}~\bibnamefont {Pinard}}, \ and\ \bibinfo {author} {\bibfnamefont
  {A.}~\bibnamefont {Remillieux}},\ }\href
  {http://stacks.iop.org/0264-9381/24/i=2/a=008} {\bibfield  {journal}
  {\bibinfo  {journal} {Classical and Quantum Gravity}\ }\textbf {\bibinfo
  {volume} {24}},\ \bibinfo {pages} {405} (\bibinfo {year} {2007})}\BibitemShut
  {NoStop}%
\bibitem [{\citenamefont {Zucker}\ and\ \citenamefont
  {Whitcomb}(1996)}]{Zucker_GAS_1996}%
  \BibitemOpen
  \bibfield  {author} {\bibinfo {author} {\bibfnamefont {M.}~\bibnamefont
  {Zucker}}\ and\ \bibinfo {author} {\bibfnamefont {S.}~\bibnamefont
  {Whitcomb}},\ }in\ \href@noop {} {\emph {\bibinfo {booktitle} {Proceedings of
  the Seventh Marcel Grossman Meeting on recent developments in theoretical and
  experimental general relativity, gravitation, and relativistic field
  theories}}}\ (\bibinfo {year} {1996})\ pp.\ \bibinfo {pages}
  {1434--1436}\BibitemShut {NoStop}%
\bibitem [{\citenamefont {{The LIGO Scientific
  Collaboration}}(2013)}]{LSC_SQUEEZING_2013}%
  \BibitemOpen
  \bibfield  {author} {\bibinfo {author} {\bibnamefont {{The LIGO Scientific
  Collaboration}}},\ }\href {\doibase 10.1038/nphoton.2013.177} {\bibfield
  {journal} {\bibinfo  {journal} {Nature Photonics}\ }\textbf {\bibinfo
  {volume} {7}},\ \bibinfo {pages} {613} (\bibinfo {year} {2013})}\BibitemShut
  {NoStop}%
\bibitem [{\citenamefont {Miao}\ \emph
  {et~al.}(2018{\natexlab{a}})\citenamefont {Miao}, \citenamefont {Yang},\ and\
  \citenamefont {Martynov}}]{Miao_kHz_2018}%
  \BibitemOpen
  \bibfield  {author} {\bibinfo {author} {\bibfnamefont {H.}~\bibnamefont
  {Miao}}, \bibinfo {author} {\bibfnamefont {H.}~\bibnamefont {Yang}}, \ and\
  \bibinfo {author} {\bibfnamefont {D.}~\bibnamefont {Martynov}},\ }\href
  {\doibase 10.1103/PhysRevD.98.044044} {\bibfield  {journal} {\bibinfo
  {journal} {Phys. Rev. D}\ }\textbf {\bibinfo {volume} {98}},\ \bibinfo
  {pages} {044044} (\bibinfo {year} {2018}{\natexlab{a}})}\BibitemShut
  {NoStop}%
\bibitem [{\citenamefont {Th{\"{u}}ring}\ \emph {et~al.}(2007)\citenamefont
  {Th{\"{u}}ring}, \citenamefont {Schnabel}, \citenamefont {L{\"{u}}ck},\ and\
  \citenamefont {Danzmann}}]{Thuring_SRC_2007}%
  \BibitemOpen
  \bibfield  {author} {\bibinfo {author} {\bibfnamefont {A.}~\bibnamefont
  {Th{\"{u}}ring}}, \bibinfo {author} {\bibfnamefont {R.}~\bibnamefont
  {Schnabel}}, \bibinfo {author} {\bibfnamefont {H.}~\bibnamefont
  {L{\"{u}}ck}}, \ and\ \bibinfo {author} {\bibfnamefont {K.}~\bibnamefont
  {Danzmann}},\ }\href {\doibase 10.1364/OL.32.000985} {\bibfield  {journal}
  {\bibinfo  {journal} {Optics letters}\ }\textbf {\bibinfo {volume} {32}},\
  \bibinfo {pages} {985} (\bibinfo {year} {2007})}\BibitemShut {NoStop}%
\bibitem [{\citenamefont {Th{\"{u}}ring}\ \emph {et~al.}(2009)\citenamefont
  {Th{\"{u}}ring}, \citenamefont {Gr{\"{a}}f}, \citenamefont {Vahlbruch},
  \citenamefont {Mehmet}, \citenamefont {Danzmann},\ and\ \citenamefont
  {Schnabel}}]{Thuering_SRC_2009}%
  \BibitemOpen
  \bibfield  {author} {\bibinfo {author} {\bibfnamefont {A.}~\bibnamefont
  {Th{\"{u}}ring}}, \bibinfo {author} {\bibfnamefont {C.}~\bibnamefont
  {Gr{\"{a}}f}}, \bibinfo {author} {\bibfnamefont {H.}~\bibnamefont
  {Vahlbruch}}, \bibinfo {author} {\bibfnamefont {M.}~\bibnamefont {Mehmet}},
  \bibinfo {author} {\bibfnamefont {K.}~\bibnamefont {Danzmann}}, \ and\
  \bibinfo {author} {\bibfnamefont {R.}~\bibnamefont {Schnabel}},\ }\href
  {http://ol.osa.org/abstract.cfm?URI=ol-34-6-824} {\bibfield  {journal}
  {\bibinfo  {journal} {Opt. Lett.}\ }\textbf {\bibinfo {volume} {34}},\
  \bibinfo {pages} {824} (\bibinfo {year} {2009})}\BibitemShut {NoStop}%
\bibitem [{\citenamefont {Gr{\"{a}}f}\ \emph {et~al.}(2013)\citenamefont
  {Gr{\"{a}}f}, \citenamefont {Th{\"{u}}ring}, \citenamefont {Vahlbruch},
  \citenamefont {Danzmann},\ and\ \citenamefont {Schnabel}}]{Graf_SRC_2013}%
  \BibitemOpen
  \bibfield  {author} {\bibinfo {author} {\bibfnamefont {C.}~\bibnamefont
  {Gr{\"{a}}f}}, \bibinfo {author} {\bibfnamefont {A.}~\bibnamefont
  {Th{\"{u}}ring}}, \bibinfo {author} {\bibfnamefont {H.}~\bibnamefont
  {Vahlbruch}}, \bibinfo {author} {\bibfnamefont {K.}~\bibnamefont {Danzmann}},
  \ and\ \bibinfo {author} {\bibfnamefont {R.}~\bibnamefont {Schnabel}},\
  }\href {\doibase 10.1364/OE.21.005287} {\bibfield  {journal} {\bibinfo
  {journal} {Optics express}\ }\textbf {\bibinfo {volume} {21}},\ \bibinfo
  {pages} {5287} (\bibinfo {year} {2013})}\BibitemShut {NoStop}%
\bibitem [{\citenamefont {Miao}\ \emph {et~al.}(2014)\citenamefont {Miao},
  \citenamefont {Yang}, \citenamefont {Adhikari},\ and\ \citenamefont
  {Chen}}]{Miao_SRC_2014}%
  \BibitemOpen
  \bibfield  {author} {\bibinfo {author} {\bibfnamefont {H.}~\bibnamefont
  {Miao}}, \bibinfo {author} {\bibfnamefont {H.}~\bibnamefont {Yang}}, \bibinfo
  {author} {\bibfnamefont {R.~X.}\ \bibnamefont {Adhikari}}, \ and\ \bibinfo
  {author} {\bibfnamefont {Y.}~\bibnamefont {Chen}},\ }\href {\doibase
  10.1088/0264-9381/31/16/165010} {\bibfield  {journal} {\bibinfo  {journal}
  {Class. Quant Grav.}\ }\textbf {\bibinfo {volume} {31}},\ \bibinfo {pages}
  {165010} (\bibinfo {year} {2014})}\BibitemShut {NoStop}%
\bibitem [{\citenamefont {Essick}\ \emph
  {et~al.}(2017{\natexlab{a}})\citenamefont {Essick}, \citenamefont {Vitale},\
  and\ \citenamefont {Evans}}]{Essick_2017}%
  \BibitemOpen
  \bibfield  {author} {\bibinfo {author} {\bibfnamefont {R.}~\bibnamefont
  {Essick}}, \bibinfo {author} {\bibfnamefont {S.}~\bibnamefont {Vitale}}, \
  and\ \bibinfo {author} {\bibfnamefont {M.}~\bibnamefont {Evans}},\ }\href
  {\doibase 10.1103/PhysRevD.96.084004} {\bibfield  {journal} {\bibinfo
  {journal} {Phys. Rev. D}\ }\textbf {\bibinfo {volume} {96}},\ \bibinfo
  {pages} {084004} (\bibinfo {year} {2017}{\natexlab{a}})}\BibitemShut
  {NoStop}%
\bibitem [{\citenamefont {Punturo}\ \emph {et~al.}(2010)\citenamefont
  {Punturo}, \citenamefont {Abernathy}, \citenamefont {Acernese}, \citenamefont
  {Allen}, \citenamefont {Andersson}, \citenamefont {Arun}, \citenamefont
  {Barone}, \citenamefont {Barr},\ and\ \citenamefont
  {Others}}]{Punturo_ET_2010}%
  \BibitemOpen
  \bibfield  {author} {\bibinfo {author} {\bibfnamefont {M.}~\bibnamefont
  {Punturo}}, \bibinfo {author} {\bibfnamefont {M.}~\bibnamefont {Abernathy}},
  \bibinfo {author} {\bibfnamefont {F.}~\bibnamefont {Acernese}}, \bibinfo
  {author} {\bibfnamefont {B.}~\bibnamefont {Allen}}, \bibinfo {author}
  {\bibfnamefont {N.}~\bibnamefont {Andersson}}, \bibinfo {author}
  {\bibfnamefont {K.}~\bibnamefont {Arun}}, \bibinfo {author} {\bibfnamefont
  {F.}~\bibnamefont {Barone}}, \bibinfo {author} {\bibfnamefont
  {B.}~\bibnamefont {Barr}}, \ and\ \bibinfo {author} {\bibnamefont {Others}},\
  }\href {http://stacks.iop.org/0264-9381/27/i=19/a=194002} {\bibfield
  {journal} {\bibinfo  {journal} {Classical and Quantum Gravity}\ }\textbf
  {\bibinfo {volume} {27}},\ \bibinfo {pages} {194002} (\bibinfo {year}
  {2010})}\BibitemShut {NoStop}%
\bibitem [{\citenamefont {{The LIGO Scientific
  Collaboration}}(2017)}]{LSC_FUTURE_2017}%
  \BibitemOpen
  \bibfield  {author} {\bibinfo {author} {\bibnamefont {{The LIGO Scientific
  Collaboration}}},\ }\href
  {http://iopscience.iop.org/article/10.1088/1361-6382/aa51f4} {\bibfield
  {journal} {\bibinfo  {journal} {Classical and Quantum Gravity}\ }\textbf
  {\bibinfo {volume} {34}},\ \bibinfo {pages} {44001} (\bibinfo {year}
  {2017})}\BibitemShut {NoStop}%
\bibitem [{\citenamefont {{The LIGO Scientific
  Collaboration}}(2014)}]{LSC_White_2014}%
  \BibitemOpen
  \bibfield  {author} {\bibinfo {author} {\bibnamefont {{The LIGO Scientific
  Collaboration}}},\ }\href
  {https://dcc.ligo.org/public/0113/T1400316/004/T1400316-v4.pdf} {\bibfield
  {journal} {\bibinfo  {journal} {LIGO DCC-T1400316}\ } (\bibinfo {year}
  {2014})}\BibitemShut {NoStop}%
\bibitem [{\citenamefont {Adhikari}\ \emph {et~al.}(2014)\citenamefont
  {Adhikari}, \citenamefont {Smith}, \citenamefont {Brooks}, \citenamefont
  {Barsotti}, \citenamefont {Shapiro}, \citenamefont {Lantz}, \citenamefont
  {McClelland}, \citenamefont {Gustafson}, \citenamefont {Martynov},
  \citenamefont {Mitrofanov}, \citenamefont {Coyne}, \citenamefont {Arai},
  \citenamefont {Torrie},\ and\ \citenamefont {Wipf}}]{Adhikari_VOYAGER_2014}%
  \BibitemOpen
  \bibfield  {author} {\bibinfo {author} {\bibfnamefont {R.~X.}\ \bibnamefont
  {Adhikari}}, \bibinfo {author} {\bibfnamefont {N.}~\bibnamefont {Smith}},
  \bibinfo {author} {\bibfnamefont {A.}~\bibnamefont {Brooks}}, \bibinfo
  {author} {\bibfnamefont {L.}~\bibnamefont {Barsotti}}, \bibinfo {author}
  {\bibfnamefont {B.}~\bibnamefont {Shapiro}}, \bibinfo {author} {\bibfnamefont
  {B.}~\bibnamefont {Lantz}}, \bibinfo {author} {\bibfnamefont
  {D.}~\bibnamefont {McClelland}}, \bibinfo {author} {\bibfnamefont {E.~K.}\
  \bibnamefont {Gustafson}}, \bibinfo {author} {\bibfnamefont {D.~V.}\
  \bibnamefont {Martynov}}, \bibinfo {author} {\bibfnamefont {V.}~\bibnamefont
  {Mitrofanov}}, \bibinfo {author} {\bibfnamefont {D.}~\bibnamefont {Coyne}},
  \bibinfo {author} {\bibfnamefont {K.}~\bibnamefont {Arai}}, \bibinfo {author}
  {\bibfnamefont {C.}~\bibnamefont {Torrie}}, \ and\ \bibinfo {author}
  {\bibfnamefont {C.}~\bibnamefont {Wipf}},\ }\href
  {https://dcc.ligo.org/DocDB/0112/T1400226/009/main.pdf} {\bibfield  {journal}
  {\bibinfo  {journal} {LIGO DCC-T1400226}\ } (\bibinfo {year}
  {2014})}\BibitemShut {NoStop}%
\bibitem [{\citenamefont {{Hild}}\ \emph {et~al.}(2007)\citenamefont {{Hild}},
  \citenamefont {{Grote}}, \citenamefont {{Hewtison}}, \citenamefont
  {{L{\"u}ck}}, \citenamefont {{Smith}}, \citenamefont {{Strain}},
  \citenamefont {{Willke}},\ and\ \citenamefont {{Danzmann}}}]{Hild_SRC_2007}%
  \BibitemOpen
  \bibfield  {author} {\bibinfo {author} {\bibfnamefont {S.}~\bibnamefont
  {{Hild}}}, \bibinfo {author} {\bibfnamefont {H.}~\bibnamefont {{Grote}}},
  \bibinfo {author} {\bibfnamefont {M.}~\bibnamefont {{Hewtison}}}, \bibinfo
  {author} {\bibfnamefont {H.}~\bibnamefont {{L{\"u}ck}}}, \bibinfo {author}
  {\bibfnamefont {J.~R.}\ \bibnamefont {{Smith}}}, \bibinfo {author}
  {\bibfnamefont {K.~A.}\ \bibnamefont {{Strain}}}, \bibinfo {author}
  {\bibfnamefont {B.}~\bibnamefont {{Willke}}}, \ and\ \bibinfo {author}
  {\bibfnamefont {K.}~\bibnamefont {{Danzmann}}},\ }\href {\doibase
  10.1088/0264-9381/24/6/009} {\bibfield  {journal} {\bibinfo  {journal}
  {Classical and Quantum Gravity}\ }\textbf {\bibinfo {volume} {24}},\ \bibinfo
  {pages} {1513} (\bibinfo {year} {2007})}\BibitemShut {NoStop}%
\bibitem [{\citenamefont {Ward}(2010)}]{Ward_THESIS_2010}%
  \BibitemOpen
  \bibfield  {author} {\bibinfo {author} {\bibfnamefont {R.}~\bibnamefont
  {Ward}},\ }\emph {\bibinfo {title} {Length Sensing and Control of a Prototype
  Advanced Interferometric Gravitational Wave Detector}},\ \href
  {https://thesis.library.caltech.edu/5836/} {Ph.D. thesis},\ \bibinfo
  {school} {Caltech} (\bibinfo {year} {2010})\BibitemShut {NoStop}%
\bibitem [{\citenamefont {Kimble}\ \emph {et~al.}(2001)\citenamefont {Kimble},
  \citenamefont {Levin}, \citenamefont {Matsko}, \citenamefont {Thorne},\ and\
  \citenamefont {Vyatchanin}}]{Kimble_FILCAV_2001}%
  \BibitemOpen
  \bibfield  {author} {\bibinfo {author} {\bibfnamefont {H.~J.}\ \bibnamefont
  {Kimble}}, \bibinfo {author} {\bibfnamefont {Y.}~\bibnamefont {Levin}},
  \bibinfo {author} {\bibfnamefont {A.~B.}\ \bibnamefont {Matsko}}, \bibinfo
  {author} {\bibfnamefont {K.~S.}\ \bibnamefont {Thorne}}, \ and\ \bibinfo
  {author} {\bibfnamefont {S.~P.}\ \bibnamefont {Vyatchanin}},\ }\href
  {http://link.aps.org/doi/10.1103/PhysRevD.65.022002} {\bibfield  {journal}
  {\bibinfo  {journal} {Phys. Rev. D}\ }\textbf {\bibinfo {volume} {65}},\
  \bibinfo {pages} {022002} (\bibinfo {year} {2001})}\BibitemShut {NoStop}%
\bibitem [{\citenamefont {Martynov}(2015)}]{Martynov_THESIS_2015}%
  \BibitemOpen
  \bibfield  {author} {\bibinfo {author} {\bibfnamefont {D.}~\bibnamefont
  {Martynov}},\ }\emph {\bibinfo {title} {Lock Acquisition and Sensitivity
  Analysis of Advanced LIGO Interferometers}},\ \href
  {http://resolver.caltech.edu/CaltechTHESIS:05282015-142013480} {Ph.D.
  thesis},\ \bibinfo  {school} {Caltech} (\bibinfo {year} {2015})\BibitemShut
  {NoStop}%
\bibitem [{\citenamefont {Staley}\ \emph {et~al.}(2014)\citenamefont {Staley},
  \citenamefont {Martynov}, \citenamefont {Abbott}, \citenamefont {Adhikari},
  \citenamefont {Arai}, \citenamefont {Ballmer}, \citenamefont {Barsotti},
  \citenamefont {Brooks}, \citenamefont {DeRosa}, \citenamefont {Dwyer},
  \citenamefont {Effler}, \citenamefont {Evans}, \citenamefont {Fritschel},
  \citenamefont {Frolov}, \citenamefont {Gray}, \citenamefont {Guido},
  \citenamefont {Gustafson}, \citenamefont {Heintze}, \citenamefont {Hoak},
  \citenamefont {Izumi}, \citenamefont {Kawabe}, \citenamefont {King},
  \citenamefont {Kissel}, \citenamefont {Kokeyama}, \citenamefont {Landry},
  \citenamefont {McClelland}, \citenamefont {Miller}, \citenamefont {Mullavey},
  \citenamefont {O?Reilly}, \citenamefont {Rollins}, \citenamefont {Sanders},
  \citenamefont {Schofield}, \citenamefont {Sigg}, \citenamefont {Slagmolen},
  \citenamefont {Smith-Lefebvre}, \citenamefont {Vajente}, \citenamefont
  {Ward},\ and\ \citenamefont {Wipf}}]{Staley_LOCK_2014}%
  \BibitemOpen
  \bibfield  {author} {\bibinfo {author} {\bibfnamefont {A.}~\bibnamefont
  {Staley}}, \bibinfo {author} {\bibfnamefont {D.}~\bibnamefont {Martynov}},
  \bibinfo {author} {\bibfnamefont {R.}~\bibnamefont {Abbott}}, \bibinfo
  {author} {\bibfnamefont {R.~X.}\ \bibnamefont {Adhikari}}, \bibinfo {author}
  {\bibfnamefont {K.}~\bibnamefont {Arai}}, \bibinfo {author} {\bibfnamefont
  {S.}~\bibnamefont {Ballmer}}, \bibinfo {author} {\bibfnamefont
  {L.}~\bibnamefont {Barsotti}}, \bibinfo {author} {\bibfnamefont {A.~F.}\
  \bibnamefont {Brooks}}, \bibinfo {author} {\bibfnamefont {R.~T.}\
  \bibnamefont {DeRosa}}, \bibinfo {author} {\bibfnamefont {S.}~\bibnamefont
  {Dwyer}}, \bibinfo {author} {\bibfnamefont {A.}~\bibnamefont {Effler}},
  \bibinfo {author} {\bibfnamefont {M.}~\bibnamefont {Evans}}, \bibinfo
  {author} {\bibfnamefont {P.}~\bibnamefont {Fritschel}}, \bibinfo {author}
  {\bibfnamefont {V.~V.}\ \bibnamefont {Frolov}}, \bibinfo {author}
  {\bibfnamefont {C.}~\bibnamefont {Gray}}, \bibinfo {author} {\bibfnamefont
  {C.~J.}\ \bibnamefont {Guido}}, \bibinfo {author} {\bibfnamefont
  {R.}~\bibnamefont {Gustafson}}, \bibinfo {author} {\bibfnamefont
  {M.}~\bibnamefont {Heintze}}, \bibinfo {author} {\bibfnamefont
  {D.}~\bibnamefont {Hoak}}, \bibinfo {author} {\bibfnamefont {K.}~\bibnamefont
  {Izumi}}, \bibinfo {author} {\bibfnamefont {K.}~\bibnamefont {Kawabe}},
  \bibinfo {author} {\bibfnamefont {E.~J.}\ \bibnamefont {King}}, \bibinfo
  {author} {\bibfnamefont {J.~S.}\ \bibnamefont {Kissel}}, \bibinfo {author}
  {\bibfnamefont {K.}~\bibnamefont {Kokeyama}}, \bibinfo {author}
  {\bibfnamefont {M.}~\bibnamefont {Landry}}, \bibinfo {author} {\bibfnamefont
  {D.~E.}\ \bibnamefont {McClelland}}, \bibinfo {author} {\bibfnamefont
  {J.}~\bibnamefont {Miller}}, \bibinfo {author} {\bibfnamefont
  {A.}~\bibnamefont {Mullavey}}, \bibinfo {author} {\bibfnamefont
  {B.}~\bibnamefont {O?Reilly}}, \bibinfo {author} {\bibfnamefont {J.~G.}\
  \bibnamefont {Rollins}}, \bibinfo {author} {\bibfnamefont {J.~R.}\
  \bibnamefont {Sanders}}, \bibinfo {author} {\bibfnamefont {R.~M.~S.}\
  \bibnamefont {Schofield}}, \bibinfo {author} {\bibfnamefont {D.}~\bibnamefont
  {Sigg}}, \bibinfo {author} {\bibfnamefont {B.~J.~J.}\ \bibnamefont
  {Slagmolen}}, \bibinfo {author} {\bibfnamefont {N.~D.}\ \bibnamefont
  {Smith-Lefebvre}}, \bibinfo {author} {\bibfnamefont {G.}~\bibnamefont
  {Vajente}}, \bibinfo {author} {\bibfnamefont {R.~L.}\ \bibnamefont {Ward}}, \
  and\ \bibinfo {author} {\bibfnamefont {C.}~\bibnamefont {Wipf}},\ }\href
  {http://stacks.iop.org/0264-9381/31/i=24/a=245010} {\bibfield  {journal}
  {\bibinfo  {journal} {Classical and Quantum Gravity}\ }\textbf {\bibinfo
  {volume} {31}},\ \bibinfo {pages} {245010} (\bibinfo {year}
  {2014})}\BibitemShut {NoStop}%
\bibitem [{\citenamefont {{Zhao}}\ \emph {et~al.}(2006)\citenamefont {{Zhao}},
  \citenamefont {{Degallaix}}, \citenamefont {{Ju}}, \citenamefont {{Fan}},
  \citenamefont {{Blair}}, \citenamefont {{Slagmolen}}, \citenamefont {{Gray}},
  \citenamefont {{Lowry}}, \citenamefont {{McClelland}}, \citenamefont
  {{Hosken}}, \citenamefont {{Mudge}}, \citenamefont {{Brooks}}, \citenamefont
  {{Munch}}, \citenamefont {{Veitch}}, \citenamefont {{Barton}},\ and\
  \citenamefont {{Billingsley}}}]{Zhao_TCS_2006}%
  \BibitemOpen
  \bibfield  {author} {\bibinfo {author} {\bibfnamefont {C.}~\bibnamefont
  {{Zhao}}}, \bibinfo {author} {\bibfnamefont {J.}~\bibnamefont {{Degallaix}}},
  \bibinfo {author} {\bibfnamefont {L.}~\bibnamefont {{Ju}}}, \bibinfo {author}
  {\bibfnamefont {Y.}~\bibnamefont {{Fan}}}, \bibinfo {author} {\bibfnamefont
  {D.~G.}\ \bibnamefont {{Blair}}}, \bibinfo {author} {\bibfnamefont
  {B.~J.~J.}\ \bibnamefont {{Slagmolen}}}, \bibinfo {author} {\bibfnamefont
  {M.~B.}\ \bibnamefont {{Gray}}}, \bibinfo {author} {\bibfnamefont {C.~M.~M.}\
  \bibnamefont {{Lowry}}}, \bibinfo {author} {\bibfnamefont {D.~E.}\
  \bibnamefont {{McClelland}}}, \bibinfo {author} {\bibfnamefont {D.~J.}\
  \bibnamefont {{Hosken}}}, \bibinfo {author} {\bibfnamefont {D.}~\bibnamefont
  {{Mudge}}}, \bibinfo {author} {\bibfnamefont {A.}~\bibnamefont {{Brooks}}},
  \bibinfo {author} {\bibfnamefont {J.}~\bibnamefont {{Munch}}}, \bibinfo
  {author} {\bibfnamefont {P.~J.}\ \bibnamefont {{Veitch}}}, \bibinfo {author}
  {\bibfnamefont {M.~A.}\ \bibnamefont {{Barton}}}, \ and\ \bibinfo {author}
  {\bibfnamefont {G.}~\bibnamefont {{Billingsley}}},\ }\href {\doibase
  10.1103/PhysRevLett.96.231101} {\bibfield  {journal} {\bibinfo  {journal}
  {Physical Review Letters}\ }\textbf {\bibinfo {volume} {96}},\ \bibinfo {eid}
  {231101} (\bibinfo {year} {2006})},\ \Eprint
  {http://arxiv.org/abs/gr-qc/0602096} {gr-qc/0602096} \BibitemShut {NoStop}%
\bibitem [{\citenamefont {Brooks}\ \emph {et~al.}(2016)\citenamefont {Brooks},
  \citenamefont {Abbott}, \citenamefont {Arain}, \citenamefont {Ciani},
  \citenamefont {Cole}, \citenamefont {Grabeel}, \citenamefont {Gustafson},
  \citenamefont {Guido}, \citenamefont {Heintze}, \citenamefont {Heptonstall},
  \citenamefont {Jacobson}, \citenamefont {Kim}, \citenamefont {King},
  \citenamefont {Lynch}, \citenamefont {O'Connor}, \citenamefont {Ottaway},
  \citenamefont {Mailand}, \citenamefont {Mueller}, \citenamefont {Munch},
  \citenamefont {Sannibale}, \citenamefont {Shao}, \citenamefont {Smith},
  \citenamefont {Veitch}, \citenamefont {Vo}, \citenamefont {Vorvick},\ and\
  \citenamefont {Willems}}]{Brooks_TCS_2016}%
  \BibitemOpen
  \bibfield  {author} {\bibinfo {author} {\bibfnamefont {A.~F.}\ \bibnamefont
  {Brooks}}, \bibinfo {author} {\bibfnamefont {B.}~\bibnamefont {Abbott}},
  \bibinfo {author} {\bibfnamefont {M.~A.}\ \bibnamefont {Arain}}, \bibinfo
  {author} {\bibfnamefont {G.}~\bibnamefont {Ciani}}, \bibinfo {author}
  {\bibfnamefont {A.}~\bibnamefont {Cole}}, \bibinfo {author} {\bibfnamefont
  {G.}~\bibnamefont {Grabeel}}, \bibinfo {author} {\bibfnamefont
  {E.}~\bibnamefont {Gustafson}}, \bibinfo {author} {\bibfnamefont
  {C.}~\bibnamefont {Guido}}, \bibinfo {author} {\bibfnamefont
  {M.}~\bibnamefont {Heintze}}, \bibinfo {author} {\bibfnamefont
  {A.}~\bibnamefont {Heptonstall}}, \bibinfo {author} {\bibfnamefont
  {M.}~\bibnamefont {Jacobson}}, \bibinfo {author} {\bibfnamefont
  {W.}~\bibnamefont {Kim}}, \bibinfo {author} {\bibfnamefont {E.}~\bibnamefont
  {King}}, \bibinfo {author} {\bibfnamefont {A.}~\bibnamefont {Lynch}},
  \bibinfo {author} {\bibfnamefont {S.}~\bibnamefont {O'Connor}}, \bibinfo
  {author} {\bibfnamefont {D.}~\bibnamefont {Ottaway}}, \bibinfo {author}
  {\bibfnamefont {K.}~\bibnamefont {Mailand}}, \bibinfo {author} {\bibfnamefont
  {G.}~\bibnamefont {Mueller}}, \bibinfo {author} {\bibfnamefont
  {J.}~\bibnamefont {Munch}}, \bibinfo {author} {\bibfnamefont
  {V.}~\bibnamefont {Sannibale}}, \bibinfo {author} {\bibfnamefont
  {Z.}~\bibnamefont {Shao}}, \bibinfo {author} {\bibfnamefont {M.}~\bibnamefont
  {Smith}}, \bibinfo {author} {\bibfnamefont {P.}~\bibnamefont {Veitch}},
  \bibinfo {author} {\bibfnamefont {T.}~\bibnamefont {Vo}}, \bibinfo {author}
  {\bibfnamefont {C.}~\bibnamefont {Vorvick}}, \ and\ \bibinfo {author}
  {\bibfnamefont {P.}~\bibnamefont {Willems}},\ }\href {\doibase
  10.1364/AO.55.008256} {\bibfield  {journal} {\bibinfo  {journal} {Appl.
  Opt.}\ }\textbf {\bibinfo {volume} {55}},\ \bibinfo {pages} {8256} (\bibinfo
  {year} {2016})}\BibitemShut {NoStop}%
\bibitem [{\citenamefont {{LIGO Scientific
  Collaboration}}(2011)}]{LSC_SQUEEZING_2011}%
  \BibitemOpen
  \bibfield  {author} {\bibinfo {author} {\bibnamefont {{LIGO Scientific
  Collaboration}}},\ }\href {\doibase 10.1038/nphys2083} {\bibfield  {journal}
  {\bibinfo  {journal} {Nature Physics}\ }\textbf {\bibinfo {volume} {7}},\
  \bibinfo {pages} {962} (\bibinfo {year} {2011})}\BibitemShut {NoStop}%
\bibitem [{\citenamefont {Grote}\ \emph {et~al.}(2013)\citenamefont {Grote},
  \citenamefont {Danzmann}, \citenamefont {Dooley}, \citenamefont {Schnabel},
  \citenamefont {Slutsky},\ and\ \citenamefont
  {Vahlbruch}}]{Grote_SQUEEZING_2013}%
  \BibitemOpen
  \bibfield  {author} {\bibinfo {author} {\bibfnamefont {H.}~\bibnamefont
  {Grote}}, \bibinfo {author} {\bibfnamefont {K.}~\bibnamefont {Danzmann}},
  \bibinfo {author} {\bibfnamefont {K.~L.}\ \bibnamefont {Dooley}}, \bibinfo
  {author} {\bibfnamefont {R.}~\bibnamefont {Schnabel}}, \bibinfo {author}
  {\bibfnamefont {J.}~\bibnamefont {Slutsky}}, \ and\ \bibinfo {author}
  {\bibfnamefont {H.}~\bibnamefont {Vahlbruch}},\ }\href
  {http://link.aps.org/doi/10.1103/PhysRevLett.110.181101} {\bibfield
  {journal} {\bibinfo  {journal} {Phys. Rev. Lett.}\ }\textbf {\bibinfo
  {volume} {110}},\ \bibinfo {pages} {181101} (\bibinfo {year}
  {2013})}\BibitemShut {NoStop}%
\bibitem [{\citenamefont {Caves}(1981)}]{Caves_1981}%
  \BibitemOpen
  \bibfield  {author} {\bibinfo {author} {\bibfnamefont {C.~M.}\ \bibnamefont
  {Caves}},\ }\href {\doibase 10.1103/PhysRevD.23.1693} {\bibfield  {journal}
  {\bibinfo  {journal} {Phys. Rev. D}\ }\textbf {\bibinfo {volume} {23}},\
  \bibinfo {pages} {1693} (\bibinfo {year} {1981})}\BibitemShut {NoStop}%
\bibitem [{\citenamefont {Miao}\ \emph
  {et~al.}(2018{\natexlab{b}})\citenamefont {Miao}, \citenamefont {Smith},\
  and\ \citenamefont {Evans}}]{Miao_LOSS_2018}%
  \BibitemOpen
  \bibfield  {author} {\bibinfo {author} {\bibfnamefont {H.}~\bibnamefont
  {Miao}}, \bibinfo {author} {\bibfnamefont {N.}~\bibnamefont {Smith}}, \ and\
  \bibinfo {author} {\bibfnamefont {M.}~\bibnamefont {Evans}},\ }\href@noop {}
  {\  (\bibinfo {year} {2018}{\natexlab{b}})},\ \Eprint
  {http://arxiv.org/abs/1807.11734} {arXiv:1807.11734 [quant-ph]} \BibitemShut
  {NoStop}%
\bibitem [{\citenamefont {Bauswein}\ \emph {et~al.}(2018)\citenamefont
  {Bauswein}, \citenamefont {Bastian}, \citenamefont {Blaschke}, \citenamefont
  {Chatziioannou}, \citenamefont {Clark}, \citenamefont {Fischer},\ and\
  \citenamefont {Oertel}}]{Bauswein_NS_2018}%
  \BibitemOpen
  \bibfield  {author} {\bibinfo {author} {\bibfnamefont {A.}~\bibnamefont
  {Bauswein}}, \bibinfo {author} {\bibfnamefont {N.-U.}\ \bibnamefont
  {Bastian}}, \bibinfo {author} {\bibfnamefont {D.}~\bibnamefont {Blaschke}},
  \bibinfo {author} {\bibfnamefont {K.}~\bibnamefont {Chatziioannou}}, \bibinfo
  {author} {\bibfnamefont {J.}~\bibnamefont {Clark}}, \bibinfo {author}
  {\bibfnamefont {T.}~\bibnamefont {Fischer}}, \ and\ \bibinfo {author}
  {\bibfnamefont {M.}~\bibnamefont {Oertel}},\ }\href@noop {} {\  (\bibinfo
  {year} {2018})},\ \Eprint {http://arxiv.org/abs/1809.01116} {arXiv:1809.01116
  [astro-ph]} \BibitemShut {NoStop}%
\bibitem [{\citenamefont {Yang}\ \emph {et~al.}(2018)\citenamefont {Yang},
  \citenamefont {Paschalidis}, \citenamefont {Yagi}, \citenamefont {Lehner},
  \citenamefont {Pretorius},\ and\ \citenamefont {Yunes}}]{Yang:2017xlf}%
  \BibitemOpen
  \bibfield  {author} {\bibinfo {author} {\bibfnamefont {H.}~\bibnamefont
  {Yang}}, \bibinfo {author} {\bibfnamefont {V.}~\bibnamefont {Paschalidis}},
  \bibinfo {author} {\bibfnamefont {K.}~\bibnamefont {Yagi}}, \bibinfo {author}
  {\bibfnamefont {L.}~\bibnamefont {Lehner}}, \bibinfo {author} {\bibfnamefont
  {F.}~\bibnamefont {Pretorius}}, \ and\ \bibinfo {author} {\bibfnamefont
  {N.}~\bibnamefont {Yunes}},\ }\href {\doibase 10.1103/PhysRevD.97.024049}
  {\bibfield  {journal} {\bibinfo  {journal} {Phys. Rev.}\ }\textbf {\bibinfo
  {volume} {D97}},\ \bibinfo {pages} {024049} (\bibinfo {year} {2018})},\
  \Eprint {http://arxiv.org/abs/1707.00207} {arXiv:1707.00207 [gr-qc]}
  \BibitemShut {NoStop}%
\bibitem [{\citenamefont {Kiziltan}\ \emph {et~al.}(2013)\citenamefont
  {Kiziltan}, \citenamefont {Kottas}, \citenamefont {Yoreo},\ and\
  \citenamefont {Thorsett}}]{Kiziltan_NS_2013}%
  \BibitemOpen
  \bibfield  {author} {\bibinfo {author} {\bibfnamefont {B.}~\bibnamefont
  {Kiziltan}}, \bibinfo {author} {\bibfnamefont {A.}~\bibnamefont {Kottas}},
  \bibinfo {author} {\bibfnamefont {M.~D.}\ \bibnamefont {Yoreo}}, \ and\
  \bibinfo {author} {\bibfnamefont {S.~E.}\ \bibnamefont {Thorsett}},\ }\href
  {http://stacks.iop.org/0004-637X/778/i=1/a=66} {\bibfield  {journal}
  {\bibinfo  {journal} {The Astrophysical Journal}\ }\textbf {\bibinfo {volume}
  {778}},\ \bibinfo {pages} {66} (\bibinfo {year} {2013})}\BibitemShut
  {NoStop}%
\bibitem [{\citenamefont {Acernese}\ \emph {et~al.}(2015)\citenamefont
  {Acernese}, \citenamefont {Agathos}, \citenamefont {Agatsuma}, \citenamefont
  {Aisa}, \citenamefont {Allemandou}, \citenamefont {Allocca}, \citenamefont
  {Amarni}, \citenamefont {Astone}, \citenamefont {Balestri} \emph
  {et~al.}}]{Acernese_aVIRGO_2015}%
  \BibitemOpen
  \bibfield  {author} {\bibinfo {author} {\bibfnamefont {F.}~\bibnamefont
  {Acernese}}, \bibinfo {author} {\bibfnamefont {M.}~\bibnamefont {Agathos}},
  \bibinfo {author} {\bibfnamefont {K.}~\bibnamefont {Agatsuma}}, \bibinfo
  {author} {\bibfnamefont {D.}~\bibnamefont {Aisa}}, \bibinfo {author}
  {\bibfnamefont {N.}~\bibnamefont {Allemandou}}, \bibinfo {author}
  {\bibfnamefont {A.}~\bibnamefont {Allocca}}, \bibinfo {author} {\bibfnamefont
  {J.}~\bibnamefont {Amarni}}, \bibinfo {author} {\bibfnamefont
  {P.}~\bibnamefont {Astone}}, \bibinfo {author} {\bibfnamefont
  {G.}~\bibnamefont {Balestri}},  \emph {et~al.},\ }\href
  {http://stacks.iop.org/0264-9381/32/i=2/a=024001} {\bibfield  {journal}
  {\bibinfo  {journal} {Classical and Quantum Gravity}\ }\textbf {\bibinfo
  {volume} {32}},\ \bibinfo {pages} {024001} (\bibinfo {year}
  {2015})}\BibitemShut {NoStop}%
\bibitem [{\citenamefont {Somiya}(2012)}]{Somiya_Kagra_2012}%
  \BibitemOpen
  \bibfield  {author} {\bibinfo {author} {\bibfnamefont {K.}~\bibnamefont
  {Somiya}},\ }\href {http://stacks.iop.org/0264-9381/29/i=12/a=124007}
  {\bibfield  {journal} {\bibinfo  {journal} {Classical and Quantum Gravity}\
  }\textbf {\bibinfo {volume} {29}},\ \bibinfo {pages} {124007} (\bibinfo
  {year} {2012})}\BibitemShut {NoStop}%
\bibitem [{\citenamefont {Martynov}\ \emph {et~al.}(2017)\citenamefont
  {Martynov}, \citenamefont {Frolov}, \citenamefont {Kandhasamy}, \citenamefont
  {Izumi}, \citenamefont {Miao}, \citenamefont {Mavalvala}, \citenamefont
  {Hall}, \citenamefont {Lanza}, \citenamefont {Abbott}, \citenamefont
  {Abbott}, \citenamefont {Abbott}, \citenamefont {Adams}, \citenamefont
  {Adhikari}, \citenamefont {Anderson}, \citenamefont {Ananyeva}, \citenamefont
  {Appert}, \citenamefont {Arai}, \citenamefont {Aston}, \citenamefont
  {Ballmer}, \citenamefont {Barker}, \citenamefont {Barr}, \citenamefont
  {Barsotti}, \citenamefont {Bartlett}, \citenamefont {Bartos}, \citenamefont
  {Batch}, \citenamefont {Bell}, \citenamefont {Betzwieser}, \citenamefont
  {Billingsley}, \citenamefont {Birch}, \citenamefont {Biscans}, \citenamefont
  {Biwer}, \citenamefont {Blair}, \citenamefont {Bork}, \citenamefont {Brooks},
  \citenamefont {Ciani}, \citenamefont {Clara}, \citenamefont {Countryman},
  \citenamefont {Cowart}, \citenamefont {Coyne}, \citenamefont {Cumming},
  \citenamefont {Cunningham}, \citenamefont {Danzmann}, \citenamefont
  {Da~Silva~Costa}, \citenamefont {Daw}, \citenamefont {DeBra}, \citenamefont
  {DeRosa}, \citenamefont {DeSalvo}, \citenamefont {Dooley}, \citenamefont
  {Doravari}, \citenamefont {Driggers}, \citenamefont {Dwyer}, \citenamefont
  {Effler}, \citenamefont {Etzel}, \citenamefont {Evans}, \citenamefont
  {Evans}, \citenamefont {Factourovich}, \citenamefont {Fair}, \citenamefont
  {Fern\'andez~Galiana}, \citenamefont {Fisher}, \citenamefont {Fritschel},
  \citenamefont {Fulda}, \citenamefont {Fyffe}, \citenamefont {Giaime},
  \citenamefont {Giardina}, \citenamefont {Goetz}, \citenamefont {Goetz},
  \citenamefont {Gras}, \citenamefont {Gray}, \citenamefont {Grote},
  \citenamefont {Gushwa}, \citenamefont {Gustafson}, \citenamefont {Gustafson},
  \citenamefont {Hammond}, \citenamefont {Hanks}, \citenamefont {Hanson},
  \citenamefont {Hardwick}, \citenamefont {Harry}, \citenamefont {Heintze},
  \citenamefont {Heptonstall}, \citenamefont {Hough}, \citenamefont {Jones},
  \citenamefont {Karki}, \citenamefont {Kasprzack}, \citenamefont {Kaufer},
  \citenamefont {Kawabe}, \citenamefont {Kijbunchoo}, \citenamefont {King},
  \citenamefont {King}, \citenamefont {Kissel}, \citenamefont {Korth},
  \citenamefont {Kuehn}, \citenamefont {Landry}, \citenamefont {Lantz},
  \citenamefont {Lockerbie}, \citenamefont {Lormand}, \citenamefont {Lundgren},
  \citenamefont {MacInnis}, \citenamefont {Macleod}, \citenamefont {M\'arka},
  \citenamefont {M\'arka}, \citenamefont {Markosyan}, \citenamefont {Maros},
  \citenamefont {Martin}, \citenamefont {Mason}, \citenamefont {Massinger},
  \citenamefont {Matichard}, \citenamefont {McCarthy}, \citenamefont
  {McClelland}, \citenamefont {McCormick}, \citenamefont {McIntyre},
  \citenamefont {McIver}, \citenamefont {Mendell}, \citenamefont {Merilh},
  \citenamefont {Meyers}, \citenamefont {Miller}, \citenamefont {Mittleman},
  \citenamefont {Moreno}, \citenamefont {Mueller}, \citenamefont {Mullavey},
  \citenamefont {Munch}, \citenamefont {Nuttall}, \citenamefont {Oberling},
  \citenamefont {Oppermann}, \citenamefont {Oram}, \citenamefont {O'Reilly},
  \citenamefont {Ottaway}, \citenamefont {Overmier}, \citenamefont {Palamos},
  \citenamefont {Paris}, \citenamefont {Parker}, \citenamefont {Pele},
  \citenamefont {Penn}, \citenamefont {Phelps}, \citenamefont {Pierro},
  \citenamefont {Pinto}, \citenamefont {Principe}, \citenamefont {Prokhorov},
  \citenamefont {Puncken}, \citenamefont {Quetschke}, \citenamefont {Quintero},
  \citenamefont {Raab}, \citenamefont {Radkins}, \citenamefont {Raffai},
  \citenamefont {Reid}, \citenamefont {Reitze}, \citenamefont {Robertson},
  \citenamefont {Rollins}, \citenamefont {Roma}, \citenamefont {Romie},
  \citenamefont {Rowan}, \citenamefont {Ryan}, \citenamefont {Sadecki},
  \citenamefont {Sanchez}, \citenamefont {Sandberg}, \citenamefont {Savage},
  \citenamefont {Schofield}, \citenamefont {Sellers}, \citenamefont {Shaddock},
  \citenamefont {Shaffer}, \citenamefont {Shapiro}, \citenamefont {Shawhan},
  \citenamefont {Shoemaker}, \citenamefont {Sigg}, \citenamefont {Slagmolen},
  \citenamefont {Smith}, \citenamefont {Smith}, \citenamefont {Sorazu},
  \citenamefont {Staley}, \citenamefont {Strain}, \citenamefont {Tanner},
  \citenamefont {Taylor}, \citenamefont {Thomas}, \citenamefont {Thomas},
  \citenamefont {Thorne}, \citenamefont {Thrane}, \citenamefont {Torrie},
  \citenamefont {Traylor}, \citenamefont {Vajente}, \citenamefont {Valdes},
  \citenamefont {van Veggel}, \citenamefont {Vecchio}, \citenamefont {Veitch},
  \citenamefont {Venkateswara}, \citenamefont {Vo}, \citenamefont {Vorvick},
  \citenamefont {Walker}, \citenamefont {Ward}, \citenamefont {Warner},
  \citenamefont {Weaver}, \citenamefont {Weiss}, \citenamefont {We\ss{}els},
  \citenamefont {Willke}, \citenamefont {Wipf}, \citenamefont {Worden},
  \citenamefont {Wu}, \citenamefont {Yamamoto}, \citenamefont {Yancey},
  \citenamefont {Yu}, \citenamefont {Yu}, \citenamefont {Zhang}, \citenamefont
  {Zucker},\ and\ \citenamefont {Zweizig}}]{Martynov_QUCORR_2017}%
  \BibitemOpen
  \bibfield  {author} {\bibinfo {author} {\bibfnamefont {D.~V.}\ \bibnamefont
  {Martynov}}, \bibinfo {author} {\bibfnamefont {V.~V.}\ \bibnamefont
  {Frolov}}, \bibinfo {author} {\bibfnamefont {S.}~\bibnamefont {Kandhasamy}},
  \bibinfo {author} {\bibfnamefont {K.}~\bibnamefont {Izumi}}, \bibinfo
  {author} {\bibfnamefont {H.}~\bibnamefont {Miao}}, \bibinfo {author}
  {\bibfnamefont {N.}~\bibnamefont {Mavalvala}}, \bibinfo {author}
  {\bibfnamefont {E.~D.}\ \bibnamefont {Hall}}, \bibinfo {author}
  {\bibfnamefont {R.}~\bibnamefont {Lanza}}, \bibinfo {author} {\bibfnamefont
  {B.~P.}\ \bibnamefont {Abbott}}, \bibinfo {author} {\bibfnamefont
  {R.}~\bibnamefont {Abbott}}, \bibinfo {author} {\bibfnamefont {T.~D.}\
  \bibnamefont {Abbott}}, \bibinfo {author} {\bibfnamefont {C.}~\bibnamefont
  {Adams}}, \bibinfo {author} {\bibfnamefont {R.~X.}\ \bibnamefont {Adhikari}},
  \bibinfo {author} {\bibfnamefont {S.~B.}\ \bibnamefont {Anderson}}, \bibinfo
  {author} {\bibfnamefont {A.}~\bibnamefont {Ananyeva}}, \bibinfo {author}
  {\bibfnamefont {S.}~\bibnamefont {Appert}}, \bibinfo {author} {\bibfnamefont
  {K.}~\bibnamefont {Arai}}, \bibinfo {author} {\bibfnamefont {S.~M.}\
  \bibnamefont {Aston}}, \bibinfo {author} {\bibfnamefont {S.~W.}\ \bibnamefont
  {Ballmer}}, \bibinfo {author} {\bibfnamefont {D.}~\bibnamefont {Barker}},
  \bibinfo {author} {\bibfnamefont {B.}~\bibnamefont {Barr}}, \bibinfo {author}
  {\bibfnamefont {L.}~\bibnamefont {Barsotti}}, \bibinfo {author}
  {\bibfnamefont {J.}~\bibnamefont {Bartlett}}, \bibinfo {author}
  {\bibfnamefont {I.}~\bibnamefont {Bartos}}, \bibinfo {author} {\bibfnamefont
  {J.~C.}\ \bibnamefont {Batch}}, \bibinfo {author} {\bibfnamefont {A.~S.}\
  \bibnamefont {Bell}}, \bibinfo {author} {\bibfnamefont {J.}~\bibnamefont
  {Betzwieser}}, \bibinfo {author} {\bibfnamefont {G.}~\bibnamefont
  {Billingsley}}, \bibinfo {author} {\bibfnamefont {J.}~\bibnamefont {Birch}},
  \bibinfo {author} {\bibfnamefont {S.}~\bibnamefont {Biscans}}, \bibinfo
  {author} {\bibfnamefont {C.}~\bibnamefont {Biwer}}, \bibinfo {author}
  {\bibfnamefont {C.~D.}\ \bibnamefont {Blair}}, \bibinfo {author}
  {\bibfnamefont {R.}~\bibnamefont {Bork}}, \bibinfo {author} {\bibfnamefont
  {A.~F.}\ \bibnamefont {Brooks}}, \bibinfo {author} {\bibfnamefont
  {G.}~\bibnamefont {Ciani}}, \bibinfo {author} {\bibfnamefont
  {F.}~\bibnamefont {Clara}}, \bibinfo {author} {\bibfnamefont {S.~T.}\
  \bibnamefont {Countryman}}, \bibinfo {author} {\bibfnamefont {M.~J.}\
  \bibnamefont {Cowart}}, \bibinfo {author} {\bibfnamefont {D.~C.}\
  \bibnamefont {Coyne}}, \bibinfo {author} {\bibfnamefont {A.}~\bibnamefont
  {Cumming}}, \bibinfo {author} {\bibfnamefont {L.}~\bibnamefont {Cunningham}},
  \bibinfo {author} {\bibfnamefont {K.}~\bibnamefont {Danzmann}}, \bibinfo
  {author} {\bibfnamefont {C.~F.}\ \bibnamefont {Da~Silva~Costa}}, \bibinfo
  {author} {\bibfnamefont {E.~J.}\ \bibnamefont {Daw}}, \bibinfo {author}
  {\bibfnamefont {D.}~\bibnamefont {DeBra}}, \bibinfo {author} {\bibfnamefont
  {R.~T.}\ \bibnamefont {DeRosa}}, \bibinfo {author} {\bibfnamefont
  {R.}~\bibnamefont {DeSalvo}}, \bibinfo {author} {\bibfnamefont {K.~L.}\
  \bibnamefont {Dooley}}, \bibinfo {author} {\bibfnamefont {S.}~\bibnamefont
  {Doravari}}, \bibinfo {author} {\bibfnamefont {J.~C.}\ \bibnamefont
  {Driggers}}, \bibinfo {author} {\bibfnamefont {S.~E.}\ \bibnamefont {Dwyer}},
  \bibinfo {author} {\bibfnamefont {A.}~\bibnamefont {Effler}}, \bibinfo
  {author} {\bibfnamefont {T.}~\bibnamefont {Etzel}}, \bibinfo {author}
  {\bibfnamefont {M.}~\bibnamefont {Evans}}, \bibinfo {author} {\bibfnamefont
  {T.~M.}\ \bibnamefont {Evans}}, \bibinfo {author} {\bibfnamefont
  {M.}~\bibnamefont {Factourovich}}, \bibinfo {author} {\bibfnamefont
  {H.}~\bibnamefont {Fair}}, \bibinfo {author} {\bibfnamefont {A.}~\bibnamefont
  {Fern\'andez~Galiana}}, \bibinfo {author} {\bibfnamefont {R.~P.}\
  \bibnamefont {Fisher}}, \bibinfo {author} {\bibfnamefont {P.}~\bibnamefont
  {Fritschel}}, \bibinfo {author} {\bibfnamefont {P.}~\bibnamefont {Fulda}},
  \bibinfo {author} {\bibfnamefont {M.}~\bibnamefont {Fyffe}}, \bibinfo
  {author} {\bibfnamefont {J.~A.}\ \bibnamefont {Giaime}}, \bibinfo {author}
  {\bibfnamefont {K.~D.}\ \bibnamefont {Giardina}}, \bibinfo {author}
  {\bibfnamefont {E.}~\bibnamefont {Goetz}}, \bibinfo {author} {\bibfnamefont
  {R.}~\bibnamefont {Goetz}}, \bibinfo {author} {\bibfnamefont
  {S.}~\bibnamefont {Gras}}, \bibinfo {author} {\bibfnamefont {C.}~\bibnamefont
  {Gray}}, \bibinfo {author} {\bibfnamefont {H.}~\bibnamefont {Grote}},
  \bibinfo {author} {\bibfnamefont {K.~E.}\ \bibnamefont {Gushwa}}, \bibinfo
  {author} {\bibfnamefont {E.~K.}\ \bibnamefont {Gustafson}}, \bibinfo {author}
  {\bibfnamefont {R.}~\bibnamefont {Gustafson}}, \bibinfo {author}
  {\bibfnamefont {G.}~\bibnamefont {Hammond}}, \bibinfo {author} {\bibfnamefont
  {J.}~\bibnamefont {Hanks}}, \bibinfo {author} {\bibfnamefont
  {J.}~\bibnamefont {Hanson}}, \bibinfo {author} {\bibfnamefont
  {T.}~\bibnamefont {Hardwick}}, \bibinfo {author} {\bibfnamefont {G.~M.}\
  \bibnamefont {Harry}}, \bibinfo {author} {\bibfnamefont {M.~C.}\ \bibnamefont
  {Heintze}}, \bibinfo {author} {\bibfnamefont {A.~W.}\ \bibnamefont
  {Heptonstall}}, \bibinfo {author} {\bibfnamefont {J.}~\bibnamefont {Hough}},
  \bibinfo {author} {\bibfnamefont {R.}~\bibnamefont {Jones}}, \bibinfo
  {author} {\bibfnamefont {S.}~\bibnamefont {Karki}}, \bibinfo {author}
  {\bibfnamefont {M.}~\bibnamefont {Kasprzack}}, \bibinfo {author}
  {\bibfnamefont {S.}~\bibnamefont {Kaufer}}, \bibinfo {author} {\bibfnamefont
  {K.}~\bibnamefont {Kawabe}}, \bibinfo {author} {\bibfnamefont
  {N.}~\bibnamefont {Kijbunchoo}}, \bibinfo {author} {\bibfnamefont {E.~J.}\
  \bibnamefont {King}}, \bibinfo {author} {\bibfnamefont {P.~J.}\ \bibnamefont
  {King}}, \bibinfo {author} {\bibfnamefont {J.~S.}\ \bibnamefont {Kissel}},
  \bibinfo {author} {\bibfnamefont {W.~Z.}\ \bibnamefont {Korth}}, \bibinfo
  {author} {\bibfnamefont {G.}~\bibnamefont {Kuehn}}, \bibinfo {author}
  {\bibfnamefont {M.}~\bibnamefont {Landry}}, \bibinfo {author} {\bibfnamefont
  {B.}~\bibnamefont {Lantz}}, \bibinfo {author} {\bibfnamefont {N.~A.}\
  \bibnamefont {Lockerbie}}, \bibinfo {author} {\bibfnamefont {M.}~\bibnamefont
  {Lormand}}, \bibinfo {author} {\bibfnamefont {A.~P.}\ \bibnamefont
  {Lundgren}}, \bibinfo {author} {\bibfnamefont {M.}~\bibnamefont {MacInnis}},
  \bibinfo {author} {\bibfnamefont {D.~M.}\ \bibnamefont {Macleod}}, \bibinfo
  {author} {\bibfnamefont {S.}~\bibnamefont {M\'arka}}, \bibinfo {author}
  {\bibfnamefont {Z.}~\bibnamefont {M\'arka}}, \bibinfo {author} {\bibfnamefont
  {A.~S.}\ \bibnamefont {Markosyan}}, \bibinfo {author} {\bibfnamefont
  {E.}~\bibnamefont {Maros}}, \bibinfo {author} {\bibfnamefont {I.~W.}\
  \bibnamefont {Martin}}, \bibinfo {author} {\bibfnamefont {K.}~\bibnamefont
  {Mason}}, \bibinfo {author} {\bibfnamefont {T.~J.}\ \bibnamefont
  {Massinger}}, \bibinfo {author} {\bibfnamefont {F.}~\bibnamefont
  {Matichard}}, \bibinfo {author} {\bibfnamefont {R.}~\bibnamefont {McCarthy}},
  \bibinfo {author} {\bibfnamefont {D.~E.}\ \bibnamefont {McClelland}},
  \bibinfo {author} {\bibfnamefont {S.}~\bibnamefont {McCormick}}, \bibinfo
  {author} {\bibfnamefont {G.}~\bibnamefont {McIntyre}}, \bibinfo {author}
  {\bibfnamefont {J.}~\bibnamefont {McIver}}, \bibinfo {author} {\bibfnamefont
  {G.}~\bibnamefont {Mendell}}, \bibinfo {author} {\bibfnamefont {E.~L.}\
  \bibnamefont {Merilh}}, \bibinfo {author} {\bibfnamefont {P.~M.}\
  \bibnamefont {Meyers}}, \bibinfo {author} {\bibfnamefont {J.}~\bibnamefont
  {Miller}}, \bibinfo {author} {\bibfnamefont {R.}~\bibnamefont {Mittleman}},
  \bibinfo {author} {\bibfnamefont {G.}~\bibnamefont {Moreno}}, \bibinfo
  {author} {\bibfnamefont {G.}~\bibnamefont {Mueller}}, \bibinfo {author}
  {\bibfnamefont {A.}~\bibnamefont {Mullavey}}, \bibinfo {author}
  {\bibfnamefont {J.}~\bibnamefont {Munch}}, \bibinfo {author} {\bibfnamefont
  {L.~K.}\ \bibnamefont {Nuttall}}, \bibinfo {author} {\bibfnamefont
  {J.}~\bibnamefont {Oberling}}, \bibinfo {author} {\bibfnamefont
  {P.}~\bibnamefont {Oppermann}}, \bibinfo {author} {\bibfnamefont {R.~J.}\
  \bibnamefont {Oram}}, \bibinfo {author} {\bibfnamefont {B.}~\bibnamefont
  {O'Reilly}}, \bibinfo {author} {\bibfnamefont {D.~J.}\ \bibnamefont
  {Ottaway}}, \bibinfo {author} {\bibfnamefont {H.}~\bibnamefont {Overmier}},
  \bibinfo {author} {\bibfnamefont {J.~R.}\ \bibnamefont {Palamos}}, \bibinfo
  {author} {\bibfnamefont {H.~R.}\ \bibnamefont {Paris}}, \bibinfo {author}
  {\bibfnamefont {W.}~\bibnamefont {Parker}}, \bibinfo {author} {\bibfnamefont
  {A.}~\bibnamefont {Pele}}, \bibinfo {author} {\bibfnamefont {S.}~\bibnamefont
  {Penn}}, \bibinfo {author} {\bibfnamefont {M.}~\bibnamefont {Phelps}},
  \bibinfo {author} {\bibfnamefont {V.}~\bibnamefont {Pierro}}, \bibinfo
  {author} {\bibfnamefont {I.}~\bibnamefont {Pinto}}, \bibinfo {author}
  {\bibfnamefont {M.}~\bibnamefont {Principe}}, \bibinfo {author}
  {\bibfnamefont {L.~G.}\ \bibnamefont {Prokhorov}}, \bibinfo {author}
  {\bibfnamefont {O.}~\bibnamefont {Puncken}}, \bibinfo {author} {\bibfnamefont
  {V.}~\bibnamefont {Quetschke}}, \bibinfo {author} {\bibfnamefont {E.~A.}\
  \bibnamefont {Quintero}}, \bibinfo {author} {\bibfnamefont {F.~J.}\
  \bibnamefont {Raab}}, \bibinfo {author} {\bibfnamefont {H.}~\bibnamefont
  {Radkins}}, \bibinfo {author} {\bibfnamefont {P.}~\bibnamefont {Raffai}},
  \bibinfo {author} {\bibfnamefont {S.}~\bibnamefont {Reid}}, \bibinfo {author}
  {\bibfnamefont {D.~H.}\ \bibnamefont {Reitze}}, \bibinfo {author}
  {\bibfnamefont {N.~A.}\ \bibnamefont {Robertson}}, \bibinfo {author}
  {\bibfnamefont {J.~G.}\ \bibnamefont {Rollins}}, \bibinfo {author}
  {\bibfnamefont {V.~J.}\ \bibnamefont {Roma}}, \bibinfo {author}
  {\bibfnamefont {J.~H.}\ \bibnamefont {Romie}}, \bibinfo {author}
  {\bibfnamefont {S.}~\bibnamefont {Rowan}}, \bibinfo {author} {\bibfnamefont
  {K.}~\bibnamefont {Ryan}}, \bibinfo {author} {\bibfnamefont {T.}~\bibnamefont
  {Sadecki}}, \bibinfo {author} {\bibfnamefont {E.~J.}\ \bibnamefont
  {Sanchez}}, \bibinfo {author} {\bibfnamefont {V.}~\bibnamefont {Sandberg}},
  \bibinfo {author} {\bibfnamefont {R.~L.}\ \bibnamefont {Savage}}, \bibinfo
  {author} {\bibfnamefont {R.~M.~S.}\ \bibnamefont {Schofield}}, \bibinfo
  {author} {\bibfnamefont {D.}~\bibnamefont {Sellers}}, \bibinfo {author}
  {\bibfnamefont {D.~A.}\ \bibnamefont {Shaddock}}, \bibinfo {author}
  {\bibfnamefont {T.~J.}\ \bibnamefont {Shaffer}}, \bibinfo {author}
  {\bibfnamefont {B.}~\bibnamefont {Shapiro}}, \bibinfo {author} {\bibfnamefont
  {P.}~\bibnamefont {Shawhan}}, \bibinfo {author} {\bibfnamefont {D.~H.}\
  \bibnamefont {Shoemaker}}, \bibinfo {author} {\bibfnamefont {D.}~\bibnamefont
  {Sigg}}, \bibinfo {author} {\bibfnamefont {B.~J.~J.}\ \bibnamefont
  {Slagmolen}}, \bibinfo {author} {\bibfnamefont {B.}~\bibnamefont {Smith}},
  \bibinfo {author} {\bibfnamefont {J.~R.}\ \bibnamefont {Smith}}, \bibinfo
  {author} {\bibfnamefont {B.}~\bibnamefont {Sorazu}}, \bibinfo {author}
  {\bibfnamefont {A.}~\bibnamefont {Staley}}, \bibinfo {author} {\bibfnamefont
  {K.~A.}\ \bibnamefont {Strain}}, \bibinfo {author} {\bibfnamefont {D.~B.}\
  \bibnamefont {Tanner}}, \bibinfo {author} {\bibfnamefont {R.}~\bibnamefont
  {Taylor}}, \bibinfo {author} {\bibfnamefont {M.}~\bibnamefont {Thomas}},
  \bibinfo {author} {\bibfnamefont {P.}~\bibnamefont {Thomas}}, \bibinfo
  {author} {\bibfnamefont {K.~A.}\ \bibnamefont {Thorne}}, \bibinfo {author}
  {\bibfnamefont {E.}~\bibnamefont {Thrane}}, \bibinfo {author} {\bibfnamefont
  {C.~I.}\ \bibnamefont {Torrie}}, \bibinfo {author} {\bibfnamefont
  {G.}~\bibnamefont {Traylor}}, \bibinfo {author} {\bibfnamefont
  {G.}~\bibnamefont {Vajente}}, \bibinfo {author} {\bibfnamefont
  {G.}~\bibnamefont {Valdes}}, \bibinfo {author} {\bibfnamefont {A.~A.}\
  \bibnamefont {van Veggel}}, \bibinfo {author} {\bibfnamefont
  {A.}~\bibnamefont {Vecchio}}, \bibinfo {author} {\bibfnamefont {P.~J.}\
  \bibnamefont {Veitch}}, \bibinfo {author} {\bibfnamefont {K.}~\bibnamefont
  {Venkateswara}}, \bibinfo {author} {\bibfnamefont {T.}~\bibnamefont {Vo}},
  \bibinfo {author} {\bibfnamefont {C.}~\bibnamefont {Vorvick}}, \bibinfo
  {author} {\bibfnamefont {M.}~\bibnamefont {Walker}}, \bibinfo {author}
  {\bibfnamefont {R.~L.}\ \bibnamefont {Ward}}, \bibinfo {author}
  {\bibfnamefont {J.}~\bibnamefont {Warner}}, \bibinfo {author} {\bibfnamefont
  {B.}~\bibnamefont {Weaver}}, \bibinfo {author} {\bibfnamefont
  {R.}~\bibnamefont {Weiss}}, \bibinfo {author} {\bibfnamefont
  {P.}~\bibnamefont {We\ss{}els}}, \bibinfo {author} {\bibfnamefont
  {B.}~\bibnamefont {Willke}}, \bibinfo {author} {\bibfnamefont {C.~C.}\
  \bibnamefont {Wipf}}, \bibinfo {author} {\bibfnamefont {J.}~\bibnamefont
  {Worden}}, \bibinfo {author} {\bibfnamefont {G.}~\bibnamefont {Wu}}, \bibinfo
  {author} {\bibfnamefont {H.}~\bibnamefont {Yamamoto}}, \bibinfo {author}
  {\bibfnamefont {C.~C.}\ \bibnamefont {Yancey}}, \bibinfo {author}
  {\bibfnamefont {H.}~\bibnamefont {Yu}}, \bibinfo {author} {\bibfnamefont
  {H.}~\bibnamefont {Yu}}, \bibinfo {author} {\bibfnamefont {L.}~\bibnamefont
  {Zhang}}, \bibinfo {author} {\bibfnamefont {M.~E.}\ \bibnamefont {Zucker}}, \
  and\ \bibinfo {author} {\bibfnamefont {J.}~\bibnamefont {Zweizig}} (\bibinfo
  {collaboration} {LSC Instrument Authors}),\ }\href {\doibase
  10.1103/PhysRevA.95.043831} {\bibfield  {journal} {\bibinfo  {journal} {Phys.
  Rev. A}\ }\textbf {\bibinfo {volume} {95}},\ \bibinfo {pages} {043831}
  (\bibinfo {year} {2017})}\BibitemShut {NoStop}%
\bibitem [{\citenamefont {Harry}\ \emph {et~al.}(2012)\citenamefont {Harry},
  \citenamefont {Bodiya},\ and\ \citenamefont {DeSalvo}}]{Harry_Thermal_2012}%
  \BibitemOpen
  \bibfield  {author} {\bibinfo {author} {\bibfnamefont {G.}~\bibnamefont
  {Harry}}, \bibinfo {author} {\bibfnamefont {T.}~\bibnamefont {Bodiya}}, \
  and\ \bibinfo {author} {\bibfnamefont {R.}~\bibnamefont {DeSalvo}},\ }\href
  {https://books.google.ru/books?id=HBz2ngEACAAJ} {\emph {\bibinfo {title}
  {Optical Coatings and Thermal Noise in Precision Measurement}}}\ (\bibinfo
  {publisher} {Cambridge University Press},\ \bibinfo {year}
  {2012})\BibitemShut {NoStop}%
\bibitem [{\citenamefont {Gras}\ \emph {et~al.}(2017)\citenamefont {Gras},
  \citenamefont {Yu}, \citenamefont {Yam}, \citenamefont {Martynov},\ and\
  \citenamefont {Evans}}]{Gras_CTN_2017}%
  \BibitemOpen
  \bibfield  {author} {\bibinfo {author} {\bibfnamefont {S.}~\bibnamefont
  {Gras}}, \bibinfo {author} {\bibfnamefont {H.}~\bibnamefont {Yu}}, \bibinfo
  {author} {\bibfnamefont {W.}~\bibnamefont {Yam}}, \bibinfo {author}
  {\bibfnamefont {D.}~\bibnamefont {Martynov}}, \ and\ \bibinfo {author}
  {\bibfnamefont {M.}~\bibnamefont {Evans}},\ }\href {\doibase
  10.1103/PhysRevD.95.022001} {\bibfield  {journal} {\bibinfo  {journal} {Phys.
  Rev. D}\ }\textbf {\bibinfo {volume} {95}},\ \bibinfo {pages} {022001}
  (\bibinfo {year} {2017})}\BibitemShut {NoStop}%
\bibitem [{\citenamefont {Driggers}\ \emph {et~al.}(2012)\citenamefont
  {Driggers}, \citenamefont {Harms},\ and\ \citenamefont
  {Adhikari}}]{Driggers_NN_2012}%
  \BibitemOpen
  \bibfield  {author} {\bibinfo {author} {\bibfnamefont {J.}~\bibnamefont
  {Driggers}}, \bibinfo {author} {\bibfnamefont {J.}~\bibnamefont {Harms}}, \
  and\ \bibinfo {author} {\bibfnamefont {R.}~\bibnamefont {Adhikari}},\ }\href
  {http://journals.aps.org/prd/abstract/10.1103/PhysRevD.86.102001} {\bibfield
  {journal} {\bibinfo  {journal} {Phys. Rev. D}\ }\textbf {\bibinfo {volume}
  {86}} (\bibinfo {year} {2012})}\BibitemShut {NoStop}%
\bibitem [{\citenamefont {GonzÃ¡lez}(2000)}]{Gonzalez_SUS_2000}%
  \BibitemOpen
  \bibfield  {author} {\bibinfo {author} {\bibfnamefont {G.}~\bibnamefont
  {GonzÃ¡lez}},\ }\href {http://stacks.iop.org/0264-9381/17/i=21/a=305}
  {\bibfield  {journal} {\bibinfo  {journal} {Classical and Quantum Gravity}\
  }\textbf {\bibinfo {volume} {17}},\ \bibinfo {pages} {4409} (\bibinfo {year}
  {2000})}\BibitemShut {NoStop}%
\bibitem [{\citenamefont {Ballmer}\ and\ \citenamefont
  {Ottaway}(2013)}]{Ballmer_FOLD_2013}%
  \BibitemOpen
  \bibfield  {author} {\bibinfo {author} {\bibfnamefont {S.~W.}\ \bibnamefont
  {Ballmer}}\ and\ \bibinfo {author} {\bibfnamefont {D.~J.}\ \bibnamefont
  {Ottaway}},\ }\href {\doibase 10.1103/PhysRevD.88.062004} {\bibfield
  {journal} {\bibinfo  {journal} {Phys. Rev. D}\ }\textbf {\bibinfo {volume}
  {88}},\ \bibinfo {pages} {062004} (\bibinfo {year} {2013})}\BibitemShut
  {NoStop}%
\bibitem [{\citenamefont {Heinert}\ \emph {et~al.}(2014)\citenamefont
  {Heinert}, \citenamefont {Craig}, \citenamefont {Grote}, \citenamefont
  {Hild}, \citenamefont {L\"uck}, \citenamefont {Nawrodt}, \citenamefont
  {Simakov}, \citenamefont {Vasilyev}, \citenamefont {Vyatchanin},\ and\
  \citenamefont {Wittel}}]{Heinert_FOLD_2014}%
  \BibitemOpen
  \bibfield  {author} {\bibinfo {author} {\bibfnamefont {D.}~\bibnamefont
  {Heinert}}, \bibinfo {author} {\bibfnamefont {K.}~\bibnamefont {Craig}},
  \bibinfo {author} {\bibfnamefont {H.}~\bibnamefont {Grote}}, \bibinfo
  {author} {\bibfnamefont {S.}~\bibnamefont {Hild}}, \bibinfo {author}
  {\bibfnamefont {H.}~\bibnamefont {L\"uck}}, \bibinfo {author} {\bibfnamefont
  {R.}~\bibnamefont {Nawrodt}}, \bibinfo {author} {\bibfnamefont {D.~A.}\
  \bibnamefont {Simakov}}, \bibinfo {author} {\bibfnamefont {D.~V.}\
  \bibnamefont {Vasilyev}}, \bibinfo {author} {\bibfnamefont {S.~P.}\
  \bibnamefont {Vyatchanin}}, \ and\ \bibinfo {author} {\bibfnamefont
  {H.}~\bibnamefont {Wittel}},\ }\href {\doibase 10.1103/PhysRevD.90.042001}
  {\bibfield  {journal} {\bibinfo  {journal} {Phys. Rev. D}\ }\textbf {\bibinfo
  {volume} {90}},\ \bibinfo {pages} {042001} (\bibinfo {year}
  {2014})}\BibitemShut {NoStop}%
\bibitem [{\citenamefont {Essick}\ \emph
  {et~al.}(2017{\natexlab{b}})\citenamefont {Essick}, \citenamefont {Vitale},\
  and\ \citenamefont {Evans}}]{Essick:2017}%
  \BibitemOpen
  \bibfield  {author} {\bibinfo {author} {\bibfnamefont {R.}~\bibnamefont
  {Essick}}, \bibinfo {author} {\bibfnamefont {S.}~\bibnamefont {Vitale}}, \
  and\ \bibinfo {author} {\bibfnamefont {M.}~\bibnamefont {Evans}},\ }\href
  {\doibase 10.1103/PhysRevD.96.084004} {\bibfield  {journal} {\bibinfo
  {journal} {Phys. Rev. D}\ }\textbf {\bibinfo {volume} {96}},\ \bibinfo
  {pages} {084004} (\bibinfo {year} {2017}{\natexlab{b}})}\BibitemShut
  {NoStop}%
\bibitem [{\citenamefont {Arain}\ and\ \citenamefont
  {Mueller}(2008)}]{Arain_Recycling_2008}%
  \BibitemOpen
  \bibfield  {author} {\bibinfo {author} {\bibfnamefont {M.~A.}\ \bibnamefont
  {Arain}}\ and\ \bibinfo {author} {\bibfnamefont {G.}~\bibnamefont
  {Mueller}},\ }\href {\doibase 10.1364/OE.16.010018} {\bibfield  {journal}
  {\bibinfo  {journal} {Optics Express}\ }\textbf {\bibinfo {volume} {16}},\
  \bibinfo {pages} {10018} (\bibinfo {year} {2008})}\BibitemShut {NoStop}%
\bibitem [{\citenamefont {Bose}\ \emph {et~al.}(2018)\citenamefont {Bose},
  \citenamefont {Chakravarti}, \citenamefont {Rezzolla}, \citenamefont
  {Sathyaprakash},\ and\ \citenamefont {Takami}}]{Bose:2017jvk}%
  \BibitemOpen
  \bibfield  {author} {\bibinfo {author} {\bibfnamefont {S.}~\bibnamefont
  {Bose}}, \bibinfo {author} {\bibfnamefont {K.}~\bibnamefont {Chakravarti}},
  \bibinfo {author} {\bibfnamefont {L.}~\bibnamefont {Rezzolla}}, \bibinfo
  {author} {\bibfnamefont {B.~S.}\ \bibnamefont {Sathyaprakash}}, \ and\
  \bibinfo {author} {\bibfnamefont {K.}~\bibnamefont {Takami}},\ }\href
  {\doibase 10.1103/PhysRevLett.120.031102} {\bibfield  {journal} {\bibinfo
  {journal} {Phys. Rev. Lett.}\ }\textbf {\bibinfo {volume} {120}},\ \bibinfo
  {pages} {031102} (\bibinfo {year} {2018})},\ \Eprint
  {http://arxiv.org/abs/1705.10850} {arXiv:1705.10850 [gr-qc]} \BibitemShut
  {NoStop}%
\bibitem [{\citenamefont {Chatziioannou}\ \emph {et~al.}(2017)\citenamefont
  {Chatziioannou}, \citenamefont {Clark}, \citenamefont {Bauswein},
  \citenamefont {Millhouse}, \citenamefont {Littenberg},\ and\ \citenamefont
  {Cornish}}]{Katerina_2017}%
  \BibitemOpen
  \bibfield  {author} {\bibinfo {author} {\bibfnamefont {K.}~\bibnamefont
  {Chatziioannou}}, \bibinfo {author} {\bibfnamefont {J.~A.}\ \bibnamefont
  {Clark}}, \bibinfo {author} {\bibfnamefont {A.}~\bibnamefont {Bauswein}},
  \bibinfo {author} {\bibfnamefont {M.}~\bibnamefont {Millhouse}}, \bibinfo
  {author} {\bibfnamefont {T.~B.}\ \bibnamefont {Littenberg}}, \ and\ \bibinfo
  {author} {\bibfnamefont {N.}~\bibnamefont {Cornish}},\ }\href {\doibase
  10.1103/PhysRevD.96.124035} {\bibfield  {journal} {\bibinfo  {journal} {Phys.
  Rev.}\ }\textbf {\bibinfo {volume} {D96}},\ \bibinfo {pages} {124035}
  (\bibinfo {year} {2017})},\ \Eprint {http://arxiv.org/abs/1711.00040}
  {arXiv:1711.00040 [gr-qc]} \BibitemShut {NoStop}%
\bibitem [{\citenamefont {Damour}\ \emph {et~al.}(2012)\citenamefont {Damour},
  \citenamefont {Nagar},\ and\ \citenamefont {Villain}}]{Damour:2012yf}%
  \BibitemOpen
  \bibfield  {author} {\bibinfo {author} {\bibfnamefont {T.}~\bibnamefont
  {Damour}}, \bibinfo {author} {\bibfnamefont {A.}~\bibnamefont {Nagar}}, \
  and\ \bibinfo {author} {\bibfnamefont {L.}~\bibnamefont {Villain}},\ }\href
  {\doibase 10.1103/PhysRevD.85.123007} {\bibfield  {journal} {\bibinfo
  {journal} {Phys. Rev.}\ }\textbf {\bibinfo {volume} {D85}},\ \bibinfo {pages}
  {123007} (\bibinfo {year} {2012})},\ \Eprint {http://arxiv.org/abs/1203.4352}
  {arXiv:1203.4352 [gr-qc]} \BibitemShut {NoStop}%
\bibitem [{\citenamefont {Takami}\ \emph {et~al.}(2015)\citenamefont {Takami},
  \citenamefont {Rezzolla},\ and\ \citenamefont
  {Baiotti}}]{Takami_Merger_2015}%
  \BibitemOpen
  \bibfield  {author} {\bibinfo {author} {\bibfnamefont {K.}~\bibnamefont
  {Takami}}, \bibinfo {author} {\bibfnamefont {L.}~\bibnamefont {Rezzolla}}, \
  and\ \bibinfo {author} {\bibfnamefont {L.}~\bibnamefont {Baiotti}},\ }\href
  {\doibase 10.1103/PhysRevD.91.064001} {\bibfield  {journal} {\bibinfo
  {journal} {Phys. Rev. D}\ }\textbf {\bibinfo {volume} {91}},\ \bibinfo
  {pages} {064001} (\bibinfo {year} {2015})}\BibitemShut {NoStop}%
\bibitem [{\citenamefont {Steiner}\ \emph {et~al.}(2013)\citenamefont
  {Steiner}, \citenamefont {Hempel},\ and\ \citenamefont
  {Fischer}}]{Steiner2013}%
  \BibitemOpen
  \bibfield  {author} {\bibinfo {author} {\bibfnamefont {A.~W.}\ \bibnamefont
  {Steiner}}, \bibinfo {author} {\bibfnamefont {M.}~\bibnamefont {Hempel}}, \
  and\ \bibinfo {author} {\bibfnamefont {T.}~\bibnamefont {Fischer}},\ }\href
  {\doibase 10.1088/0004-637X/774/1/17} {\bibfield  {journal} {\bibinfo
  {journal} {\apj}\ }\textbf {\bibinfo {volume} {774}},\ \bibinfo {eid} {17}
  (\bibinfo {year} {2013})},\ \Eprint {http://arxiv.org/abs/1207.2184}
  {arXiv:1207.2184 [astro-ph.SR]} \BibitemShut {NoStop}%
\bibitem [{\citenamefont {{Hempel}}\ \emph {et~al.}(2012)\citenamefont
  {{Hempel}}, \citenamefont {{Fischer}}, \citenamefont {{Schaffner-Bielich}},\
  and\ \citenamefont {{Liebend{\"o}rfer}}}]{Hempel2012}%
  \BibitemOpen
  \bibfield  {author} {\bibinfo {author} {\bibfnamefont {M.}~\bibnamefont
  {{Hempel}}}, \bibinfo {author} {\bibfnamefont {T.}~\bibnamefont {{Fischer}}},
  \bibinfo {author} {\bibfnamefont {J.}~\bibnamefont {{Schaffner-Bielich}}}, \
  and\ \bibinfo {author} {\bibfnamefont {M.}~\bibnamefont
  {{Liebend{\"o}rfer}}},\ }\href {\doibase 10.1088/0004-637X/748/1/70}
  {\bibfield  {journal} {\bibinfo  {journal} {\apj}\ }\textbf {\bibinfo
  {volume} {748}},\ \bibinfo {eid} {70} (\bibinfo {year} {2012})}\BibitemShut
  {NoStop}%
\bibitem [{\citenamefont {{Douchin}}\ and\ \citenamefont
  {{Haensel}}(2001)}]{Douchin2001}%
  \BibitemOpen
  \bibfield  {author} {\bibinfo {author} {\bibfnamefont {F.}~\bibnamefont
  {{Douchin}}}\ and\ \bibinfo {author} {\bibfnamefont {P.}~\bibnamefont
  {{Haensel}}},\ }\href {\doibase 10.1051/0004-6361:20011402} {\bibfield
  {journal} {\bibinfo  {journal} {Astronomy and Astrophysics}\ }\textbf
  {\bibinfo {volume} {380}},\ \bibinfo {pages} {151} (\bibinfo {year}
  {2001})},\ \Eprint {http://arxiv.org/abs/astro-ph/0111092} {astro-ph/0111092}
  \BibitemShut {NoStop}%
\bibitem [{\citenamefont {Akmal}\ \emph {et~al.}(1998)\citenamefont {Akmal},
  \citenamefont {Pandharipande},\ and\ \citenamefont {Ravenhall}}]{Akmal1998}%
  \BibitemOpen
  \bibfield  {author} {\bibinfo {author} {\bibfnamefont {A.}~\bibnamefont
  {Akmal}}, \bibinfo {author} {\bibfnamefont {V.~R.}\ \bibnamefont
  {Pandharipande}}, \ and\ \bibinfo {author} {\bibfnamefont {D.~G.}\
  \bibnamefont {Ravenhall}},\ }\href {\doibase 10.1103/PhysRevC.58.1804}
  {\bibfield  {journal} {\bibinfo  {journal} {\prc}\ }\textbf {\bibinfo
  {volume} {58}},\ \bibinfo {pages} {1804} (\bibinfo {year} {1998})},\ \Eprint
  {http://arxiv.org/abs/nucl-th/9804027} {nucl-th/9804027} \BibitemShut
  {NoStop}%
\bibitem [{\citenamefont {Palenzuela}\ \emph {et~al.}(2015)\citenamefont
  {Palenzuela}, \citenamefont {Liebling}, \citenamefont {Neilsen},
  \citenamefont {Lehner}, \citenamefont {Caballero}, \citenamefont {O'Connor},\
  and\ \citenamefont {Anderson}}]{Palenzuela_EOS_2015}%
  \BibitemOpen
  \bibfield  {author} {\bibinfo {author} {\bibfnamefont {C.}~\bibnamefont
  {Palenzuela}}, \bibinfo {author} {\bibfnamefont {S.~L.}\ \bibnamefont
  {Liebling}}, \bibinfo {author} {\bibfnamefont {D.}~\bibnamefont {Neilsen}},
  \bibinfo {author} {\bibfnamefont {L.}~\bibnamefont {Lehner}}, \bibinfo
  {author} {\bibfnamefont {O.~L.}\ \bibnamefont {Caballero}}, \bibinfo {author}
  {\bibfnamefont {E.}~\bibnamefont {O'Connor}}, \ and\ \bibinfo {author}
  {\bibfnamefont {M.}~\bibnamefont {Anderson}},\ }\href {\doibase
  10.1103/PhysRevD.92.044045} {\bibfield  {journal} {\bibinfo  {journal} {Phys.
  Rev. D}\ }\textbf {\bibinfo {volume} {92}},\ \bibinfo {pages} {44045}
  (\bibinfo {year} {2015})}\BibitemShut {NoStop}%
\bibitem [{\citenamefont {Centrella}\ \emph {et~al.}(2001)\citenamefont
  {Centrella}, \citenamefont {New}, \citenamefont {Lowe},\ and\ \citenamefont
  {Brown}}]{Centrella2001}%
  \BibitemOpen
  \bibfield  {author} {\bibinfo {author} {\bibfnamefont {J.~M.}\ \bibnamefont
  {Centrella}}, \bibinfo {author} {\bibfnamefont {K.~C.~B.}\ \bibnamefont
  {New}}, \bibinfo {author} {\bibfnamefont {L.~L.}\ \bibnamefont {Lowe}}, \
  and\ \bibinfo {author} {\bibfnamefont {J.~D.}\ \bibnamefont {Brown}},\ }\href
  {\doibase 10.1086/319634} {\bibfield  {journal} {\bibinfo  {journal}
  {Astrophys. J.}\ }\textbf {\bibinfo {volume} {550}},\ \bibinfo {pages} {L193}
  (\bibinfo {year} {2001})},\ \Eprint {http://arxiv.org/abs/astro-ph/0010574}
  {arXiv:astro-ph/0010574 [astro-ph]} \BibitemShut {NoStop}%
\bibitem [{\citenamefont {Saijo}\ \emph {et~al.}(2002)\citenamefont {Saijo},
  \citenamefont {Baumgarte},\ and\ \citenamefont {Shapiro}}]{Saijo2003}%
  \BibitemOpen
  \bibfield  {author} {\bibinfo {author} {\bibfnamefont {M.}~\bibnamefont
  {Saijo}}, \bibinfo {author} {\bibfnamefont {T.~W.}\ \bibnamefont
  {Baumgarte}}, \ and\ \bibinfo {author} {\bibfnamefont {S.~L.}\ \bibnamefont
  {Shapiro}},\ }\href {\doibase 10.1086/377334} {\bibfield  {journal} {\bibinfo
   {journal} {Astrophys. J.}\ }\textbf {\bibinfo {volume} {595}},\ \bibinfo
  {pages} {352} (\bibinfo {year} {2002})},\ \Eprint
  {http://arxiv.org/abs/astro-ph/0302436} {arXiv:astro-ph/0302436 [astro-ph]}
  \BibitemShut {NoStop}%
\bibitem [{\citenamefont {Watts}\ \emph {et~al.}(2005)\citenamefont {Watts},
  \citenamefont {Andersson},\ and\ \citenamefont {Jones}}]{Watts2005}%
  \BibitemOpen
  \bibfield  {author} {\bibinfo {author} {\bibfnamefont {A.~L.}\ \bibnamefont
  {Watts}}, \bibinfo {author} {\bibfnamefont {N.}~\bibnamefont {Andersson}}, \
  and\ \bibinfo {author} {\bibfnamefont {D.~I.}\ \bibnamefont {Jones}},\ }\href
  {\doibase 10.1086/427653} {\bibfield  {journal} {\bibinfo  {journal}
  {Astrophys. J.}\ }\textbf {\bibinfo {volume} {618}},\ \bibinfo {pages} {L37}
  (\bibinfo {year} {2005})},\ \Eprint {http://arxiv.org/abs/astro-ph/0309554}
  {arXiv:astro-ph/0309554 [astro-ph]} \BibitemShut {NoStop}%
\bibitem [{\citenamefont {Ou}\ and\ \citenamefont {Tohline}(2006)}]{Ou2006}%
  \BibitemOpen
  \bibfield  {author} {\bibinfo {author} {\bibfnamefont {S.}~\bibnamefont
  {Ou}}\ and\ \bibinfo {author} {\bibfnamefont {J.}~\bibnamefont {Tohline}},\
  }\href {\doibase 10.1086/507597} {\bibfield  {journal} {\bibinfo  {journal}
  {Astrophys. J.}\ }\textbf {\bibinfo {volume} {651}},\ \bibinfo {pages} {1068}
  (\bibinfo {year} {2006})},\ \Eprint {http://arxiv.org/abs/astro-ph/0604099}
  {arXiv:astro-ph/0604099 [astro-ph]} \BibitemShut {NoStop}%
\bibitem [{\citenamefont {Corvino}\ \emph {et~al.}(2010)\citenamefont
  {Corvino}, \citenamefont {Rezzolla}, \citenamefont {Bernuzzi}, \citenamefont
  {De~Pietri},\ and\ \citenamefont {Giacomazzo}}]{Corvino2010}%
  \BibitemOpen
  \bibfield  {author} {\bibinfo {author} {\bibfnamefont {G.}~\bibnamefont
  {Corvino}}, \bibinfo {author} {\bibfnamefont {L.}~\bibnamefont {Rezzolla}},
  \bibinfo {author} {\bibfnamefont {S.}~\bibnamefont {Bernuzzi}}, \bibinfo
  {author} {\bibfnamefont {R.}~\bibnamefont {De~Pietri}}, \ and\ \bibinfo
  {author} {\bibfnamefont {B.}~\bibnamefont {Giacomazzo}},\ }\bibfield
  {booktitle} {\emph {\bibinfo {booktitle} {{Microphysics in computational
  relativistic astrophysics. Proceedings, Workshop, MICRA2009, Copenhagen,
  Dennmark, August 24-28, 2009}}},\ }\href {\doibase
  10.1088/0264-9381/27/11/114104} {\bibfield  {journal} {\bibinfo  {journal}
  {Class. Quant. Grav.}\ }\textbf {\bibinfo {volume} {27}},\ \bibinfo {pages}
  {114104} (\bibinfo {year} {2010})},\ \Eprint {http://arxiv.org/abs/1001.5281}
  {arXiv:1001.5281 [gr-qc]} \BibitemShut {NoStop}%
\bibitem [{\citenamefont {Ott}\ \emph {et~al.}(2005)\citenamefont {Ott},
  \citenamefont {Ou}, \citenamefont {Tohline},\ and\ \citenamefont
  {Burrows}}]{Ott2005}%
  \BibitemOpen
  \bibfield  {author} {\bibinfo {author} {\bibfnamefont {C.~D.}\ \bibnamefont
  {Ott}}, \bibinfo {author} {\bibfnamefont {S.}~\bibnamefont {Ou}}, \bibinfo
  {author} {\bibfnamefont {J.~E.}\ \bibnamefont {Tohline}}, \ and\ \bibinfo
  {author} {\bibfnamefont {A.}~\bibnamefont {Burrows}},\ }\href {\doibase
  10.1086/431305} {\bibfield  {journal} {\bibinfo  {journal} {Astrophys. J.}\
  }\textbf {\bibinfo {volume} {625}},\ \bibinfo {pages} {L119} (\bibinfo {year}
  {2005})},\ \Eprint {http://arxiv.org/abs/astro-ph/0503187}
  {arXiv:astro-ph/0503187 [astro-ph]} \BibitemShut {NoStop}%
\bibitem [{\citenamefont {Ott}\ \emph {et~al.}(2007)\citenamefont {Ott},
  \citenamefont {Dimmelmeier}, \citenamefont {Marek}, \citenamefont {Janka},
  \citenamefont {Hawke}, \citenamefont {Zink},\ and\ \citenamefont
  {Schnetter}}]{Ott2007PhRvL}%
  \BibitemOpen
  \bibfield  {author} {\bibinfo {author} {\bibfnamefont {C.~D.}\ \bibnamefont
  {Ott}}, \bibinfo {author} {\bibfnamefont {H.}~\bibnamefont {Dimmelmeier}},
  \bibinfo {author} {\bibfnamefont {A.}~\bibnamefont {Marek}}, \bibinfo
  {author} {\bibfnamefont {H.~T.}\ \bibnamefont {Janka}}, \bibinfo {author}
  {\bibfnamefont {I.}~\bibnamefont {Hawke}}, \bibinfo {author} {\bibfnamefont
  {B.}~\bibnamefont {Zink}}, \ and\ \bibinfo {author} {\bibfnamefont
  {E.}~\bibnamefont {Schnetter}},\ }\href {\doibase
  10.1103/PhysRevLett.98.261101} {\bibfield  {journal} {\bibinfo  {journal}
  {Phys. Rev. Lett.}\ }\textbf {\bibinfo {volume} {98}},\ \bibinfo {pages}
  {261101} (\bibinfo {year} {2007})},\ \Eprint
  {http://arxiv.org/abs/astro-ph/0609819} {arXiv:astro-ph/0609819 [astro-ph]}
  \BibitemShut {NoStop}%
\bibitem [{\citenamefont {Kuroda}\ \emph {et~al.}(2014)\citenamefont {Kuroda},
  \citenamefont {Takiwaki},\ and\ \citenamefont {Kotake}}]{Kuroda2014}%
  \BibitemOpen
  \bibfield  {author} {\bibinfo {author} {\bibfnamefont {T.}~\bibnamefont
  {Kuroda}}, \bibinfo {author} {\bibfnamefont {T.}~\bibnamefont {Takiwaki}}, \
  and\ \bibinfo {author} {\bibfnamefont {K.}~\bibnamefont {Kotake}},\ }\href
  {\doibase 10.1103/PhysRevD.89.044011} {\bibfield  {journal} {\bibinfo
  {journal} {Phys. Rev.}\ }\textbf {\bibinfo {volume} {D89}},\ \bibinfo {pages}
  {044011} (\bibinfo {year} {2014})},\ \Eprint {http://arxiv.org/abs/1304.4372}
  {arXiv:1304.4372 [astro-ph.HE]} \BibitemShut {NoStop}%
\bibitem [{\citenamefont {East}\ \emph
  {et~al.}(2016{\natexlab{a}})\citenamefont {East}, \citenamefont
  {Paschalidis}, \citenamefont {Pretorius},\ and\ \citenamefont
  {Shapiro}}]{East:2015vix}%
  \BibitemOpen
  \bibfield  {author} {\bibinfo {author} {\bibfnamefont {W.~E.}\ \bibnamefont
  {East}}, \bibinfo {author} {\bibfnamefont {V.}~\bibnamefont {Paschalidis}},
  \bibinfo {author} {\bibfnamefont {F.}~\bibnamefont {Pretorius}}, \ and\
  \bibinfo {author} {\bibfnamefont {S.~L.}\ \bibnamefont {Shapiro}},\ }\href
  {\doibase 10.1103/PhysRevD.93.024011} {\bibfield  {journal} {\bibinfo
  {journal} {Phys. Rev.}\ }\textbf {\bibinfo {volume} {D93}},\ \bibinfo {pages}
  {024011} (\bibinfo {year} {2016}{\natexlab{a}})},\ \Eprint
  {http://arxiv.org/abs/1511.01093} {arXiv:1511.01093 [astro-ph.HE]}
  \BibitemShut {NoStop}%
\bibitem [{\citenamefont {East}\ \emph
  {et~al.}(2016{\natexlab{b}})\citenamefont {East}, \citenamefont
  {Paschalidis},\ and\ \citenamefont {Pretorius}}]{East:2016zvv}%
  \BibitemOpen
  \bibfield  {author} {\bibinfo {author} {\bibfnamefont {W.~E.}\ \bibnamefont
  {East}}, \bibinfo {author} {\bibfnamefont {V.}~\bibnamefont {Paschalidis}}, \
  and\ \bibinfo {author} {\bibfnamefont {F.}~\bibnamefont {Pretorius}},\ }\href
  {\doibase 10.1088/0264-9381/33/24/244004} {\bibfield  {journal} {\bibinfo
  {journal} {Class. Quant. Grav.}\ }\textbf {\bibinfo {volume} {33}},\ \bibinfo
  {pages} {244004} (\bibinfo {year} {2016}{\natexlab{b}})},\ \Eprint
  {http://arxiv.org/abs/1609.00725} {arXiv:1609.00725 [astro-ph.HE]}
  \BibitemShut {NoStop}%
\bibitem [{\citenamefont {Radice}\ \emph {et~al.}(2016)\citenamefont {Radice},
  \citenamefont {Bernuzzi},\ and\ \citenamefont {Ott}}]{Radice:2016gym}%
  \BibitemOpen
  \bibfield  {author} {\bibinfo {author} {\bibfnamefont {D.}~\bibnamefont
  {Radice}}, \bibinfo {author} {\bibfnamefont {S.}~\bibnamefont {Bernuzzi}}, \
  and\ \bibinfo {author} {\bibfnamefont {C.~D.}\ \bibnamefont {Ott}},\ }\href
  {\doibase 10.1103/PhysRevD.94.064011} {\bibfield  {journal} {\bibinfo
  {journal} {Phys. Rev.}\ }\textbf {\bibinfo {volume} {D94}},\ \bibinfo {pages}
  {064011} (\bibinfo {year} {2016})},\ \Eprint
  {http://arxiv.org/abs/1603.05726} {arXiv:1603.05726 [gr-qc]} \BibitemShut
  {NoStop}%
\bibitem [{\citenamefont {Lehner}\ \emph {et~al.}(2016)\citenamefont {Lehner},
  \citenamefont {Liebling}, \citenamefont {Palenzuela},\ and\ \citenamefont
  {Motl}}]{Lehner:2016wjg}%
  \BibitemOpen
  \bibfield  {author} {\bibinfo {author} {\bibfnamefont {L.}~\bibnamefont
  {Lehner}}, \bibinfo {author} {\bibfnamefont {S.~L.}\ \bibnamefont
  {Liebling}}, \bibinfo {author} {\bibfnamefont {C.}~\bibnamefont
  {Palenzuela}}, \ and\ \bibinfo {author} {\bibfnamefont {P.~M.}\ \bibnamefont
  {Motl}},\ }\href {\doibase 10.1103/PhysRevD.94.043003} {\bibfield  {journal}
  {\bibinfo  {journal} {Phys. Rev.}\ }\textbf {\bibinfo {volume} {D94}},\
  \bibinfo {pages} {043003} (\bibinfo {year} {2016})},\ \Eprint
  {http://arxiv.org/abs/1605.02369} {arXiv:1605.02369 [gr-qc]} \BibitemShut
  {NoStop}%
\bibitem [{\citenamefont {Jeffreys}(1961)}]{Jeffreys:61}%
  \BibitemOpen
  \bibfield  {author} {\bibinfo {author} {\bibfnamefont {H.}~\bibnamefont
  {Jeffreys}},\ }\href@noop {} {\emph {\bibinfo {title} {Theory of
  Probability}}},\ \bibinfo {edition} {3rd}\ ed.\ (\bibinfo  {publisher}
  {Oxford},\ \bibinfo {address} {Oxford, England},\ \bibinfo {year}
  {1961})\BibitemShut {NoStop}%
\bibitem [{\citenamefont {Flanagan}\ and\ \citenamefont
  {Hinderer}(2008)}]{Flanagan_2008}%
  \BibitemOpen
  \bibfield  {author} {\bibinfo {author} {\bibfnamefont {E.~E.}\ \bibnamefont
  {Flanagan}}\ and\ \bibinfo {author} {\bibfnamefont {T.}~\bibnamefont
  {Hinderer}},\ }\href {\doibase 10.1103/PhysRevD.77.021502} {\bibfield
  {journal} {\bibinfo  {journal} {Phys. Rev. D}\ }\textbf {\bibinfo {volume}
  {77}},\ \bibinfo {pages} {021502} (\bibinfo {year} {2008})}\BibitemShut
  {NoStop}%
\bibitem [{\citenamefont {Wade}\ \emph {et~al.}(2014)\citenamefont {Wade},
  \citenamefont {Creighton}, \citenamefont {Ochsner}, \citenamefont {Lackey},
  \citenamefont {Farr}, \citenamefont {Littenberg},\ and\ \citenamefont
  {Raymond}}]{Wade_2014}%
  \BibitemOpen
  \bibfield  {author} {\bibinfo {author} {\bibfnamefont {L.}~\bibnamefont
  {Wade}}, \bibinfo {author} {\bibfnamefont {J.~D.~E.}\ \bibnamefont
  {Creighton}}, \bibinfo {author} {\bibfnamefont {E.}~\bibnamefont {Ochsner}},
  \bibinfo {author} {\bibfnamefont {B.~D.}\ \bibnamefont {Lackey}}, \bibinfo
  {author} {\bibfnamefont {B.~F.}\ \bibnamefont {Farr}}, \bibinfo {author}
  {\bibfnamefont {T.~B.}\ \bibnamefont {Littenberg}}, \ and\ \bibinfo {author}
  {\bibfnamefont {V.}~\bibnamefont {Raymond}},\ }\href {\doibase
  10.1103/PhysRevD.89.103012} {\bibfield  {journal} {\bibinfo  {journal} {Phys.
  Rev. D}\ }\textbf {\bibinfo {volume} {89}},\ \bibinfo {pages} {103012}
  (\bibinfo {year} {2014})}\BibitemShut {NoStop}%
\bibitem [{\citenamefont {Hinderer}\ \emph {et~al.}(2010)\citenamefont
  {Hinderer}, \citenamefont {Lackey}, \citenamefont {Lang},\ and\ \citenamefont
  {Read}}]{Hinderer_2010}%
  \BibitemOpen
  \bibfield  {author} {\bibinfo {author} {\bibfnamefont {T.}~\bibnamefont
  {Hinderer}}, \bibinfo {author} {\bibfnamefont {B.~D.}\ \bibnamefont
  {Lackey}}, \bibinfo {author} {\bibfnamefont {R.~N.}\ \bibnamefont {Lang}}, \
  and\ \bibinfo {author} {\bibfnamefont {J.~S.}\ \bibnamefont {Read}},\ }\href
  {\doibase 10.1103/PhysRevD.81.123016} {\bibfield  {journal} {\bibinfo
  {journal} {Phys. Rev. D}\ }\textbf {\bibinfo {volume} {81}},\ \bibinfo
  {pages} {123016} (\bibinfo {year} {2010})}\BibitemShut {NoStop}%
\bibitem [{\citenamefont {Husa}\ \emph {et~al.}(2016)\citenamefont {Husa},
  \citenamefont {Khan}, \citenamefont {Hannam}, \citenamefont {P\"urrer},
  \citenamefont {Ohme}, \citenamefont {Forteza},\ and\ \citenamefont
  {Boh\'e}}]{Husa:2016}%
  \BibitemOpen
  \bibfield  {author} {\bibinfo {author} {\bibfnamefont {S.}~\bibnamefont
  {Husa}}, \bibinfo {author} {\bibfnamefont {S.}~\bibnamefont {Khan}}, \bibinfo
  {author} {\bibfnamefont {M.}~\bibnamefont {Hannam}}, \bibinfo {author}
  {\bibfnamefont {M.}~\bibnamefont {P\"urrer}}, \bibinfo {author}
  {\bibfnamefont {F.}~\bibnamefont {Ohme}}, \bibinfo {author} {\bibfnamefont
  {X.~J.}\ \bibnamefont {Forteza}}, \ and\ \bibinfo {author} {\bibfnamefont
  {A.}~\bibnamefont {Boh\'e}},\ }\href {\doibase 10.1103/PhysRevD.93.044006}
  {\bibfield  {journal} {\bibinfo  {journal} {Phys. Rev. D}\ }\textbf {\bibinfo
  {volume} {93}},\ \bibinfo {pages} {044006} (\bibinfo {year}
  {2016})}\BibitemShut {NoStop}%
\bibitem [{\citenamefont {Khan}\ \emph {et~al.}(2016)\citenamefont {Khan},
  \citenamefont {Husa}, \citenamefont {Hannam}, \citenamefont {Ohme},
  \citenamefont {P\"urrer}, \citenamefont {Forteza},\ and\ \citenamefont
  {Boh\'e}}]{Khan:2016}%
  \BibitemOpen
  \bibfield  {author} {\bibinfo {author} {\bibfnamefont {S.}~\bibnamefont
  {Khan}}, \bibinfo {author} {\bibfnamefont {S.}~\bibnamefont {Husa}}, \bibinfo
  {author} {\bibfnamefont {M.}~\bibnamefont {Hannam}}, \bibinfo {author}
  {\bibfnamefont {F.}~\bibnamefont {Ohme}}, \bibinfo {author} {\bibfnamefont
  {M.}~\bibnamefont {P\"urrer}}, \bibinfo {author} {\bibfnamefont {X.~J.}\
  \bibnamefont {Forteza}}, \ and\ \bibinfo {author} {\bibfnamefont
  {A.}~\bibnamefont {Boh\'e}},\ }\href {\doibase 10.1103/PhysRevD.93.044007}
  {\bibfield  {journal} {\bibinfo  {journal} {Phys. Rev. D}\ }\textbf {\bibinfo
  {volume} {93}},\ \bibinfo {pages} {044007} (\bibinfo {year}
  {2016})}\BibitemShut {NoStop}%
\bibitem [{\citenamefont {Dietrich}\ \emph {et~al.}(2017)\citenamefont
  {Dietrich}, \citenamefont {Bernuzzi},\ and\ \citenamefont
  {Tichy}}]{Dietrich:2017}%
  \BibitemOpen
  \bibfield  {author} {\bibinfo {author} {\bibfnamefont {T.}~\bibnamefont
  {Dietrich}}, \bibinfo {author} {\bibfnamefont {S.}~\bibnamefont {Bernuzzi}},
  \ and\ \bibinfo {author} {\bibfnamefont {W.}~\bibnamefont {Tichy}},\ }\href
  {\doibase 10.1103/PhysRevD.96.121501} {\bibfield  {journal} {\bibinfo
  {journal} {Phys. Rev. D}\ }\textbf {\bibinfo {volume} {96}},\ \bibinfo
  {pages} {121501} (\bibinfo {year} {2017})}\BibitemShut {NoStop}%
\bibitem [{\citenamefont {{Planck Collaboration}}(2016)}]{Planck:2015}%
  \BibitemOpen
  \bibfield  {author} {\bibinfo {author} {\bibnamefont {{Planck
  Collaboration}}},\ }\href {\doibase 10.1051/0004-6361/201525830} {\bibfield
  {journal} {\bibinfo  {journal} {A\&A}\ }\textbf {\bibinfo {volume} {594}},\
  \bibinfo {pages} {A13} (\bibinfo {year} {2016})}\BibitemShut {NoStop}%
\bibitem [{\citenamefont {Vitale}\ and\ \citenamefont
  {Chen}(2018)}]{Vitale:2018}%
  \BibitemOpen
  \bibfield  {author} {\bibinfo {author} {\bibfnamefont {S.}~\bibnamefont
  {Vitale}}\ and\ \bibinfo {author} {\bibfnamefont {H.-Y.}\ \bibnamefont
  {Chen}},\ }\href {\doibase 10.1103/PhysRevLett.121.021303} {\bibfield
  {journal} {\bibinfo  {journal} {Phys. Rev. Lett.}\ }\textbf {\bibinfo
  {volume} {121}},\ \bibinfo {pages} {021303} (\bibinfo {year}
  {2018})}\BibitemShut {NoStop}%
\bibitem [{\citenamefont {Abbott}\ \emph {et~al.}(2017)\citenamefont {Abbott}
  \emph {et~al.}}]{Abbott:2017xzu}%
  \BibitemOpen
  \bibfield  {author} {\bibinfo {author} {\bibfnamefont {B.~P.}\ \bibnamefont
  {Abbott}} \emph {et~al.} (\bibinfo {collaboration} {LIGO Scientific,
  VINROUGE, Las Cumbres Observatory, DES, DLT40, Virgo, 1M2H, Dark Energy
  Camera GW-E, MASTER}),\ }\href {\doibase 10.1038/nature24471} {\bibfield
  {journal} {\bibinfo  {journal} {Nature}\ }\textbf {\bibinfo {volume} {551}},\
  \bibinfo {pages} {85} (\bibinfo {year} {2017})},\ \Eprint
  {http://arxiv.org/abs/1710.05835} {arXiv:1710.05835 [astro-ph.CO]}
  \BibitemShut {NoStop}%
\bibitem [{\citenamefont {Shibata}\ \emph {et~al.}(2009)\citenamefont
  {Shibata}, \citenamefont {Kyutoku}, \citenamefont {Yamamoto},\ and\
  \citenamefont {Taniguchi}}]{Shibata_BHNS_2009}%
  \BibitemOpen
  \bibfield  {author} {\bibinfo {author} {\bibfnamefont {M.}~\bibnamefont
  {Shibata}}, \bibinfo {author} {\bibfnamefont {K.}~\bibnamefont {Kyutoku}},
  \bibinfo {author} {\bibfnamefont {T.}~\bibnamefont {Yamamoto}}, \ and\
  \bibinfo {author} {\bibfnamefont {K.}~\bibnamefont {Taniguchi}},\ }\href
  {\doibase 10.1103/PhysRevD.79.044030} {\bibfield  {journal} {\bibinfo
  {journal} {Phys. Rev. D}\ }\textbf {\bibinfo {volume} {79}},\ \bibinfo
  {pages} {044030} (\bibinfo {year} {2009})}\BibitemShut {NoStop}%
\bibitem [{\citenamefont {Dooley}\ \emph {et~al.}(2013)\citenamefont {Dooley},
  \citenamefont {Barsotti}, \citenamefont {Adhikari}, \citenamefont {Evans},
  \citenamefont {Fricke}, \citenamefont {Fritschel}, \citenamefont {Frolov},
  \citenamefont {Kawabe},\ and\ \citenamefont
  {Smith-Lefebvre}}]{Dooley_ASC_2013}%
  \BibitemOpen
  \bibfield  {author} {\bibinfo {author} {\bibfnamefont {K.~L.}\ \bibnamefont
  {Dooley}}, \bibinfo {author} {\bibfnamefont {L.}~\bibnamefont {Barsotti}},
  \bibinfo {author} {\bibfnamefont {R.~X.}\ \bibnamefont {Adhikari}}, \bibinfo
  {author} {\bibfnamefont {M.}~\bibnamefont {Evans}}, \bibinfo {author}
  {\bibfnamefont {T.~T.}\ \bibnamefont {Fricke}}, \bibinfo {author}
  {\bibfnamefont {P.}~\bibnamefont {Fritschel}}, \bibinfo {author}
  {\bibfnamefont {V.}~\bibnamefont {Frolov}}, \bibinfo {author} {\bibfnamefont
  {K.}~\bibnamefont {Kawabe}}, \ and\ \bibinfo {author} {\bibfnamefont
  {N.}~\bibnamefont {Smith-Lefebvre}},\ }\href {\doibase
  10.1364/JOSAA.30.002618} {\bibfield  {journal} {\bibinfo  {journal} {J. Opt.
  Soc. Am. A}\ }\textbf {\bibinfo {volume} {30}},\ \bibinfo {pages} {2618}
  (\bibinfo {year} {2013})}\BibitemShut {NoStop}%
\bibitem [{\citenamefont {Braginsky}\ \emph {et~al.}(2002)\citenamefont
  {Braginsky}, \citenamefont {Strigin},\ and\ \citenamefont
  {Vyatchanin}}]{Braginsky_PI_2002}%
  \BibitemOpen
  \bibfield  {author} {\bibinfo {author} {\bibfnamefont {V.}~\bibnamefont
  {Braginsky}}, \bibinfo {author} {\bibfnamefont {S.}~\bibnamefont {Strigin}},
  \ and\ \bibinfo {author} {\bibfnamefont {S.}~\bibnamefont {Vyatchanin}},\
  }\href {http://arxiv.org/abs/gr-qc/0209064} {\bibfield  {journal} {\bibinfo
  {journal} {Phys. Lett. A}\ }\textbf {\bibinfo {volume} {305}},\ \bibinfo
  {pages} {111} (\bibinfo {year} {2002})}\BibitemShut {NoStop}%
\bibitem [{\citenamefont {Hirose}\ \emph {et~al.}(2010)\citenamefont {Hirose},
  \citenamefont {Kawabe}, \citenamefont {Sigg}, \citenamefont {Adhikari},\ and\
  \citenamefont {Saulson}}]{Hirose_ASC_2010}%
  \BibitemOpen
  \bibfield  {author} {\bibinfo {author} {\bibfnamefont {E.}~\bibnamefont
  {Hirose}}, \bibinfo {author} {\bibfnamefont {K.}~\bibnamefont {Kawabe}},
  \bibinfo {author} {\bibfnamefont {D.}~\bibnamefont {Sigg}}, \bibinfo {author}
  {\bibfnamefont {R.}~\bibnamefont {Adhikari}}, \ and\ \bibinfo {author}
  {\bibfnamefont {P.~R.}\ \bibnamefont {Saulson}},\ }\href {\doibase
  10.1364/AO.49.003474} {\bibfield  {journal} {\bibinfo  {journal} {Appl.
  Opt.}\ }\textbf {\bibinfo {volume} {49}},\ \bibinfo {pages} {3474} (\bibinfo
  {year} {2010})}\BibitemShut {NoStop}%
\bibitem [{\citenamefont {Barsotti}\ \emph {et~al.}(2010)\citenamefont
  {Barsotti}, \citenamefont {Evans},\ and\ \citenamefont
  {Fritschel}}]{Barsotti_ASC_2010}%
  \BibitemOpen
  \bibfield  {author} {\bibinfo {author} {\bibfnamefont {L.}~\bibnamefont
  {Barsotti}}, \bibinfo {author} {\bibfnamefont {M.}~\bibnamefont {Evans}}, \
  and\ \bibinfo {author} {\bibfnamefont {P.}~\bibnamefont {Fritschel}},\ }\href
  {http://stacks.iop.org/0264-9381/27/i=8/a=084026} {\bibfield  {journal}
  {\bibinfo  {journal} {Classical and Quantum Gravity}\ }\textbf {\bibinfo
  {volume} {27}},\ \bibinfo {pages} {084026} (\bibinfo {year}
  {2010})}\BibitemShut {NoStop}%
\bibitem [{\citenamefont {Yu}\ \emph {et~al.}(2018)\citenamefont {Yu},
  \citenamefont {Martynov}, \citenamefont {Vitale}, \citenamefont {Evans},
  \citenamefont {Shoemaker}, \citenamefont {Barr}, \citenamefont {Hammond},
  \citenamefont {Hild}, \citenamefont {Hough}, \citenamefont {Huttner},
  \citenamefont {Rowan}, \citenamefont {Sorazu}, \citenamefont {Carbone},
  \citenamefont {Freise}, \citenamefont {Mow-Lowry}, \citenamefont {Dooley},
  \citenamefont {Fulda}, \citenamefont {Grote},\ and\ \citenamefont
  {Sigg}}]{Yu_5Hz_2018}%
  \BibitemOpen
  \bibfield  {author} {\bibinfo {author} {\bibfnamefont {H.}~\bibnamefont
  {Yu}}, \bibinfo {author} {\bibfnamefont {D.}~\bibnamefont {Martynov}},
  \bibinfo {author} {\bibfnamefont {S.}~\bibnamefont {Vitale}}, \bibinfo
  {author} {\bibfnamefont {M.}~\bibnamefont {Evans}}, \bibinfo {author}
  {\bibfnamefont {D.}~\bibnamefont {Shoemaker}}, \bibinfo {author}
  {\bibfnamefont {B.}~\bibnamefont {Barr}}, \bibinfo {author} {\bibfnamefont
  {G.}~\bibnamefont {Hammond}}, \bibinfo {author} {\bibfnamefont
  {S.}~\bibnamefont {Hild}}, \bibinfo {author} {\bibfnamefont {J.}~\bibnamefont
  {Hough}}, \bibinfo {author} {\bibfnamefont {S.}~\bibnamefont {Huttner}},
  \bibinfo {author} {\bibfnamefont {S.}~\bibnamefont {Rowan}}, \bibinfo
  {author} {\bibfnamefont {B.}~\bibnamefont {Sorazu}}, \bibinfo {author}
  {\bibfnamefont {L.}~\bibnamefont {Carbone}}, \bibinfo {author} {\bibfnamefont
  {A.}~\bibnamefont {Freise}}, \bibinfo {author} {\bibfnamefont
  {C.}~\bibnamefont {Mow-Lowry}}, \bibinfo {author} {\bibfnamefont {K.~L.}\
  \bibnamefont {Dooley}}, \bibinfo {author} {\bibfnamefont {P.}~\bibnamefont
  {Fulda}}, \bibinfo {author} {\bibfnamefont {H.}~\bibnamefont {Grote}}, \ and\
  \bibinfo {author} {\bibfnamefont {D.}~\bibnamefont {Sigg}},\ }\href {\doibase
  10.1103/PhysRevLett.120.141102} {\bibfield  {journal} {\bibinfo  {journal}
  {Phys. Rev. Lett.}\ }\textbf {\bibinfo {volume} {120}},\ \bibinfo {pages}
  {141102} (\bibinfo {year} {2018})}\BibitemShut {NoStop}%
\bibitem [{\citenamefont {{Evans}}\ \emph {et~al.}(2015)\citenamefont
  {{Evans}}, \citenamefont {{Gras}}, \citenamefont {{Fritschel}}, \citenamefont
  {{Miller}}, \citenamefont {{Barsotti}}, \citenamefont {{Martynov}},
  \citenamefont {{Brooks}}, \citenamefont {{Coyne}} \emph
  {et~al.}}]{Evans_PI_15}%
  \BibitemOpen
  \bibfield  {author} {\bibinfo {author} {\bibfnamefont {M.}~\bibnamefont
  {{Evans}}}, \bibinfo {author} {\bibfnamefont {S.}~\bibnamefont {{Gras}}},
  \bibinfo {author} {\bibfnamefont {P.}~\bibnamefont {{Fritschel}}}, \bibinfo
  {author} {\bibfnamefont {J.}~\bibnamefont {{Miller}}}, \bibinfo {author}
  {\bibfnamefont {L.}~\bibnamefont {{Barsotti}}}, \bibinfo {author}
  {\bibfnamefont {D.}~\bibnamefont {{Martynov}}}, \bibinfo {author}
  {\bibfnamefont {A.}~\bibnamefont {{Brooks}}}, \bibinfo {author} {\bibnamefont
  {{Coyne}}},  \emph {et~al.},\ }\href
  {http://journals.aps.org/prl/abstract/10.1103/PhysRevLett.114.161102}
  {\bibfield  {journal} {\bibinfo  {journal} {Physical Review Letters}\
  }\textbf {\bibinfo {volume} {114}} (\bibinfo {year} {2015})}\BibitemShut
  {NoStop}%
\bibitem [{\citenamefont {Blair}\ \emph {et~al.}(2017)\citenamefont {Blair},
  \citenamefont {Gras}, \citenamefont {Abbott}, \citenamefont {Aston},
  \citenamefont {Betzwieser}, \citenamefont {Blair}, \citenamefont {DeRosa},
  \citenamefont {Evans}, \citenamefont {Frolov} \emph
  {et~al.}}]{Blair_PI_2017}%
  \BibitemOpen
  \bibfield  {author} {\bibinfo {author} {\bibfnamefont {C.}~\bibnamefont
  {Blair}}, \bibinfo {author} {\bibfnamefont {S.}~\bibnamefont {Gras}},
  \bibinfo {author} {\bibfnamefont {R.}~\bibnamefont {Abbott}}, \bibinfo
  {author} {\bibfnamefont {S.}~\bibnamefont {Aston}}, \bibinfo {author}
  {\bibfnamefont {J.}~\bibnamefont {Betzwieser}}, \bibinfo {author}
  {\bibfnamefont {D.}~\bibnamefont {Blair}}, \bibinfo {author} {\bibfnamefont
  {R.}~\bibnamefont {DeRosa}}, \bibinfo {author} {\bibfnamefont
  {M.}~\bibnamefont {Evans}}, \bibinfo {author} {\bibfnamefont
  {V.}~\bibnamefont {Frolov}},  \emph {et~al.} (\bibinfo {collaboration} {LSC
  Instrument Authors}),\ }\href {\doibase 10.1103/PhysRevLett.118.151102}
  {\bibfield  {journal} {\bibinfo  {journal} {Phys. Rev. Lett.}\ }\textbf
  {\bibinfo {volume} {118}},\ \bibinfo {pages} {151102} (\bibinfo {year}
  {2017})}\BibitemShut {NoStop}%
\bibitem [{\citenamefont {Gras}\ \emph {et~al.}(2015)\citenamefont {Gras},
  \citenamefont {Fritschel}, \citenamefont {Barsotti},\ and\ \citenamefont
  {Evans}}]{Gras_PI_2015}%
  \BibitemOpen
  \bibfield  {author} {\bibinfo {author} {\bibfnamefont {S.}~\bibnamefont
  {Gras}}, \bibinfo {author} {\bibfnamefont {P.}~\bibnamefont {Fritschel}},
  \bibinfo {author} {\bibfnamefont {L.}~\bibnamefont {Barsotti}}, \ and\
  \bibinfo {author} {\bibfnamefont {M.}~\bibnamefont {Evans}},\ }\href
  {\doibase 10.1103/PhysRevD.92.082001} {\bibfield  {journal} {\bibinfo
  {journal} {Phys. Rev. D}\ }\textbf {\bibinfo {volume} {92}},\ \bibinfo
  {pages} {082001} (\bibinfo {year} {2015})}\BibitemShut {NoStop}%
\bibitem [{\citenamefont {Chen}\ \emph {et~al.}(2017)\citenamefont {Chen},
  \citenamefont {Holz}, \citenamefont {Miller}, \citenamefont {Evans},
  \citenamefont {Vitale},\ and\ \citenamefont {Creighton}}]{Chen_2017}%
  \BibitemOpen
  \bibfield  {author} {\bibinfo {author} {\bibfnamefont {H.-Y.}\ \bibnamefont
  {Chen}}, \bibinfo {author} {\bibfnamefont {D.~E.}\ \bibnamefont {Holz}},
  \bibinfo {author} {\bibfnamefont {J.}~\bibnamefont {Miller}}, \bibinfo
  {author} {\bibfnamefont {M.}~\bibnamefont {Evans}}, \bibinfo {author}
  {\bibfnamefont {S.}~\bibnamefont {Vitale}}, \ and\ \bibinfo {author}
  {\bibfnamefont {J.}~\bibnamefont {Creighton}},\ }\href@noop {} {\  (\bibinfo
  {year} {2017})},\ \Eprint {http://arxiv.org/abs/1709.08079} {arXiv:1709.08079
  [astro-ph.CO]} \BibitemShut {NoStop}%
\bibitem [{\citenamefont {Thrane}\ and\ \citenamefont
  {Romano}(2013)}]{Thrane_Stochastic_2013}%
  \BibitemOpen
  \bibfield  {author} {\bibinfo {author} {\bibfnamefont {E.}~\bibnamefont
  {Thrane}}\ and\ \bibinfo {author} {\bibfnamefont {J.~D.}\ \bibnamefont
  {Romano}},\ }\href {\doibase 10.1103/PhysRevD.88.124032} {\bibfield
  {journal} {\bibinfo  {journal} {Phys. Rev. D}\ }\textbf {\bibinfo {volume}
  {88}},\ \bibinfo {pages} {124032} (\bibinfo {year} {2013})}\BibitemShut
  {NoStop}%
\bibitem [{\citenamefont {Arvanitaki}\ \emph {et~al.}(2015)\citenamefont
  {Arvanitaki}, \citenamefont {Baryakhtar},\ and\ \citenamefont
  {Huang}}]{Arvanitaki:2014wva}%
  \BibitemOpen
  \bibfield  {author} {\bibinfo {author} {\bibfnamefont {A.}~\bibnamefont
  {Arvanitaki}}, \bibinfo {author} {\bibfnamefont {M.}~\bibnamefont
  {Baryakhtar}}, \ and\ \bibinfo {author} {\bibfnamefont {X.}~\bibnamefont
  {Huang}},\ }\href {\doibase 10.1103/PhysRevD.91.084011} {\bibfield  {journal}
  {\bibinfo  {journal} {Phys. Rev.}\ }\textbf {\bibinfo {volume} {D91}},\
  \bibinfo {pages} {084011} (\bibinfo {year} {2015})},\ \Eprint
  {http://arxiv.org/abs/1411.2263} {arXiv:1411.2263 [hep-ph]} \BibitemShut
  {NoStop}%
\bibitem [{\citenamefont {Arvanitaki}\ \emph {et~al.}(2017)\citenamefont
  {Arvanitaki}, \citenamefont {Baryakhtar}, \citenamefont {Dimopoulos},
  \citenamefont {Dubovsky},\ and\ \citenamefont
  {Lasenby}}]{Arvanitaki:2016qwi}%
  \BibitemOpen
  \bibfield  {author} {\bibinfo {author} {\bibfnamefont {A.}~\bibnamefont
  {Arvanitaki}}, \bibinfo {author} {\bibfnamefont {M.}~\bibnamefont
  {Baryakhtar}}, \bibinfo {author} {\bibfnamefont {S.}~\bibnamefont
  {Dimopoulos}}, \bibinfo {author} {\bibfnamefont {S.}~\bibnamefont
  {Dubovsky}}, \ and\ \bibinfo {author} {\bibfnamefont {R.}~\bibnamefont
  {Lasenby}},\ }\href {\doibase 10.1103/PhysRevD.95.043001} {\bibfield
  {journal} {\bibinfo  {journal} {Phys. Rev.}\ }\textbf {\bibinfo {volume}
  {D95}},\ \bibinfo {pages} {043001} (\bibinfo {year} {2017})},\ \Eprint
  {http://arxiv.org/abs/1604.03958} {arXiv:1604.03958 [hep-ph]} \BibitemShut
  {NoStop}%
\bibitem [{\citenamefont {Baryakhtar}\ \emph {et~al.}(2017)\citenamefont
  {Baryakhtar}, \citenamefont {Lasenby},\ and\ \citenamefont
  {Teo}}]{Baryakhtar:2017ngi}%
  \BibitemOpen
  \bibfield  {author} {\bibinfo {author} {\bibfnamefont {M.}~\bibnamefont
  {Baryakhtar}}, \bibinfo {author} {\bibfnamefont {R.}~\bibnamefont {Lasenby}},
  \ and\ \bibinfo {author} {\bibfnamefont {M.}~\bibnamefont {Teo}},\ }\href
  {\doibase 10.1103/PhysRevD.96.035019} {\bibfield  {journal} {\bibinfo
  {journal} {Phys. Rev.}\ }\textbf {\bibinfo {volume} {D96}},\ \bibinfo {pages}
  {035019} (\bibinfo {year} {2017})},\ \Eprint
  {http://arxiv.org/abs/1704.05081} {arXiv:1704.05081 [hep-ph]} \BibitemShut
  {NoStop}%
\bibitem [{\citenamefont {Brito}\ \emph
  {et~al.}(2017{\natexlab{a}})\citenamefont {Brito}, \citenamefont {Ghosh},
  \citenamefont {Barausse}, \citenamefont {Berti}, \citenamefont {Cardoso},
  \citenamefont {Dvorkin}, \citenamefont {Klein},\ and\ \citenamefont
  {Pani}}]{Brito:2017zvb}%
  \BibitemOpen
  \bibfield  {author} {\bibinfo {author} {\bibfnamefont {R.}~\bibnamefont
  {Brito}}, \bibinfo {author} {\bibfnamefont {S.}~\bibnamefont {Ghosh}},
  \bibinfo {author} {\bibfnamefont {E.}~\bibnamefont {Barausse}}, \bibinfo
  {author} {\bibfnamefont {E.}~\bibnamefont {Berti}}, \bibinfo {author}
  {\bibfnamefont {V.}~\bibnamefont {Cardoso}}, \bibinfo {author} {\bibfnamefont
  {I.}~\bibnamefont {Dvorkin}}, \bibinfo {author} {\bibfnamefont
  {A.}~\bibnamefont {Klein}}, \ and\ \bibinfo {author} {\bibfnamefont
  {P.}~\bibnamefont {Pani}},\ }\href {\doibase 10.1103/PhysRevD.96.064050}
  {\bibfield  {journal} {\bibinfo  {journal} {Phys. Rev.}\ }\textbf {\bibinfo
  {volume} {D96}},\ \bibinfo {pages} {064050} (\bibinfo {year}
  {2017}{\natexlab{a}})},\ \Eprint {http://arxiv.org/abs/1706.06311}
  {arXiv:1706.06311 [gr-qc]} \BibitemShut {NoStop}%
\bibitem [{\citenamefont {Brito}\ \emph
  {et~al.}(2017{\natexlab{b}})\citenamefont {Brito}, \citenamefont {Ghosh},
  \citenamefont {Barausse}, \citenamefont {Berti}, \citenamefont {Cardoso},
  \citenamefont {Dvorkin}, \citenamefont {Klein},\ and\ \citenamefont
  {Pani}}]{Brito:2017wnc}%
  \BibitemOpen
  \bibfield  {author} {\bibinfo {author} {\bibfnamefont {R.}~\bibnamefont
  {Brito}}, \bibinfo {author} {\bibfnamefont {S.}~\bibnamefont {Ghosh}},
  \bibinfo {author} {\bibfnamefont {E.}~\bibnamefont {Barausse}}, \bibinfo
  {author} {\bibfnamefont {E.}~\bibnamefont {Berti}}, \bibinfo {author}
  {\bibfnamefont {V.}~\bibnamefont {Cardoso}}, \bibinfo {author} {\bibfnamefont
  {I.}~\bibnamefont {Dvorkin}}, \bibinfo {author} {\bibfnamefont
  {A.}~\bibnamefont {Klein}}, \ and\ \bibinfo {author} {\bibfnamefont
  {P.}~\bibnamefont {Pani}},\ }\href {\doibase 10.1103/PhysRevLett.119.131101}
  {\bibfield  {journal} {\bibinfo  {journal} {Phys. Rev. Lett.}\ }\textbf
  {\bibinfo {volume} {119}},\ \bibinfo {pages} {131101} (\bibinfo {year}
  {2017}{\natexlab{b}})},\ \Eprint {http://arxiv.org/abs/1706.05097}
  {arXiv:1706.05097 [gr-qc]} \BibitemShut {NoStop}%
\bibitem [{\citenamefont {East}(2017)}]{East:2017mrj}%
  \BibitemOpen
  \bibfield  {author} {\bibinfo {author} {\bibfnamefont {W.~E.}\ \bibnamefont
  {East}},\ }\href {\doibase 10.1103/PhysRevD.96.024004} {\bibfield  {journal}
  {\bibinfo  {journal} {Phys. Rev.}\ }\textbf {\bibinfo {volume} {D96}},\
  \bibinfo {pages} {024004} (\bibinfo {year} {2017})},\ \Eprint
  {http://arxiv.org/abs/1705.01544} {arXiv:1705.01544 [gr-qc]} \BibitemShut
  {NoStop}%
\bibitem [{\citenamefont {East}(2018)}]{East:2018glu}%
  \BibitemOpen
  \bibfield  {author} {\bibinfo {author} {\bibfnamefont {W.~E.}\ \bibnamefont
  {East}},\ }\href {\doibase 10.1103/PhysRevLett.121.131104} {\bibfield
  {journal} {\bibinfo  {journal} {Phys. Rev. Lett.}\ }\textbf {\bibinfo
  {volume} {121}},\ \bibinfo {pages} {131104} (\bibinfo {year} {2018})},\
  \Eprint {http://arxiv.org/abs/1807.00043} {arXiv:1807.00043 [gr-qc]}
  \BibitemShut {NoStop}%
\bibitem [{\citenamefont {Arvanitaki}\ and\ \citenamefont
  {Dubovsky}(2011)}]{Arvanitaki:2010sy}%
  \BibitemOpen
  \bibfield  {author} {\bibinfo {author} {\bibfnamefont {A.}~\bibnamefont
  {Arvanitaki}}\ and\ \bibinfo {author} {\bibfnamefont {S.}~\bibnamefont
  {Dubovsky}},\ }\href {\doibase 10.1103/PhysRevD.83.044026} {\bibfield
  {journal} {\bibinfo  {journal} {Phys. Rev.}\ }\textbf {\bibinfo {volume}
  {D83}},\ \bibinfo {pages} {044026} (\bibinfo {year} {2011})},\ \Eprint
  {http://arxiv.org/abs/1004.3558} {arXiv:1004.3558 [hep-th]} \BibitemShut
  {NoStop}%
\bibitem [{\citenamefont {Dolan}(2013)}]{Dolan:2012yt}%
  \BibitemOpen
  \bibfield  {author} {\bibinfo {author} {\bibfnamefont {S.~R.}\ \bibnamefont
  {Dolan}},\ }\href {\doibase 10.1103/PhysRevD.87.124026} {\bibfield  {journal}
  {\bibinfo  {journal} {Phys. Rev.}\ }\textbf {\bibinfo {volume} {D87}},\
  \bibinfo {pages} {124026} (\bibinfo {year} {2013})},\ \Eprint
  {http://arxiv.org/abs/1212.1477} {arXiv:1212.1477 [gr-qc]} \BibitemShut
  {NoStop}%
\bibitem [{\citenamefont {{Brito}}\ \emph {et~al.}(2015)\citenamefont
  {{Brito}}, \citenamefont {{Cardoso}},\ and\ \citenamefont
  {{Pani}}}]{2015CQGra..32m4001B}%
  \BibitemOpen
  \bibfield  {author} {\bibinfo {author} {\bibfnamefont {R.}~\bibnamefont
  {{Brito}}}, \bibinfo {author} {\bibfnamefont {V.}~\bibnamefont {{Cardoso}}},
  \ and\ \bibinfo {author} {\bibfnamefont {P.}~\bibnamefont {{Pani}}},\ }\href
  {\doibase 10.1088/0264-9381/32/13/134001} {\bibfield  {journal} {\bibinfo
  {journal} {Classical and Quantum Gravity}\ }\textbf {\bibinfo {volume}
  {32}},\ \bibinfo {eid} {134001} (\bibinfo {year} {2015})},\ \Eprint
  {http://arxiv.org/abs/1411.0686} {arXiv:1411.0686 [gr-qc]} \BibitemShut
  {NoStop}%
\bibitem [{\citenamefont {East}\ and\ \citenamefont
  {Pretorius}(2017)}]{East:2017ovw}%
  \BibitemOpen
  \bibfield  {author} {\bibinfo {author} {\bibfnamefont {W.~E.}\ \bibnamefont
  {East}}\ and\ \bibinfo {author} {\bibfnamefont {F.}~\bibnamefont
  {Pretorius}},\ }\href {\doibase 10.1103/PhysRevLett.119.041101} {\bibfield
  {journal} {\bibinfo  {journal} {Phys. Rev. Lett.}\ }\textbf {\bibinfo
  {volume} {119}},\ \bibinfo {pages} {041101} (\bibinfo {year} {2017})},\
  \Eprint {http://arxiv.org/abs/1704.04791} {arXiv:1704.04791 [gr-qc]}
  \BibitemShut {NoStop}%
\bibitem [{\citenamefont {Weinberg}(1978)}]{Weinberg:1977ma}%
  \BibitemOpen
  \bibfield  {author} {\bibinfo {author} {\bibfnamefont {S.}~\bibnamefont
  {Weinberg}},\ }\href {\doibase 10.1103/PhysRevLett.40.223} {\bibfield
  {journal} {\bibinfo  {journal} {Phys.Rev.Lett.}\ }\textbf {\bibinfo {volume}
  {40}},\ \bibinfo {pages} {223} (\bibinfo {year} {1978})}\BibitemShut
  {NoStop}%
\bibitem [{\citenamefont {Arvanitaki}\ \emph {et~al.}(2010)\citenamefont
  {Arvanitaki}, \citenamefont {Dimopoulos}, \citenamefont {Dubovsky},
  \citenamefont {Kaloper},\ and\ \citenamefont
  {March-Russell}}]{Arvanitaki:2009fg}%
  \BibitemOpen
  \bibfield  {author} {\bibinfo {author} {\bibfnamefont {A.}~\bibnamefont
  {Arvanitaki}}, \bibinfo {author} {\bibfnamefont {S.}~\bibnamefont
  {Dimopoulos}}, \bibinfo {author} {\bibfnamefont {S.}~\bibnamefont
  {Dubovsky}}, \bibinfo {author} {\bibfnamefont {N.}~\bibnamefont {Kaloper}}, \
  and\ \bibinfo {author} {\bibfnamefont {J.}~\bibnamefont {March-Russell}},\
  }\href {\doibase 10.1103/PhysRevD.81.123530} {\bibfield  {journal} {\bibinfo
  {journal} {Phys. Rev.}\ }\textbf {\bibinfo {volume} {D81}},\ \bibinfo {pages}
  {123530} (\bibinfo {year} {2010})},\ \Eprint {http://arxiv.org/abs/0905.4720}
  {arXiv:0905.4720 [hep-th]} \BibitemShut {NoStop}%
\bibitem [{\citenamefont {Holdom}(1986)}]{Holdom:1985ag}%
  \BibitemOpen
  \bibfield  {author} {\bibinfo {author} {\bibfnamefont {B.}~\bibnamefont
  {Holdom}},\ }\href {\doibase 10.1016/0370-2693(86)91377-8} {\bibfield
  {journal} {\bibinfo  {journal} {Phys. Lett.}\ }\textbf {\bibinfo {volume}
  {B166}},\ \bibinfo {pages} {196} (\bibinfo {year} {1986})}\BibitemShut
  {NoStop}%
\bibitem [{\citenamefont {Cicoli}\ \emph {et~al.}(2011)\citenamefont {Cicoli},
  \citenamefont {Goodsell}, \citenamefont {Jaeckel},\ and\ \citenamefont
  {Ringwald}}]{Cicoli:2011yh}%
  \BibitemOpen
  \bibfield  {author} {\bibinfo {author} {\bibfnamefont {M.}~\bibnamefont
  {Cicoli}}, \bibinfo {author} {\bibfnamefont {M.}~\bibnamefont {Goodsell}},
  \bibinfo {author} {\bibfnamefont {J.}~\bibnamefont {Jaeckel}}, \ and\
  \bibinfo {author} {\bibfnamefont {A.}~\bibnamefont {Ringwald}},\ }\href
  {\doibase 10.1007/JHEP07(2011)114} {\bibfield  {journal} {\bibinfo  {journal}
  {JHEP}\ }\textbf {\bibinfo {volume} {07}},\ \bibinfo {pages} {114} (\bibinfo
  {year} {2011})},\ \Eprint {http://arxiv.org/abs/1103.3705} {arXiv:1103.3705
  [hep-th]} \BibitemShut {NoStop}%
\bibitem [{\citenamefont {Isi}\ \emph {et~al.}(2018)\citenamefont {Isi},
  \citenamefont {Sun}, \citenamefont {Brito},\ and\ \citenamefont
  {Melatos}}]{Isi:2018pzk}%
  \BibitemOpen
  \bibfield  {author} {\bibinfo {author} {\bibfnamefont {M.}~\bibnamefont
  {Isi}}, \bibinfo {author} {\bibfnamefont {L.}~\bibnamefont {Sun}}, \bibinfo
  {author} {\bibfnamefont {R.}~\bibnamefont {Brito}}, \ and\ \bibinfo {author}
  {\bibfnamefont {A.}~\bibnamefont {Melatos}},\ }\href@noop {} {\  (\bibinfo
  {year} {2018})},\ \Eprint {http://arxiv.org/abs/1810.03812} {arXiv:1810.03812
  [gr-qc]} \BibitemShut {NoStop}%
\bibitem [{\citenamefont {Goncharov}\ and\ \citenamefont
  {Thrane}(2018)}]{Goncharov:2018ufi}%
  \BibitemOpen
  \bibfield  {author} {\bibinfo {author} {\bibfnamefont {B.}~\bibnamefont
  {Goncharov}}\ and\ \bibinfo {author} {\bibfnamefont {E.}~\bibnamefont
  {Thrane}},\ }\href@noop {} {\  (\bibinfo {year} {2018})},\ \Eprint
  {http://arxiv.org/abs/1805.03761} {arXiv:1805.03761 [astro-ph.IM]}
  \BibitemShut {NoStop}%
\bibitem [{\citenamefont {Pierce}\ \emph {et~al.}(2018)\citenamefont {Pierce},
  \citenamefont {Riles},\ and\ \citenamefont {Zhao}}]{Pierce:2018xmy}%
  \BibitemOpen
  \bibfield  {author} {\bibinfo {author} {\bibfnamefont {A.}~\bibnamefont
  {Pierce}}, \bibinfo {author} {\bibfnamefont {K.}~\bibnamefont {Riles}}, \
  and\ \bibinfo {author} {\bibfnamefont {Y.}~\bibnamefont {Zhao}},\ }\href@noop
  {} {\  (\bibinfo {year} {2018})},\ \Eprint {http://arxiv.org/abs/1801.10161}
  {arXiv:1801.10161 [hep-ph]} \BibitemShut {NoStop}%
\bibitem [{\citenamefont {Yoshino}\ and\ \citenamefont
  {Kodama}(2015)}]{Yoshino:2015nsa}%
  \BibitemOpen
  \bibfield  {author} {\bibinfo {author} {\bibfnamefont {H.}~\bibnamefont
  {Yoshino}}\ and\ \bibinfo {author} {\bibfnamefont {H.}~\bibnamefont
  {Kodama}},\ }\href {\doibase 10.1088/0264-9381/32/21/214001} {\bibfield
  {journal} {\bibinfo  {journal} {Class. Quant. Grav.}\ }\textbf {\bibinfo
  {volume} {32}},\ \bibinfo {pages} {214001} (\bibinfo {year} {2015})},\
  \Eprint {http://arxiv.org/abs/1505.00714} {arXiv:1505.00714 [gr-qc]}
  \BibitemShut {NoStop}%
\bibitem [{\citenamefont {Jackiw}\ and\ \citenamefont
  {Pi}(2003)}]{Jackiw:2003pm}%
  \BibitemOpen
  \bibfield  {author} {\bibinfo {author} {\bibfnamefont {R.}~\bibnamefont
  {Jackiw}}\ and\ \bibinfo {author} {\bibfnamefont {S.~Y.}\ \bibnamefont
  {Pi}},\ }\href {\doibase 10.1103/PhysRevD.68.104012} {\bibfield  {journal}
  {\bibinfo  {journal} {Phys. Rev.}\ }\textbf {\bibinfo {volume} {D68}},\
  \bibinfo {pages} {104012} (\bibinfo {year} {2003})},\ \Eprint
  {http://arxiv.org/abs/gr-qc/0308071} {arXiv:gr-qc/0308071 [gr-qc]}
  \BibitemShut {NoStop}%
\bibitem [{\citenamefont {Moura}\ and\ \citenamefont
  {Schiappa}(2007)}]{Moura:2006pz}%
  \BibitemOpen
  \bibfield  {author} {\bibinfo {author} {\bibfnamefont {F.}~\bibnamefont
  {Moura}}\ and\ \bibinfo {author} {\bibfnamefont {R.}~\bibnamefont
  {Schiappa}},\ }\href {\doibase 10.1088/0264-9381/24/2/006} {\bibfield
  {journal} {\bibinfo  {journal} {Class. Quant. Grav.}\ }\textbf {\bibinfo
  {volume} {24}},\ \bibinfo {pages} {361} (\bibinfo {year} {2007})},\ \Eprint
  {http://arxiv.org/abs/hep-th/0605001} {arXiv:hep-th/0605001 [hep-th]}
  \BibitemShut {NoStop}%
\bibitem [{\citenamefont {Ghosh}\ \emph {et~al.}(2016)\citenamefont {Ghosh}
  \emph {et~al.}}]{Ghosh:2016qgn}%
  \BibitemOpen
  \bibfield  {author} {\bibinfo {author} {\bibfnamefont {A.}~\bibnamefont
  {Ghosh}} \emph {et~al.},\ }\href {\doibase 10.1103/PhysRevD.94.021101}
  {\bibfield  {journal} {\bibinfo  {journal} {Phys. Rev.}\ }\textbf {\bibinfo
  {volume} {D94}},\ \bibinfo {pages} {021101} (\bibinfo {year} {2016})},\
  \Eprint {http://arxiv.org/abs/1602.02453} {arXiv:1602.02453 [gr-qc]}
  \BibitemShut {NoStop}%
\end{thebibliography}%

\end{document}